\documentclass[letterpaper]{article} 
\usepackage{aaai25}  
\usepackage{times}  
\usepackage{helvet}  
\usepackage{courier}  
\usepackage[hyphens]{url}  
\usepackage{graphicx} 
\usepackage{natbib}  
\usepackage{caption} 
\usepackage{algorithm}
\usepackage{algorithmic}

\usepackage{subcaption}
\usepackage{booktabs}

\usepackage{newfloat}
\usepackage{listings}
\DeclareCaptionStyle{ruled}{labelfont=normalfont,labelsep=colon,strut=off} 
\floatstyle{ruled}
\newfloat{listing}{tb}{lst}{}
\floatname{listing}{Listing}
\title{Expressive Range Characterization of Open Text-to-Audio Models}

\author{
    Jonathan Morse\textsuperscript{\rm 1},
    Azadeh Naderi\textsuperscript{\rm 1},
    Swen Gaudl\textsuperscript{\rm 2},\\
    Mark Cartwright\textsuperscript{\rm 1},
    Amy K.\ Hoover\textsuperscript{\rm 1},
    Mark J.\ Nelson\textsuperscript{\rm 3}
}
\affiliations{
    \textsuperscript{\rm 1}New Jersey Institute of Technology, Newark, New Jersey, USA\\
    \textsuperscript{\rm 2}University of Gothenburg, Gothenburg, Sweden\\
    \textsuperscript{\rm 3}American University, Washington, D.C., USA\\
    \{jtm85,an57,mark.cartwright,ahoover\}@njit.edu, swen.gaudl@ait.gu.se, mnelson@american.edu
}

\begin{document}

\maketitle

\begin{abstract}

Text-to-audio models are a type of generative model that produces audio output in response to a given textual prompt. Although level generators and the properties of the functional content that they create (e.g., playability) dominate most discourse in procedurally generated content (PCG), games that emotionally resonate with players tend to weave together a range of creative and multimodal content (e.g., music, sounds, visuals, narrative tone), and multimodal models have begun seeing at least experimental use for this purpose. However, it remains unclear what exactly such models generate, and with what degree of variability and fidelity: \emph{audio} is an extremely broad class of output for a generative system to target.

Within the PCG community, expressive range analysis (ERA) has been used as a quantitative way to characterize generators' output space, especially for level generators. This paper adapts ERA to text-to-audio models, making the analysis tractable by looking at the expressive range of outputs for specific, fixed prompts. Experiments are conducted by prompting the models with several standardized prompts derived from the Environmental Sound Classification (ESC-50) dataset. The resulting audio is analyzed along key acoustic dimensions (e.g., pitch, loudness, and timbre). More broadly, this paper offers a framework for ERA-based exploratory evaluation of generative audio models. 
\end{abstract}

\section{Introduction}

\begin{figure}
\includegraphics[scale=.4]{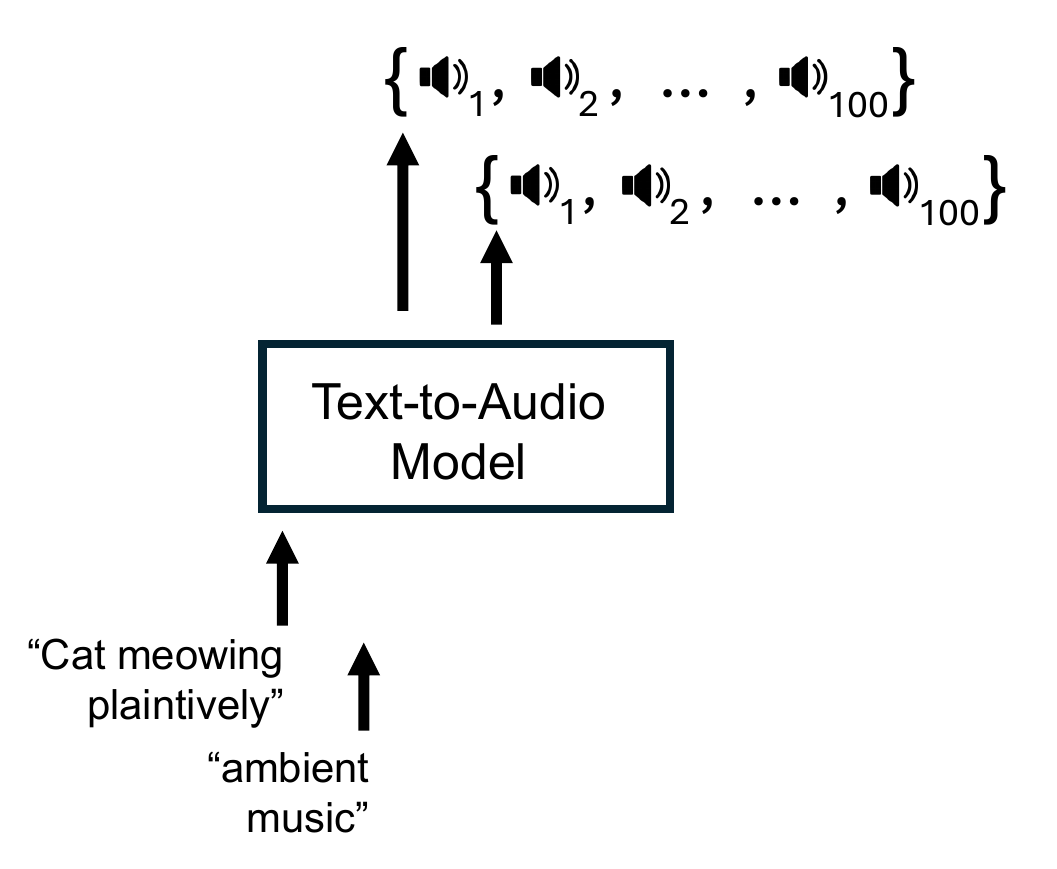}
\caption{\textbf{Querying a Text-to-Audio Model}. \label{fig:text-to-audio} Text-to-audio models operate by taking text prompts as input, mapping them to a joint embedding space that relates text and audio, and generating audio clips as output. This schematic illustrates generating 100 audio outputs each for the prompts ``cat meowing plaintively'' and ``ambient music''. The audio for this and all other figures in the paper is available at \url{https://doi.org/10.5281/zenodo.16998750}.}
\end{figure}

Joint embedding spaces that map multi-modal content into a shared coordinate system -- such as text and images (CLIP; \cite{radford:pmlr21}) or text and audio (CLAP; \cite{wu:icasp23, elizalde:icasp23}) -- enable generative models to procedurally generate domain-specific content aligned with the natural language vocabulary a game designer would need to explore them. There are a growing number of \emph{text-to-audio} models available in this category. After some initial setup, game designers could add audio samples from models like StableAudioOpen \cite{stableaudio} or MMAudio \cite{mmaudio} that can be directly queried for audio, which is nearly instantly generated to suit their needs. Figure~\ref{fig:text-to-audio} schematically illustrates the prompts ``cat meowing plaintively'' and ``ambient music'' being sent to a text-to-audio model, resulting in 100 generated examples of each.

A key question in evaluating generative systems is the range of outputs that can be generated. Do we always get essentially the \emph{same} plaintive cat meow, or many interesting yet realistic variations? And what do we mean by ``interesting'' variations? While one might hope for some kind of ``best'' generator, there is a sense in some creative communities that perhaps ``diversity is all you need'' \cite{eysenbach:arxiv18,lehman:alife08}, with the goal being a diverse array of high-quality content that differs along dimensions of diversity specified by the content-creator \cite{lehman:gecco11,mouret:arxiv15,fontaine:gecco20,fontaine:aaai21,meyerson:telo2024}. 

But how diverse are the outputs of current text-to-audio models?
Such models take their generative domain to be the very general concept of \emph{audio}. In practice, this means anything that can be recorded in a digital audio file. This flattens the rich cultural specificity of sound into a single, homogenized latent space. In addition to cat meows and ambient music, the catch-all category of \emph{audio} includes folk and world music, classical symphonies, high-school band performances, avant-garde experimental music, dramatically shattering glass from a movie, beeps and bloops from computer interface programming, human voices in many languages, and so on. It is not likely that everything is equally well represented in this flattened space of all-possible-audio.

We hypothesize that parts of the latent audio space are \emph{much} more densely modeled than others, in the sense of being able to produce a wide variety of meaningfully different examples of audio in that part of the space. In this paper, we carry out an expressive range analysis with fixed prompts and three open-source text-to-audio models -- Stable Audio Open \cite{stableaudio}, MMAudio \cite{mmaudio}, and AudioLDM 2 \cite{audioldm2-2024} -- to begin to explore that hypothesis.

Our experimental approach is to choose fixed prompts, and view a text-to-audio generator with a fixed prompt as itself a generator (a subset of the full generator). For example, Figure~\ref{fig:text-to-audio} shows two generators in this view, one for ``cat meowing plaintively'' and one for ``ambient music''. This method lets us separately probe specific parts of the large generative space that is otherwise difficult to get a handle on. We characterize the generative spaces of each of these fixed-prompt generators using expressive range analysis, with a set of prompts drawn from audio sound-effect research.

Although we also draw on audio research (specifically for the choice of sound effects to generate, and for the expressive range metrics), we situate this paper primarily in the literature on AI and interactive digital entertainment for two reasons:
\begin{itemize}
    \item Text-to-audio models are increasingly of interest to interactive media developers for sound effects, backing audio, etc., but there is not yet much analysis of what their generative output looks like.
    \item We think expressive range analysis from the procedural content generation (PCG) community \cite{smith2010analyzing} provides a particularly suitable framework for exploratory data analysis of these models as generative spaces, as we hope to explain below.
\end{itemize}

\section{Motivating Example: Thunder}

\begin{figure*}[ht]
\centering
\includegraphics[width=\textwidth]{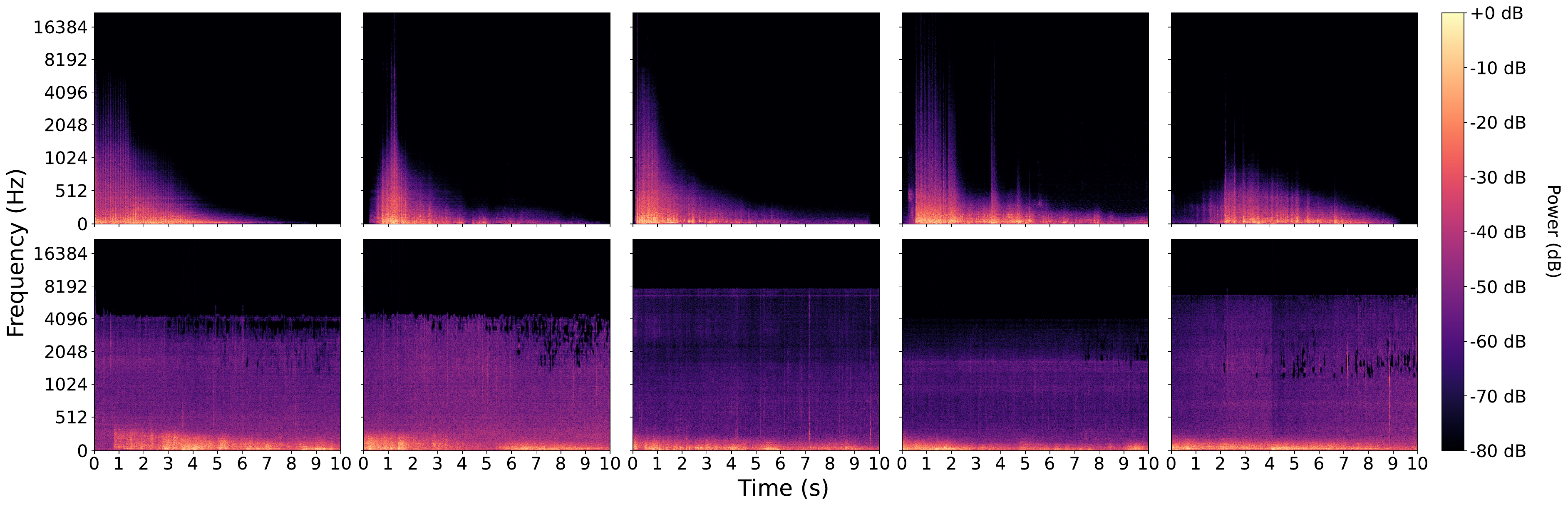}
\caption{\textbf{Spectrograms of Audio Generated with `Thunder' Prompt.} Spectrograms of five 10-second output samples for the single-word prompt \emph{thunder}. Stable Audio Open in the top row; MMAudio in the bottom row.}
\label{fig:thunder_spectrograms}
\end{figure*}

\begin{figure*}
\centering
\includegraphics[width=\textwidth]{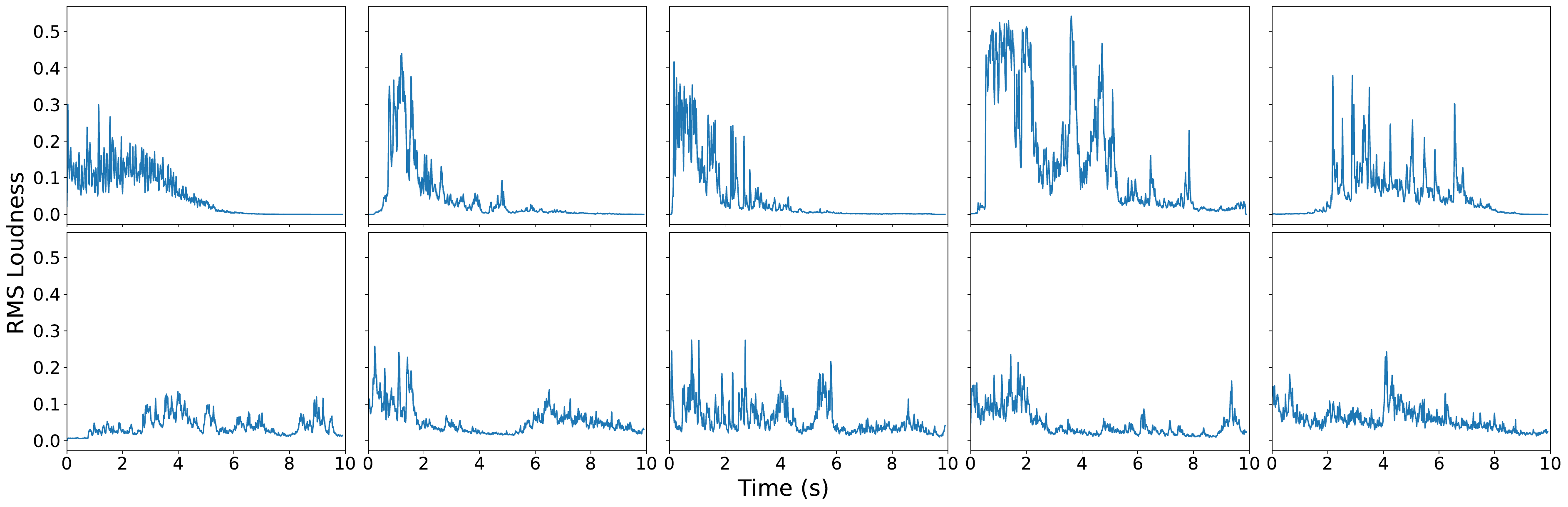}
\caption{\textbf{Loudness of Audio Generated with `Thunder' Prompt.} RMS loudness vs time plots for the same samples that were shown as spectrograms in Figure~\ref{fig:thunder_spectrograms}. Stable Audio Open in the top row; MMAudio in the bottom row.}
\label{fig:thunder_loudness}
\end{figure*}

Consider the single-word prompt \emph{thunder}. If we repeatedly feed this prompt to a text-to-audio model, each time we will get one example of generated audio that (hopefully) sounds like thunder of some kind. What we'd like to understand is how these audio outputs vary for multiple samples of a model with a fixed prompt, both within a given model and between models. Does a given model, asked simply for ``thunder'', have a characteristic type of thunder it produces by default? Are there other sounds besides thunder audible in the sample? Is there much variation between samples, or are they close variants of each other?

\subsection{Individual generative outputs}

A way to start characterizing the output is to simply generate a few sample outputs and listen to them. To visualize these audio outputs for the purpose of inclusion in a paper, Figure~\ref{fig:thunder_spectrograms} shows a total of 10 spectrograms of 10-second audio samples generated for the prompt \emph{thunder}, five each from two different models. The top row shows the outputs from Stable Audio Open, and the bottom row from MMAudio.

From both listening to the audio and looking at the spectrograms, we can make a few observations:
\begin{itemize}
    \item The Stable Audio Open samples (top row) appear to vary more across the 10 seconds than the MMAudio samples (bottom row) do.
    \item Listening to the audio samples, the reason for MMAudio's output having a wide range of frequencies (up to about 13000 Hz) fairly constant across all 10 seconds of the clips becomes clear: MMAudio also generates the sound of \emph{rain} along with the thunder, while Stable Audio Open doesn't.
    \item Stable Audio Open generates at least one distinct thunderclap in all five samples, with somewhat different timing and shape. In the first three, there is a single thunderclap near the beginning of the audio that trails off. In the fourth sample, there are several additional thunderclaps. In the fifth, the thunderclap is later and more muffled. MMAudio, by contrast, generates more muffled, rumbling thunder, only somewhat audible above the rain.
\end{itemize}

Several of these observations we made refer to the presence and timing of thunderclaps. We can look at that more explicitly. To more clearly visualize the location of the thunderclaps (if any), we can compute the total energy of the audio signal over time. This takes the full time--frequency representation given in the spectrograms, and summarizes just the energy content over time, losing the data about which energy is in which frequencies.

Figure~\ref{fig:thunder_loudness} shows the same 10 audio samples, this time with the $y$ axis giving total energy content (specifically, root-mean-square or RMS loudness). For Stable Audio Open (top row), the loudness peaks were already fairly easy to read from the spectrogram, so they simply confirm our previous observations. For MMAudio (bottom row), however, we can now more clearly see the locations of some of the muffled thunderclaps that are obscured by the rain frequencies in the spectrogram; for example, note the thunderclap about halfway through the 5th sample.

More generally, we have chosen one summary quantity (RMS loudness) and looked at its change over time. This emphasizes one of many possible changes over time that one might care about in an audio clip. In the case of thunder, the loudness peaks usually correspond to a specific semantically meaningful event, the thunderclap, though that may not be true for other types of audio.

\subsection{Expressive range analysis}

\begin{figure}
\centering
\includegraphics[width=0.9\columnwidth]{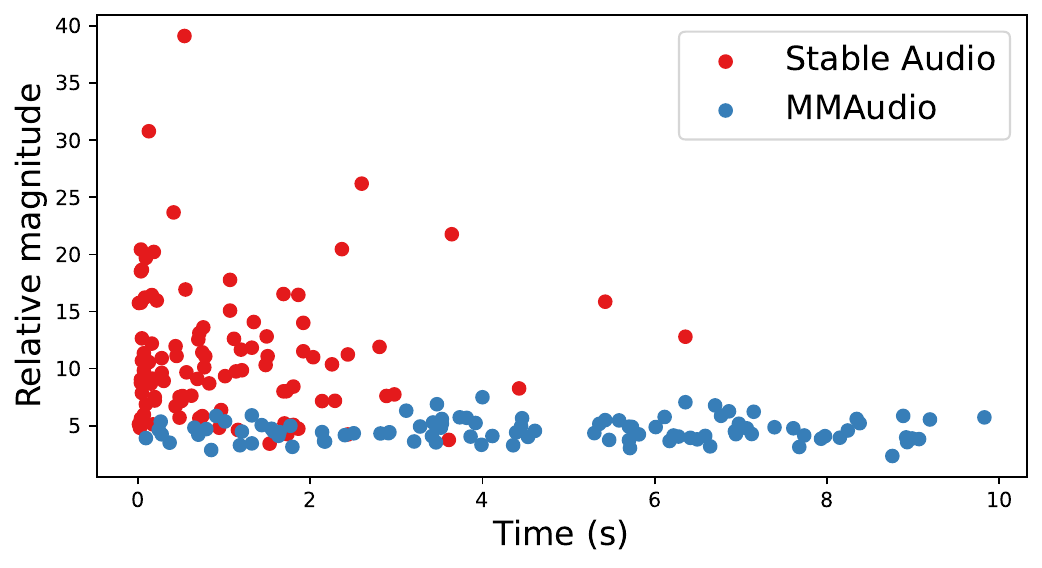}
\caption{\textbf{Expressive Range of Thunderclap Timing and Magnitude for the `Thunder' Prompt.} Relative magnitude vs.\ timing of the RMS loudness peak for 100 samples generated by each model for the prompt \emph{thunder}. Relative magnitude is defined as peak loudness divided by average loudness across the 10-second clip. From this, we can see that Stable Audio, but not MMAudio, tends to generate audio with a distinct thunderclap in the first few seconds.}
\label{fig:thunder_loudness_peak_ERA}
\end{figure}

So far, we have looked at a handful of samples, small enough to consider each one individually. To understand variation of model output, however, and especially as we consider more prompts, we will need a way to summarize a larger number of outputs in one visualization. That is where expressive range analysis comes in. If we summarize each generative output by, say, two numbers, we can plot the density of outputs on a two-dimensional plot, where each output is positioned at one point in the plot. (In general, an expressive range plot can be an $n$-dimensional plot using $n$ summary metrics, but 2D plots are most common for interpretability.)

For the specific case of thunder, we can get down to two dimensions by further summarizing the loudness-over-time plots from Figure~\ref{fig:thunder_loudness} into just into two numbers that provide one possible summary of the loudness trajectory: the location of the loudness peak (where it falls in the audio clip between 0.0 and 10.0 seconds) and the peak's relative magnitude (peak loudness divided by the clip's average loudness). Note that this kind of summarization does inevitably lose some information; for example, about secondary peaks.

From that two-number summary, we can produce an expressive range diagram: Figure~\ref{fig:thunder_loudness_peak_ERA} shows the expressive range of two text-to-audio models prompted with \emph{thunder} as the prompt, where expressive range is measured by these two metrics summarizing the loudness peaks' location and magnitude (100 samples per model).\footnote{Side note on a data presentation decision: all expressive range diagrams in this paper are shown as scatterplots, with multiple generators' outputs on each diagram, color-coded by generator. In other work, generator outputs are more often binned into 2D density histograms, with one diagram for each generator. Since we use no more than 100 samples per generator, binning the outputs is unnecessary, and plotting each point separately in a scatterplot lets us overlay multiple generators on the same diagram.}

From the expressive-range diagram, we can more confidently conclude that there is a fairly consistent difference in how the two models' generative space differs for thunder: Stable Audio Open usually (but not always) has a pronounced thunderclap in the first two seconds of the clip, while MMAudio rarely produces a pronounced thunderclap, and whatever muffled one might exist isn't clustered at any particular location in the clip.

Even with this simple example, there are a number of other things we could investigate besides the timing and loudness of the thunderclap. For example, when there is a constant noise like rain, is it always at the same pitch? Does that differ between models? For the purposes of this paper, however, we will depart from our motivating thunder example and move on to investigating a larger set of sound effects.

\section{Expressive Range With ESC-50 Labels}

\begin{figure*}[ht]
\centering
\begin{subcaptionblock}[T]{0.33\textwidth}
\includegraphics[width=\textwidth]{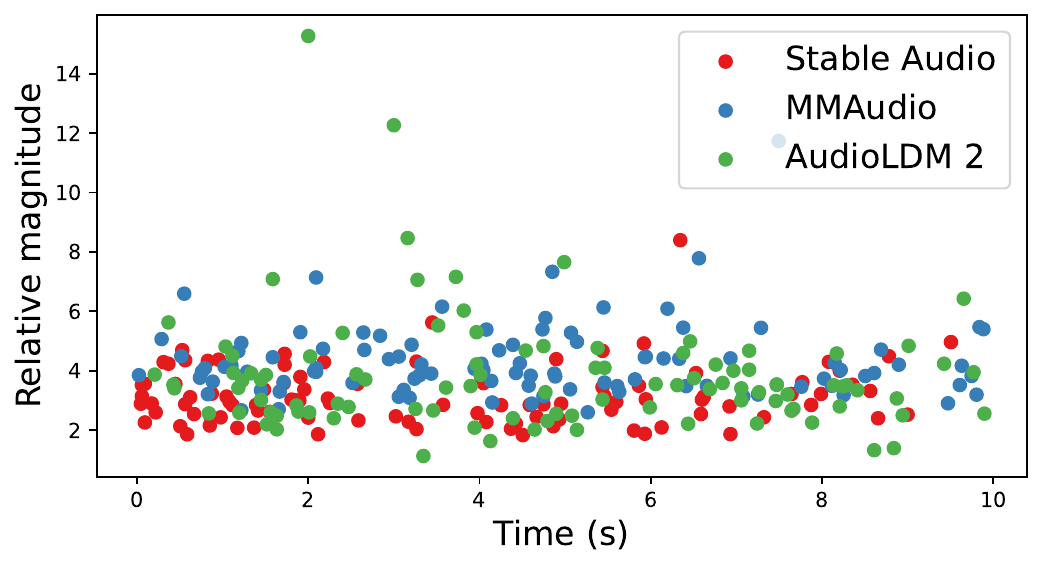}
\caption{crying baby}
\end{subcaptionblock}
\begin{subcaptionblock}[T]{0.33\textwidth}
\includegraphics[width=\textwidth]{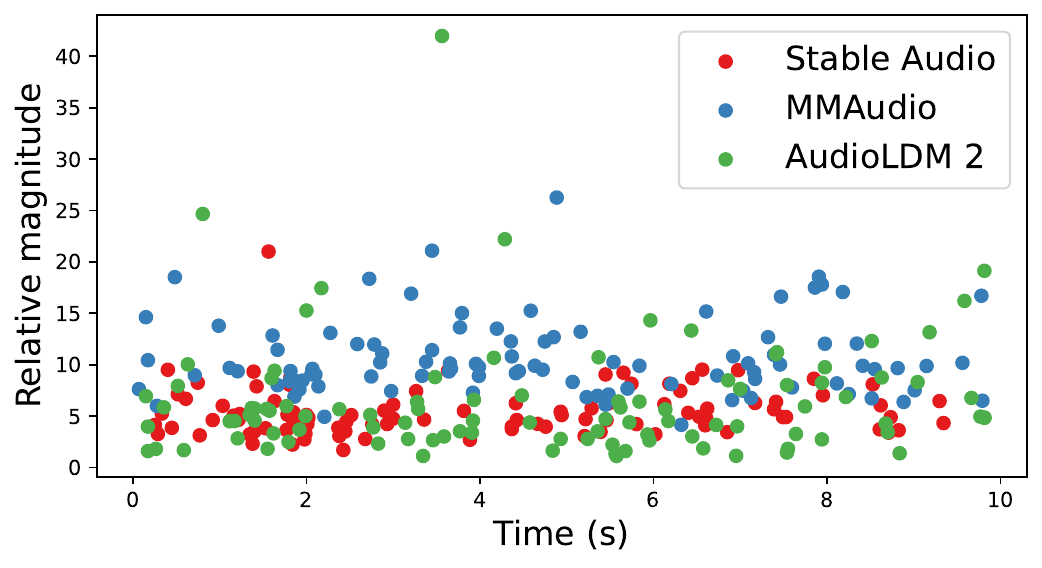}
\caption{dog}
\end{subcaptionblock}
\begin{subcaptionblock}[T]{0.33\textwidth}
\includegraphics[width=\textwidth]{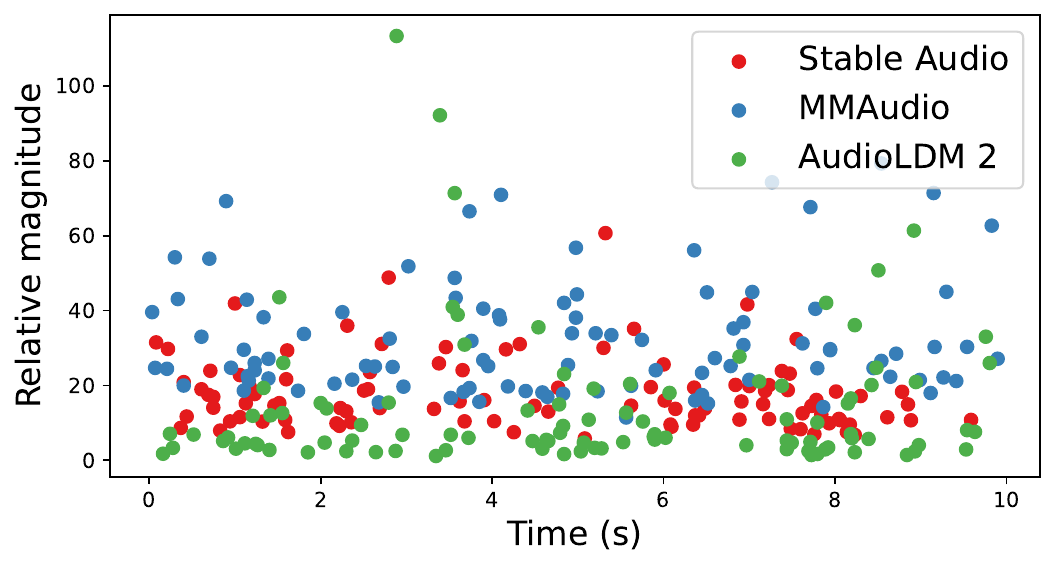}
\caption{door knock}
\par\vspace{0.5cm}
\end{subcaptionblock}
\begin{subcaptionblock}[T]{0.33\textwidth}
\includegraphics[width=\textwidth]{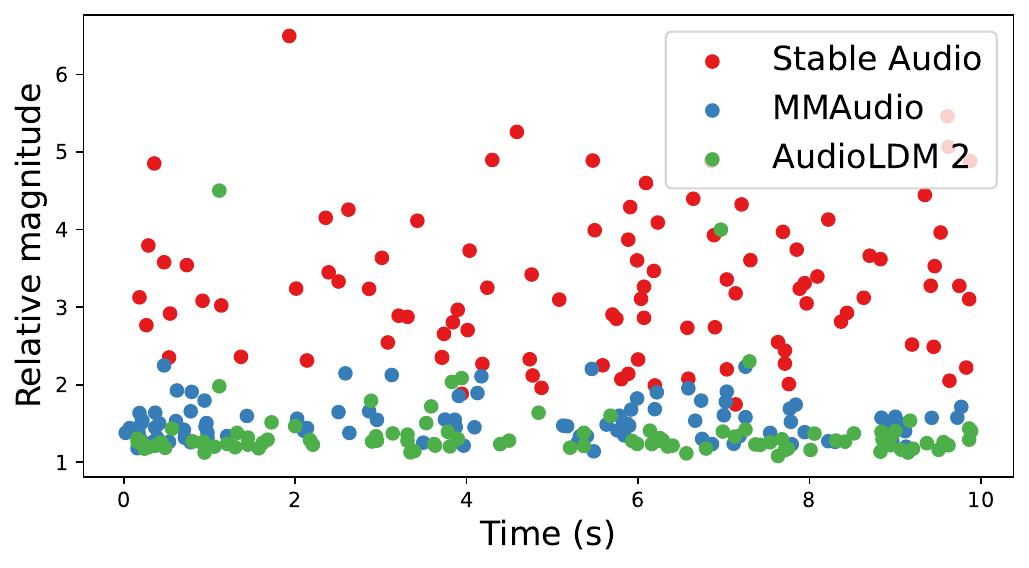}
\caption{helicopter}
\end{subcaptionblock}
\begin{subcaptionblock}[T]{0.33\textwidth}
\includegraphics[width=\textwidth]{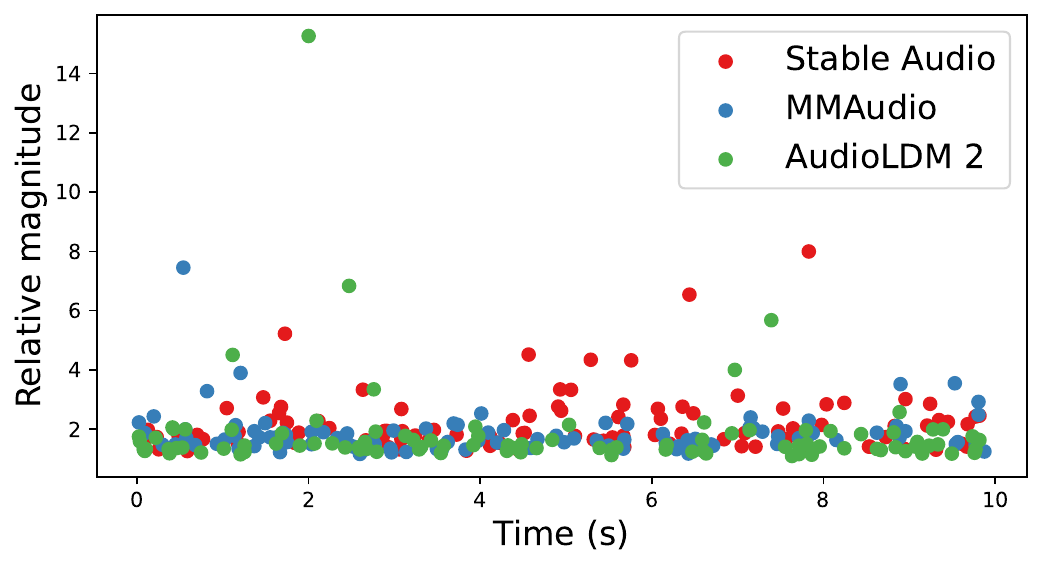}
\caption{rain}
\end{subcaptionblock}
\caption{\textbf{Expressive Range of Timing and Magnitude Peak for Five ESC-50 Prompts.} 
Expressive range diagrams showing relative magnitude and location of the RMS loudness peak for 100 samples generated by each model for each of five ESC-50 labels. (Note that the subplots' y axes cover different ranges.)}
\label{fig:esc50_peak_ERA}
\end{figure*}

\begin{figure*}
\centering
\begin{subcaptionblock}[T]{0.33\textwidth}
\includegraphics[width=\textwidth]{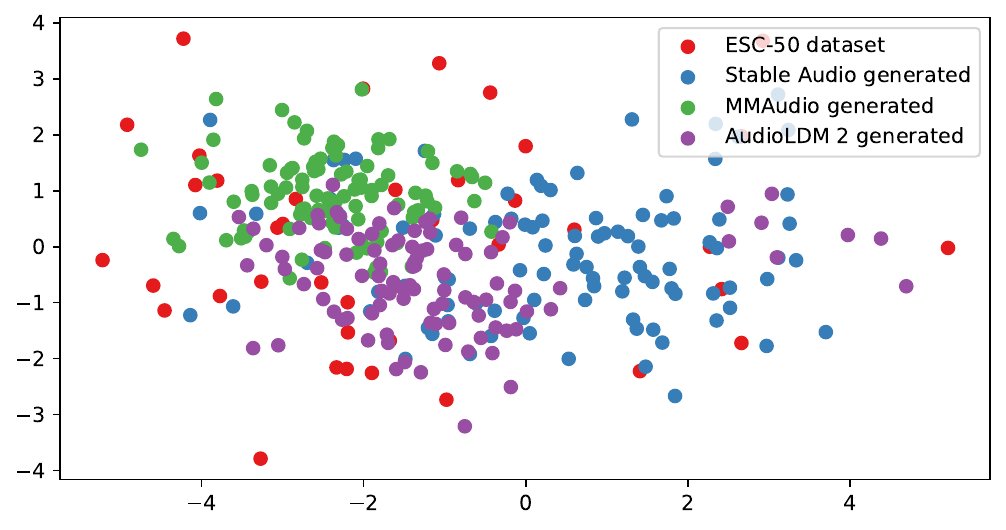}
\caption{crying baby -- loudness}
\end{subcaptionblock}
\begin{subcaptionblock}[T]{0.33\textwidth}
\includegraphics[width=\textwidth]{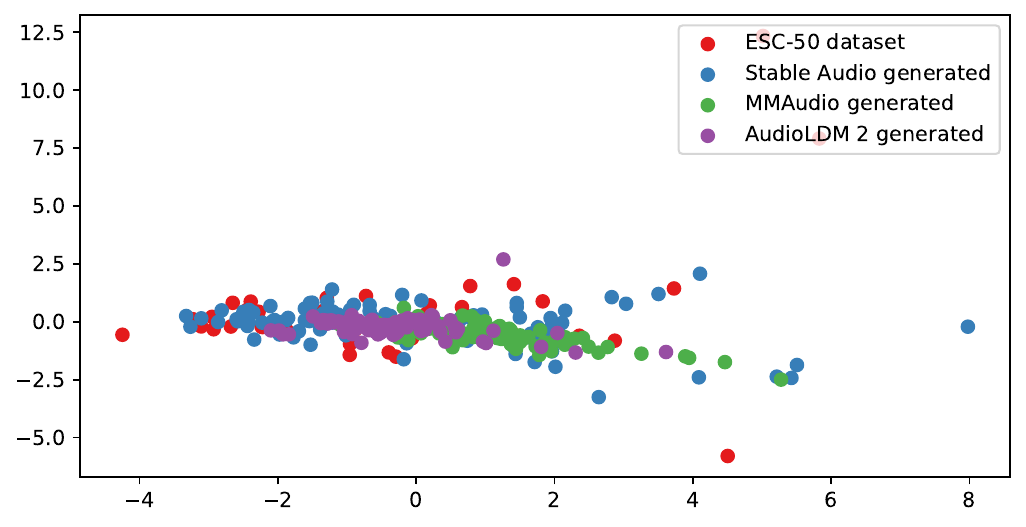}
\caption{crying baby -- pitch}
\end{subcaptionblock}
\begin{subcaptionblock}[T]{0.33\textwidth}
\includegraphics[width=\textwidth]{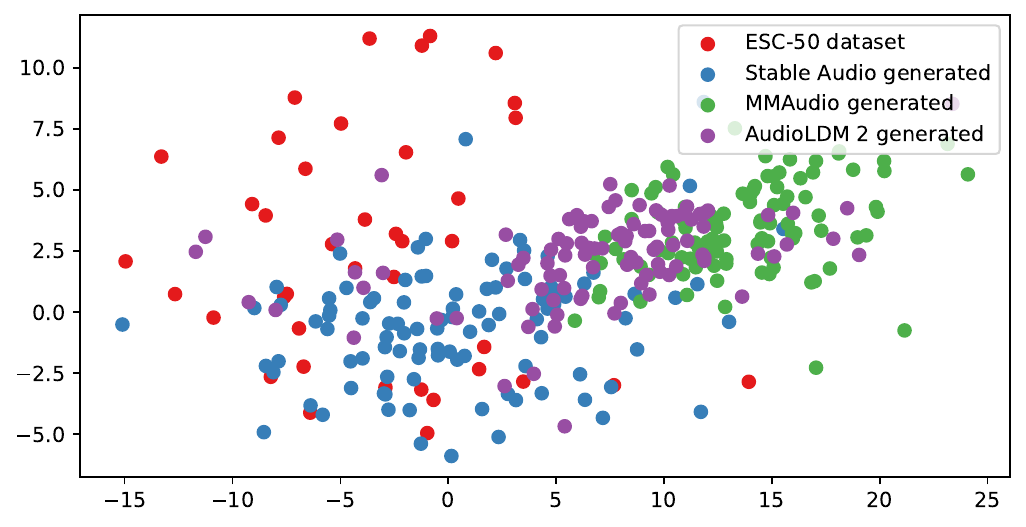}
\caption{crying baby -- timbre}
\end{subcaptionblock}
\par\vspace{0.5cm}
\begin{subcaptionblock}[T]{0.33\textwidth}
\includegraphics[width=\textwidth]{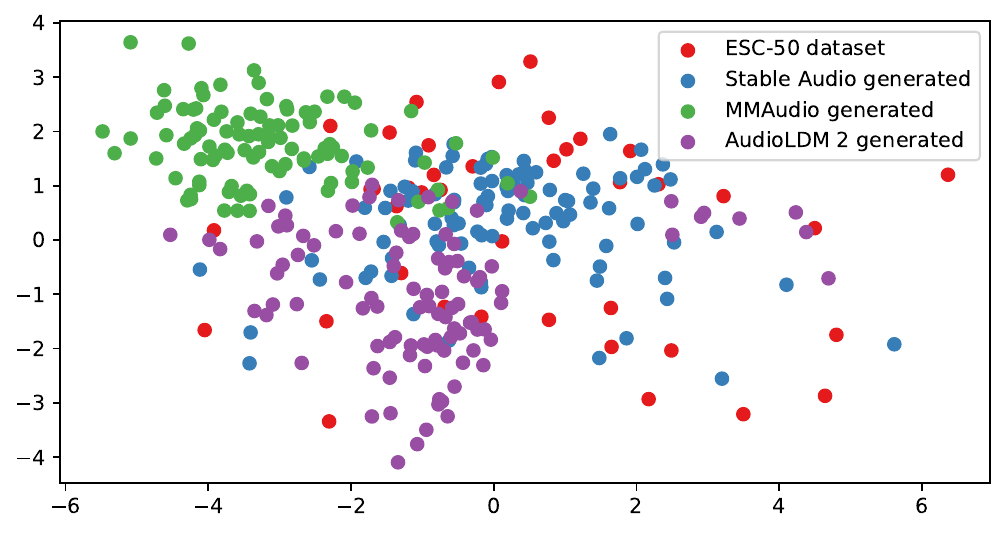}
\caption{dog -- loudness}
\end{subcaptionblock}
\begin{subcaptionblock}[T]{0.33\textwidth}
\includegraphics[width=\textwidth]{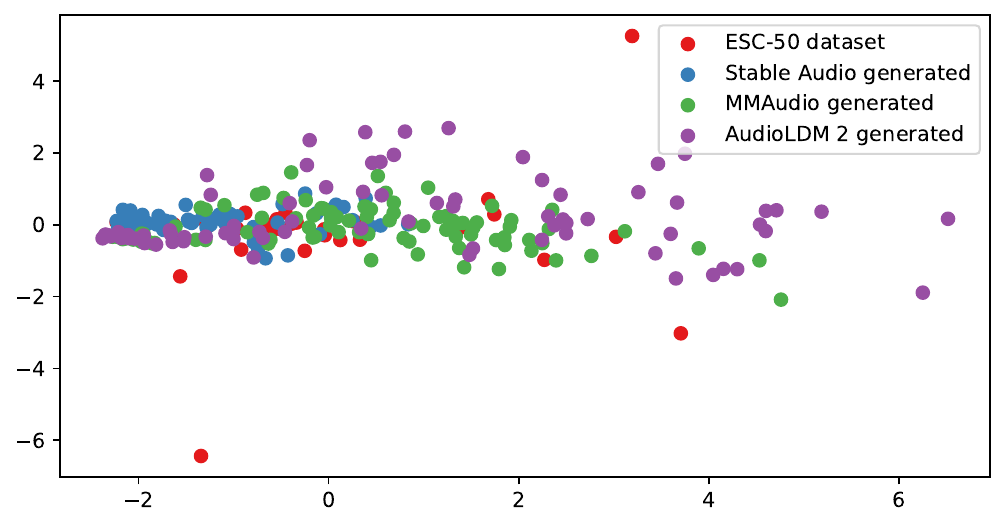}
\caption{dog -- pitch}
\end{subcaptionblock}
\begin{subcaptionblock}[T]{0.33\textwidth}
\includegraphics[width=\textwidth]{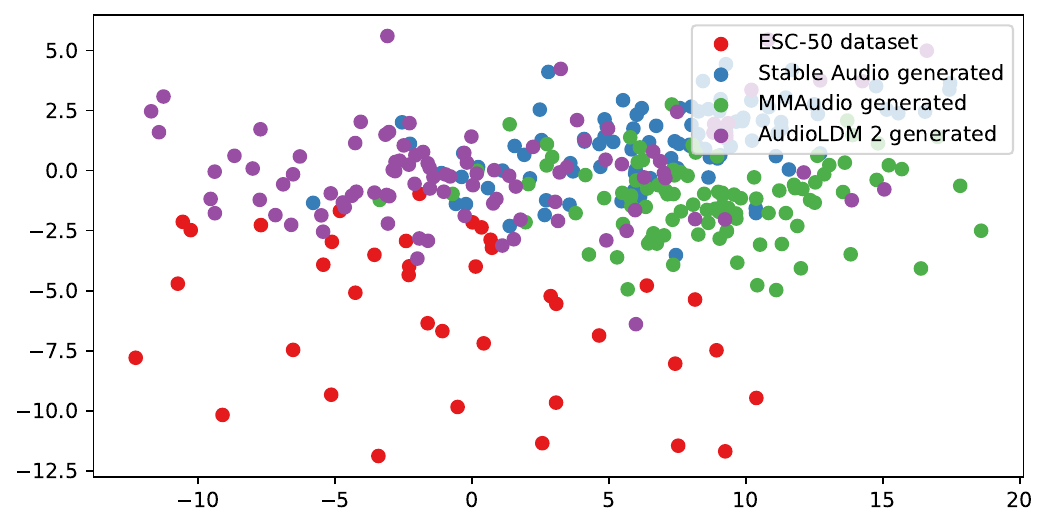}
\caption{dog -- timbre}
\end{subcaptionblock}
\par\vspace{0.5cm}
\begin{subcaptionblock}[T]{0.33\textwidth}
\includegraphics[width=\textwidth]{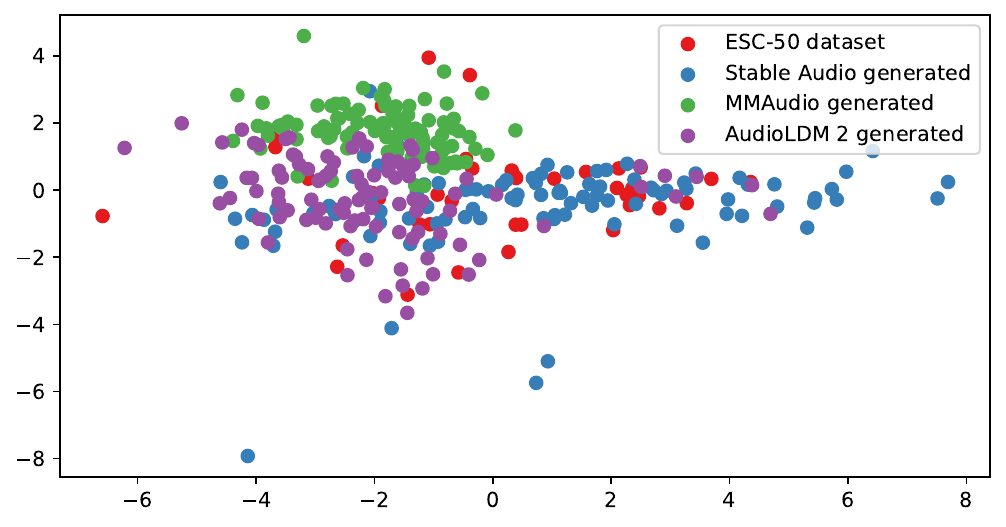}
\caption{door knock -- loudness}
\end{subcaptionblock}
\begin{subcaptionblock}[T]{0.33\textwidth}
\includegraphics[width=\textwidth]{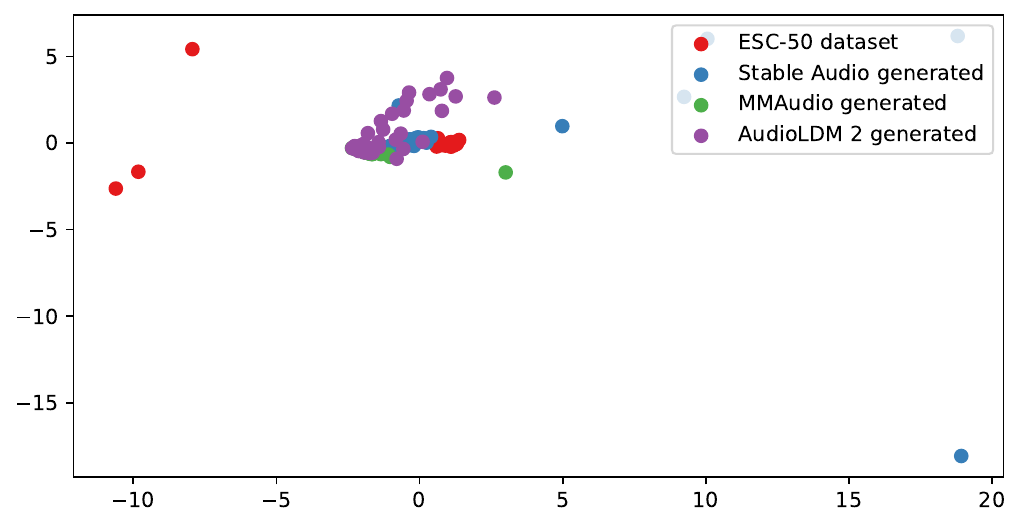}
\caption{door knock -- pitch}
\end{subcaptionblock}
\begin{subcaptionblock}[T]{0.33\textwidth}
\includegraphics[width=\textwidth]{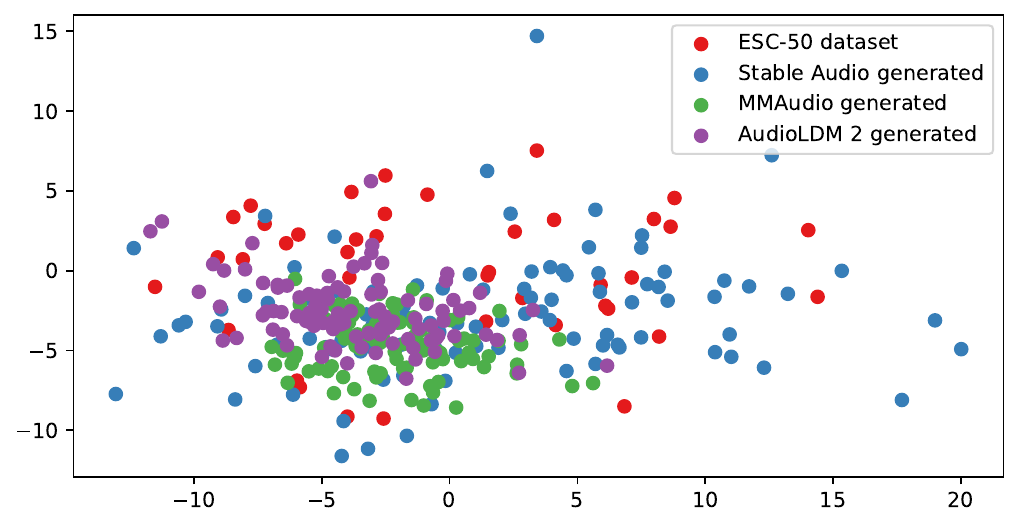}
\caption{door knock -- timbre}
\end{subcaptionblock}
\par\vspace{0.5cm}
\begin{subcaptionblock}[T]{0.33\textwidth}
\includegraphics[width=\textwidth]{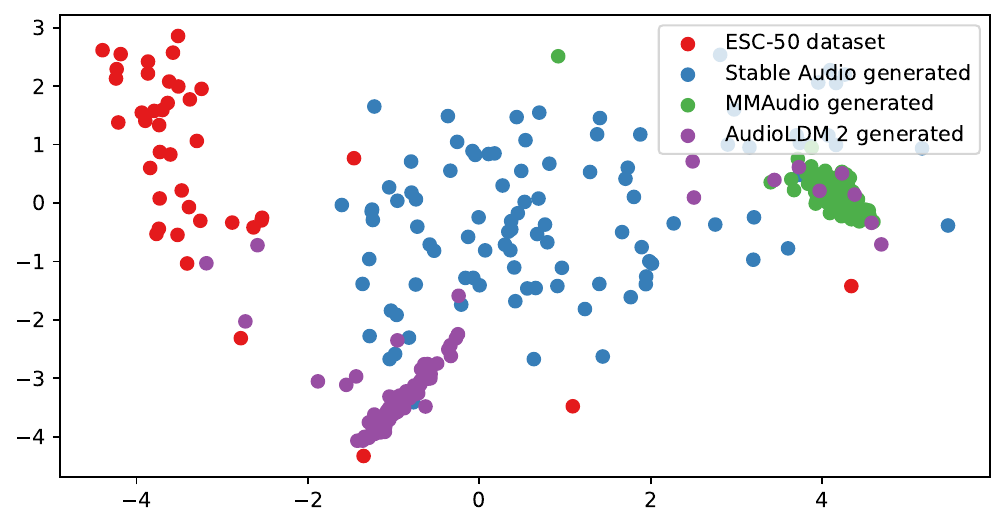}
\caption{helicopter -- loudness}
\end{subcaptionblock}
\begin{subcaptionblock}[T]{0.33\textwidth}
\includegraphics[width=\textwidth]{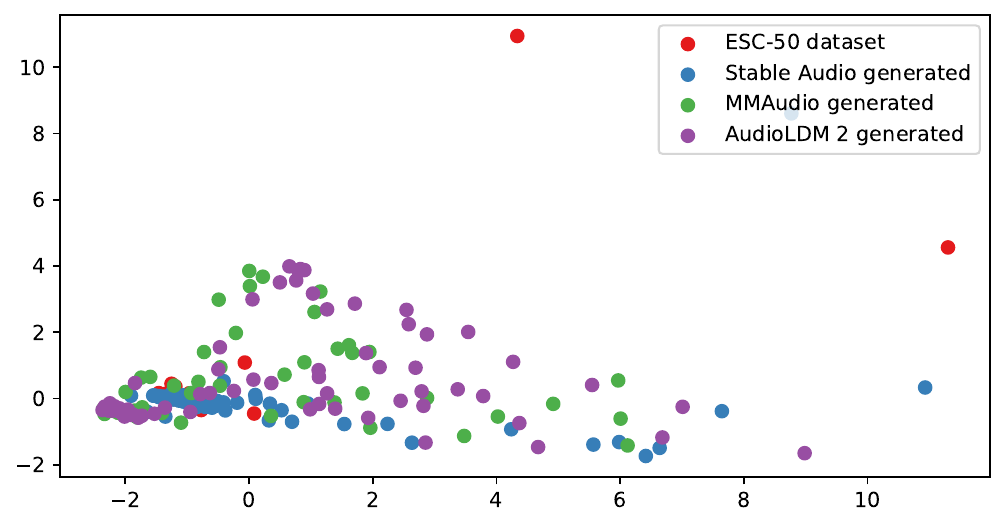}
\caption{helicopter -- pitch}
\end{subcaptionblock}
\begin{subcaptionblock}[T]{0.33\textwidth}
\includegraphics[width=\textwidth]{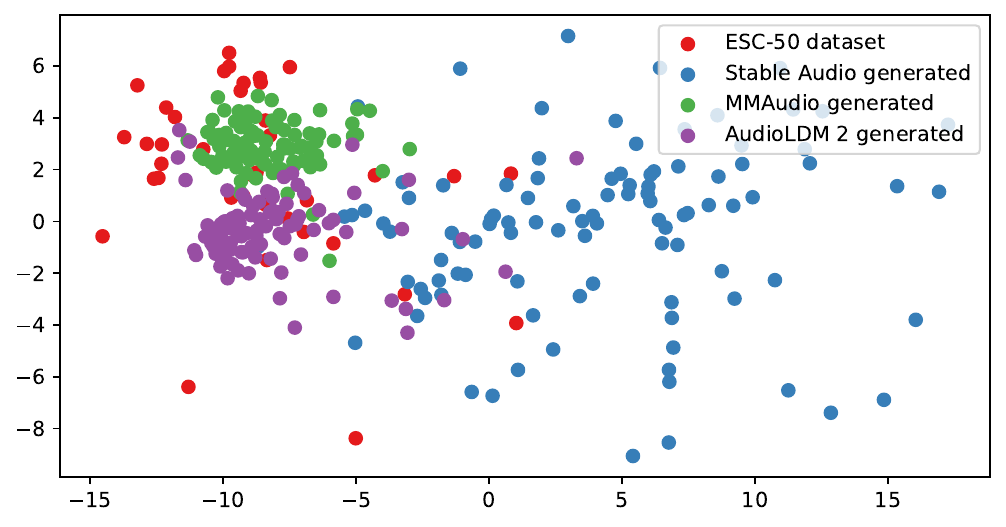}
\caption{helicopter -- timbre}
\end{subcaptionblock}
\par\vspace{0.5cm}
\begin{subcaptionblock}[T]{0.33\textwidth}
\includegraphics[width=\textwidth]{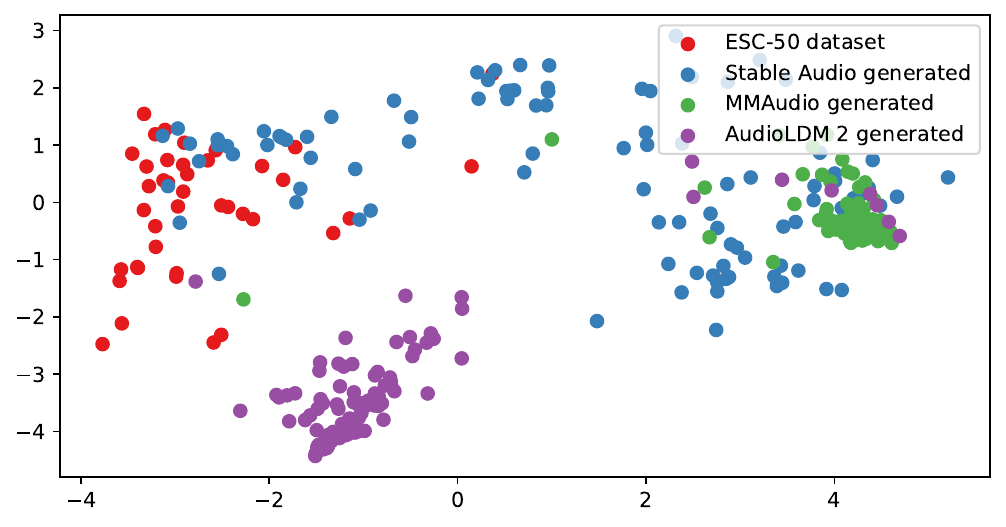}
\caption{rain -- loudness}
\end{subcaptionblock}
\begin{subcaptionblock}[T]{0.33\textwidth}
\includegraphics[width=\textwidth]{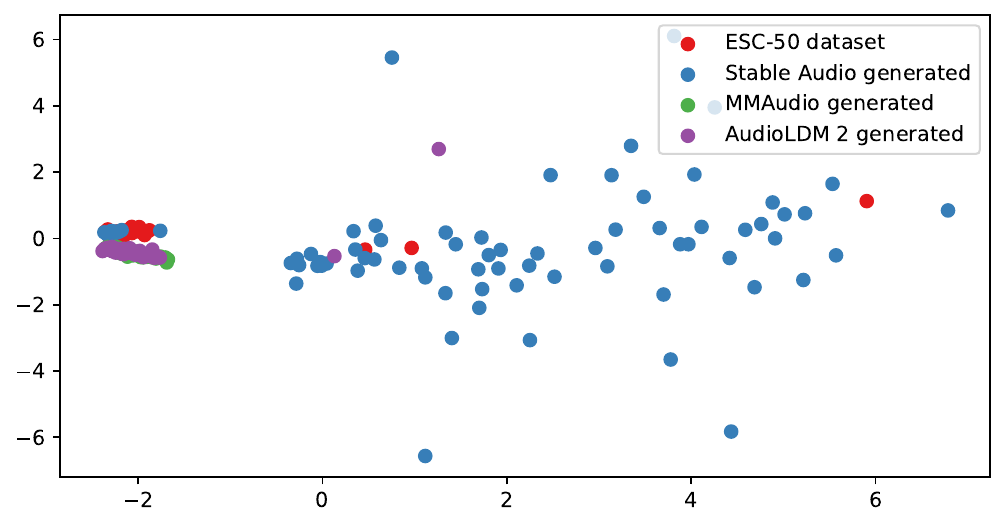}
\caption{rain -- pitch}
\end{subcaptionblock}
\begin{subcaptionblock}[T]{0.33\textwidth}
\includegraphics[width=\textwidth]{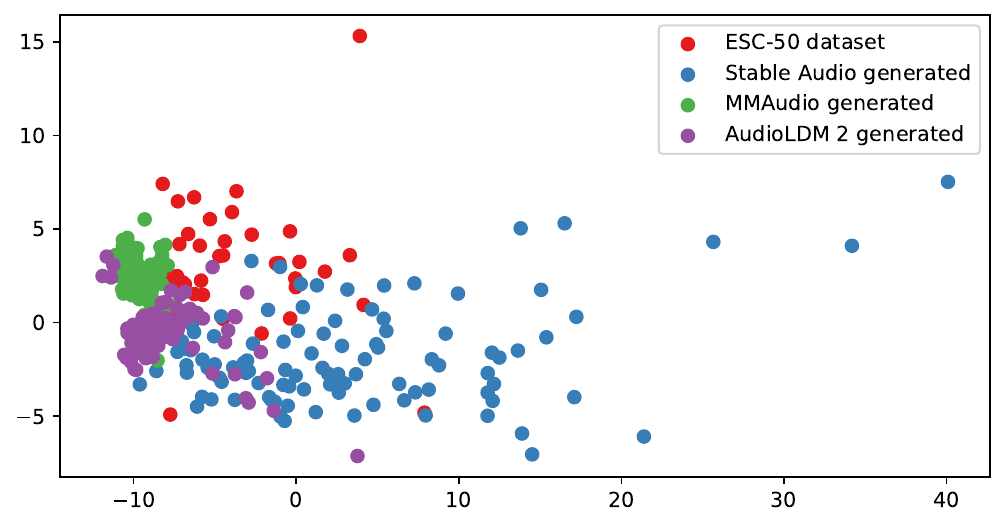}
\caption{rain -- timbre}
\end{subcaptionblock}
\par\vspace{0.5cm}
\caption{\textbf{Expressive Range of PCA-Reduced Audio Features for Five ESC-50 Prompts.}
Expressive range diagrams showing the first two principal components (x and y axes, respectively) of higher-dimensional audio metrics intended to capture samples' loudness, pitch, and timbre characteristics (see text for details). Each row shows data for one label from ESC-50. For each label, the ESC-50 class represents data for the 40 curated samples in the ESC-50 dataset; the remaining three classes represent data for 100 audio samples generated from each text-to-audio model.}
\label{fig:esc50_pca_ERA}
\end{figure*}

Next, we expanded our investigation beyond the single ``thunder'' example to the 50 labels from the ESC-50 dataset for environmental sound \cite{piczak2015dataset}.

The ESC-50 dataset consists of a total of 2000 hand-curated audio samples, with 40 samples for each of the 50 labels. The 50 labels are all short descriptions of between one and three words each, such as \emph{rain}, \emph{crying baby}, and \emph{can opening}. They are grouped into 5 categories, with 10 labels in each category: ``Animals'', ``Natural soundscapes \& water sounds'', ``Human, non-speech sounds'', ``Interior/domestic sounds'', and ``Exterior/urban noises''.

Although originally developed for sound classification rather than generation, using ESC-50 gives us a set of target labels to investigate, each of which contains real, non-generated reference sounds, to which the generative range of each model's outputs for that label can be compared.

\paragraph{Prompting:} We prompted each model with \emph{Sound of [label]}, where \emph{[label]} was replaced with each label from ESC-50; for example, \emph{Sound of can opening}. Prompt construction isn't the focus of this paper, and this was a simple prompt that, nonetheless, in spot-checks still produced reasonable outputs; but we note that it is common in practice to use much more elaborate prompts.

We generated 100 10-second audio clips for each label, using each of three models: Stable Audio Open, MMAudio, and AudioLDM 2. These 15,000 generated audio clips are available in the Zenodo archive accompanying this paper: \url{https://doi.org/10.5281/zenodo.16998750}.

\subsection{Which expressive range metrics?}

The analysis of the \emph{thunder} prompt above used a set of metrics developed bottom-up by listening to example outputs. After listening to some examples, we determined that the timing and magnitude of a thunderclap was a particularly salient feature, and then plotted an expressive range diagram illustrating the range of two models' generative outputs on those two axes.

We might hypothesize that, at least for sound-effect types of audio, these two metrics (location and magnitude of a loudness peak) would be interesting in general, for many kinds of sound effects. However, in our initial analysis, that seems not to be the case. Figure~\ref{fig:esc50_peak_ERA} shows loudness-peak expressive range diagrams for five labels from ESC-50; to avoid cherry-picking, these are the alphabetically first labels in each of the five ESC-50 categories. With one exception, they do not seem to illuminate much about either the individual models or the differences between the models. The one exception is that for the label ``helicopter'' (prompt: \emph{sound of helicopter}), Stable Audio Open consistently produces a loudness peak (peak loudness 3-5x average loudness), while the other two models don't. Listening to the samples, this appears to be because Stable Audio Open's helicopter clips tend to produce helicopter fly-bys, where the sound gets louder and then softer again, while the other two models don't (however, it has no clear pattern to whether the fly-by is early or late in the clip).

We could nonetheless follow the basic methodology outlined in the ``thunder'' motivating example: for a given label, listen to some example clips (at least a few from each model), write some qualitative observations about what seems to vary both within and between models, and then come up with bespoke metrics that illustrate these semantically meaningful variations in generative output. In fact, we believe doing so is a promising form of exploratory data analysis, and we especially recommend following such an approach when a sound designer intends to use a particular type of audio output repeatedly. However, to be able to summarize the generative range of models across all 50 labels in the ESC-50 dataset, we need something more general.

\subsection{Acoustic diversity analysis}

To apply a single set of metrics across the entire ESC-50 set, we represent audio outputs in terms of perceptual sound attributes, using relatively standard methods from the audio analysis literature. These are designed to be comparable across prompts, but each axis will not have as simple an interpretation as metrics like the loudness peaks in the previous figures. We use summarized representations of pitch, loudness, and timbre characteristics intended to capture their overall perceptual qualities, as well as variation in such qualities over the sample duration.

\paragraph{Audio metric computation:} We first converted all audio files to mono and the same sampling rate (22050 Hz). We then extracted signal processing features that are strongly correlated with these perceptual attributes of sound, namely fundamental frequency (for pitch), A-weighted RMS energy (for loudness), and Mel-frequency cepstral coefficients or MFCCs (for timbre).

To extract fundamental frequency (f0), we used the pYIN algorithm \cite{mauch2014pyin} as implemented in the \texttt{librosa} audio analysis Python library \cite{mcfee2015librosa}, using a hop size of 512 samples. To compute A-weighted RMS energy, we used a frame length of 20248 samples and a hop size of 512 samples. To compute MFCCs, we used a mel spectrogram with 128 bins, a frame length of 2048 samples, a hop size of 512 samples, and retained the first 13 MFCCs. We then augmented each representation with first- and second-order differences (i.e., `deltas' and `delta deltas') and summarized these features over time by computing the mean, standard deviation, minimum, and maximum. This resulted in three feature vectors for each audio file, one relating to each perceptual attribute (pitch (f0 statistics - 12 dimensions), loudness (a-weighted rms statistics - 12 dimensions), and timbre (MFCC statistics - 156 dimensions)). Note that when calculating statistics for f0, only voiced (i.e., pitched) frames are included, and if a file does not have any voiced frames, this file is excluded from the f0 analysis. 

Finally, for the purpose of plotting these multi-dimensional representations of loudness, pitch, and timbre as two-dimensional expressive range diagrams for each prompt, we used principal components analysis (PCA) to extract the first two principal components for each (pooled across all three generators' outputs plus the ESC-50 samples corresponding to each prompt).\footnote{For a previous example of using PCA for expressive range analysis, see \cite{herve2023exploring}.}

\subsection{Results}

Figure~\ref{fig:esc50_pca_ERA} plots the first two principal components of our loudness, pitch, and timbre metrics for five labels from ESC-50 (the same five labels as in Figure~\ref{fig:esc50_peak_ERA}). There are 340 samples plotted on each scatterplot: 100 each from the three generative models being investigated, plus the 40 hand-cured examples in that category from ESC-50. Note that since the two axes are the result of a PCA transformation, they do not have any direct interpretation, unlike previous figures. However, relative positions are comparable, such as whether the dot cloud for two models overlaps, or is more or less tightly clustered.

We argue that in contrast to Figure~\ref{fig:esc50_peak_ERA}, we do begin to see useful distinctions in the Figure~\ref{fig:esc50_pca_ERA} diagrams, suggesting that these are useful general metrics (not designed for a specific prompt) for exploratory data analysis across prompts. As one example, MMAudio (in green) shows outputs fairly tightly clustered in all three metrics for \emph{rain}, and in two metrics (loudness and timbre) for \emph{helicopter}, indicating relatively little diversity of generative outputs for those two labels. On the other hand, Stable Audio does have a diverse range of outputs by these metrics for \emph{helicopter} -- but they are not very close to the reference outputs in the hand-curated ESC-50 samples.

While our primary goal is per-prompt exploratory data analysis with expressive range diagrams, it may also be useful to calculate some overall summary statistics. Table~\ref{table:variation} shows a summary of the variance in generative model outputs on these metrics, aggregated across all 50 labels from ESC-50 (i.e., 5000 generated samples per model, and the 2000 dataset samples for ESC-50).

\paragraph{Variance computation:} For each of the audio metrics -- loudness, pitch, and timbre -- we 1) pooled together the feature vectors from all the model outputs, along with those from the ESC-50 dataset; 2) computed the principal components analysis (PCA) transformation matrix, retaining the components that represent 95\% of the variation in the pooled dataset; 3) computed the trace of the covariance matrix of the PCA-transformed data for each subset of interest to quantify the total variance of the subset; and 4) normalized the total variance value for each subset by the total variance of the ESC-50 subset to scale the values to a reference.

\begin{table}
    \centering
    \begin{tabular}{llr}
        \toprule
        \textbf{Audio source} & \textbf{Metric} & \textbf{Normalized} \\
         &  & \textbf{Total Variance} \\
        \midrule
        ESC-50 dataset & loudness & 1.00 \\
        Stable Audio generated & loudness & 0.73 \\
        MMAudio generated & loudness & 0.66 \\
        AudioLDM 2 generated & loudness & 0.42 \\
        ESC-50 dataset & pitch & 1.00 \\
        Stable Audio generated & pitch & 1.69 \\
        MMAudio generated & pitch & 1.26 \\
        AudioLDM 2 generated & pitch & 1.12 \\
        ESC-50 dataset & timbre & 1.00 \\
        Stable Audio generated & timbre & 0.83 \\
        MMAudio generated & timbre & 0.67 \\
        AudioLDM 2 generated & timbre & 0.43 \\
        \midrule
        ESC-50 dataset & mean & 1.00 \\
        Stable Audio generated & mean & 1.08 \\
        MMAudio generated & mean & 0.86 \\
        AudioLDM 2 generated & mean & 0.65 \\
        \bottomrule
    \end{tabular}
    \caption{Summary of Output Variance}
    \label{table:variation}
\end{table}

If we look at the mean of these values over audio features in Table~\ref{table:variation}, we can see Stable Audio has the most overall variation, even surpassing ESC-50, which is then followed by MMaudio and AudioLDM 2. The ranking of generative models by total variation is maintained over all three audio features. Interestingly, all of the generative models have more variation in pitch than ESC-50, but less variation in loudness and timbre. 

\section{Conclusions}

We have carried out an exploratory analysis of the generative range of text-to-audio models by looking at generative output variation for specific prompts, connecting this to the expressive-range analysis methodology from the procedural content generation (PCG) community.

One difference from much of the existing expressive range analysis in PCG is that we do a analysis conditioned on a prompt, treating ``StableAudio with fixed prompt [x]'' as a generator, and then looking at this for different fixed prompts. Much of the prior PCG work looks at the entire generative space of a given generator; e.g., all possible levels that a level generator can produce.

Why start with fixed prompts? The PCG systems to which expressive range analysis is normally applied have tended to produce artifacts that are more directly comparable with each other, most commonly videogame levels \cite{smith2010analyzing,herve2023exploring}. The entire generative space of a text-to-audio model is a jumble of all possible audio -- or at least whatever subset of all possible audio it has successfully modeled -- which is itself part of the problem we wish to unpack.\footnote{\citeauthor{boucher2024era}~\shortcite{boucher2024era} also discuss this issue.} For example, music often has rhythm and a scale in which it's composed, while environmental sound effects typically don't. We believe it is more feasible to start opening up the block box of this jumbled generative space by digging in detail into specific parts of it. A text-to-audio model with a fixed prompt (but not a fixed random seed) is still itself a generator, with a generative space, and can be analyzed as such.

In future work we are interested in the range of audio that a given model can produce more generally, and how it relates to the prompts, which are the main control mechanism. For example, it would be interesting to know the expressive range of a model's output of thunder-like sounds in general, not just those produced for the specific prompt \emph{thunder}. One could start by investigating prompt variations (``rolling thunder'', ``dramatic thunder like in a Western film'', etc.); but thunder sounds might also be found under other prompts that don't contain the word \emph{thunder} verbatim.

In addition, the reverse is sometimes true: models vary in prompt fidelity, and sometimes produce sounds that \emph{don't} closely resemble the description in the prompt. Our summary metrics here quantify range and variation, but quantifying expressive range based on dimensionality-reduced audio features doesn't distinguish the reasons for variation. For example, a model that sometimes produces non-dog sounds for the ``sound of dog'' prompt would appear to have an increased expressive range, but probably not in the way that a user looking for a diverse set of dog sounds would have wanted. Untangling all these questions of prompt--audio relationship would require different experiments.

Although we do present summary results for the generative range of three models on the ESC-50 dataset, our primary goal is not to conclude that a given model is better or worse, but to present a case study and outline of a methodology for using expressive range analysis (ERA) as an exploratory data analysis framework to understand text-to-audio models. We believe that this is a useful way of systematically thinking about the output of such models. Specifically, we demonstrated two ways to construct ERA metrics and diagrams.
\begin{itemize}
    \item First, a bottom-up method that starts by listening to individual audio samples' qualitative variations (here, of thunder), and building from that to quantitative analyses of generative output range on metrics suitable for that particular type of audio and identified relevant features (here, thunderclap timing and magnitude).
    \item Second, more of a big-picture shotgun approach that starts by computing a battery of general-purpose audio features such as pitch, loudness, and timbre on many prompts, and visualizes them using dimensionality reduction based ERA plots (here, on the ESC-50 classes).
\end{itemize}

\section*{Acknowledgments}

S.\ Gaudl, A.K.\ Hoover, and M.J.\ Nelson were partly supported by Vinnova international collaboration grants No.\ 2023-02015 and 2024-02061.

\newpage

\bibliography{references}

\appendix
\onecolumn
\raggedbottom

\section{Appendix}
Expanded version of Figure~\ref{fig:esc50_pca_ERA} with all fifty ESC-50
classes. PCA-reduced measures of loudness, pitch, and timbre are plotted
as in Figure~\ref{fig:esc50_pca_ERA}. All 15,000 generated audio samples
available at: \url{https://doi.org/10.5281/zenodo.16998749}

\subsection{Prompt: Sound of airplane}\centering
\includegraphics[width=0.68\textwidth]{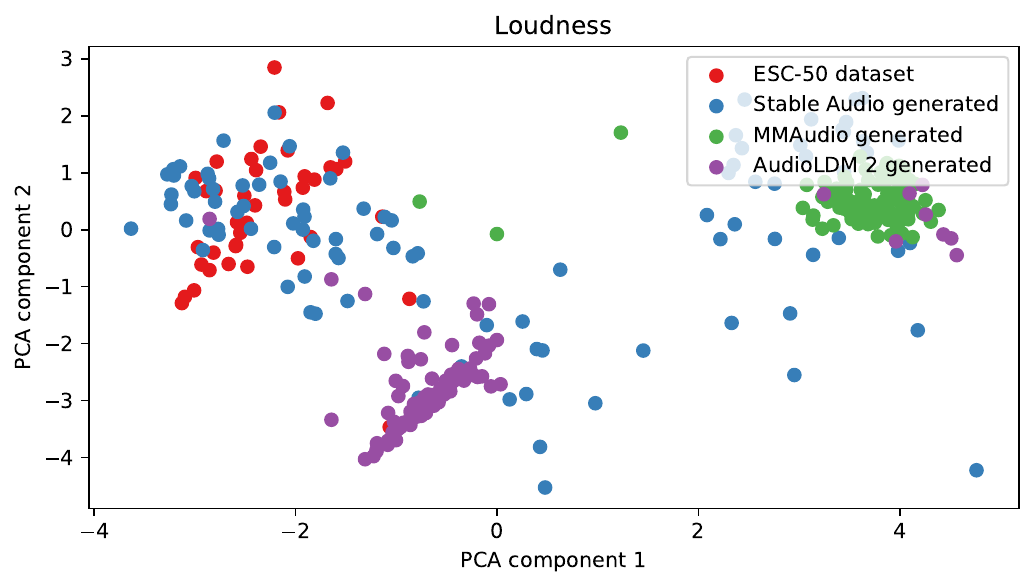}
\includegraphics[width=0.68\textwidth]{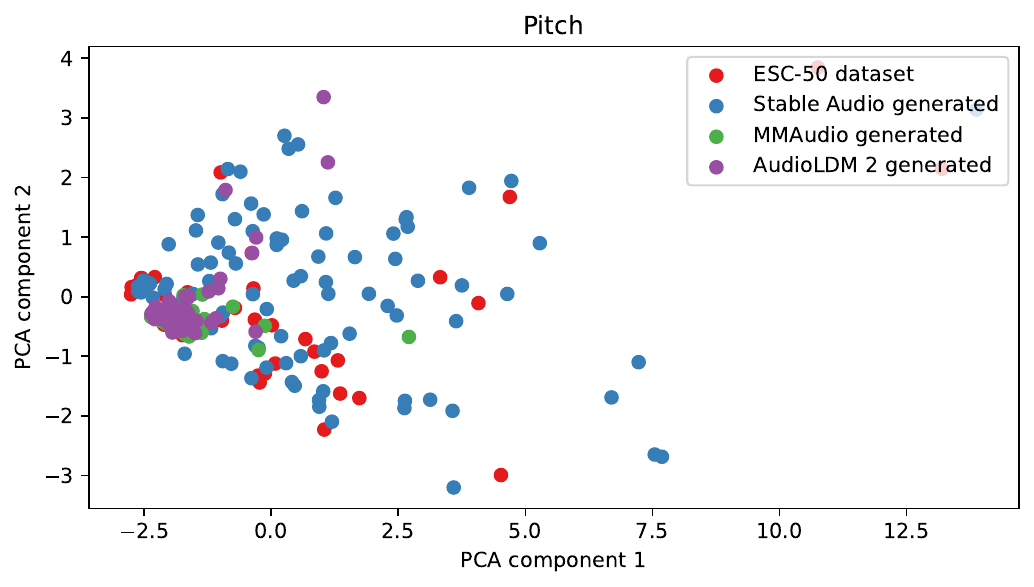}
\includegraphics[width=0.68\textwidth]{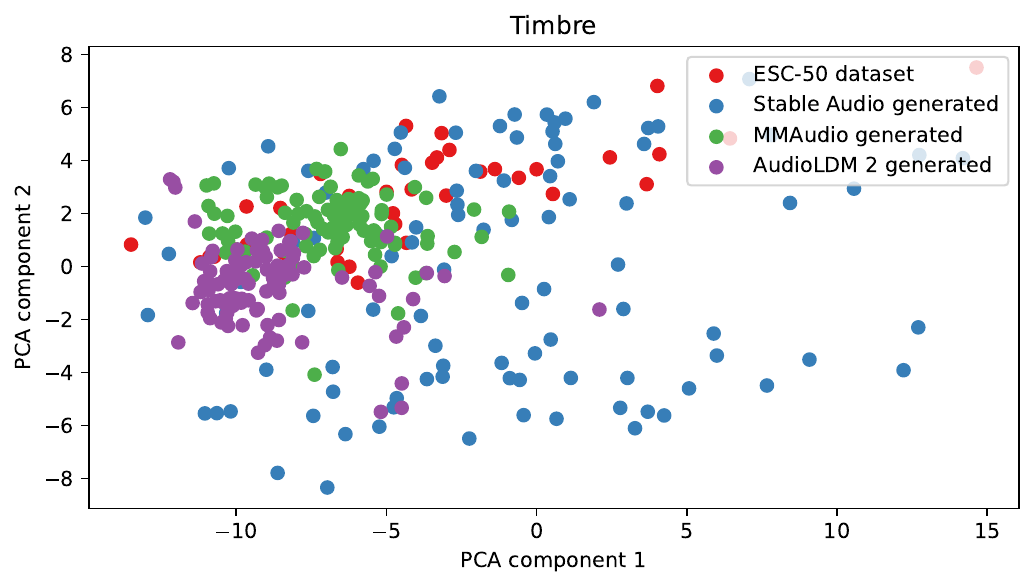}
\subsection{Prompt: Sound of breathing}\centering
\includegraphics[width=0.68\textwidth]{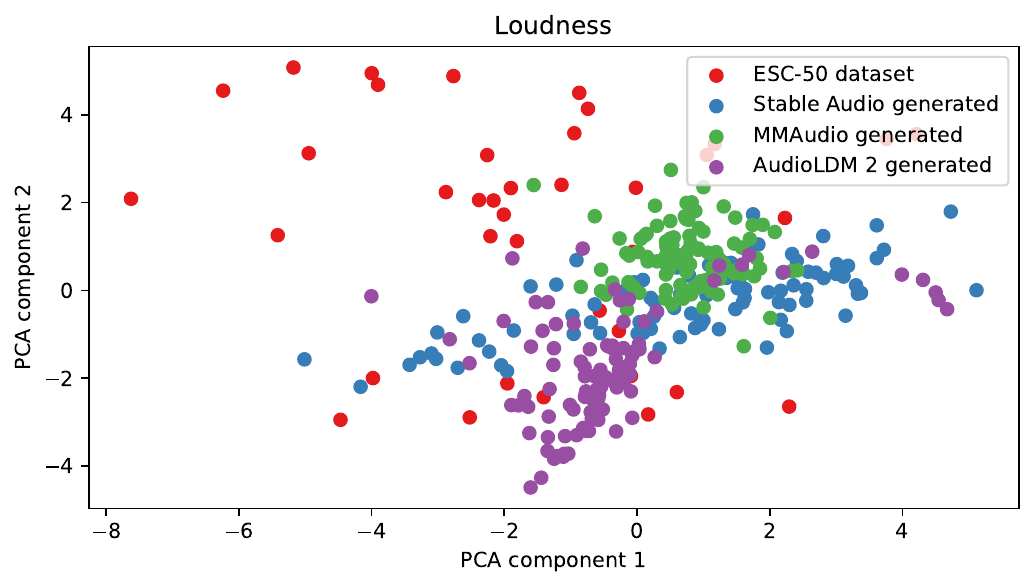}
\includegraphics[width=0.68\textwidth]{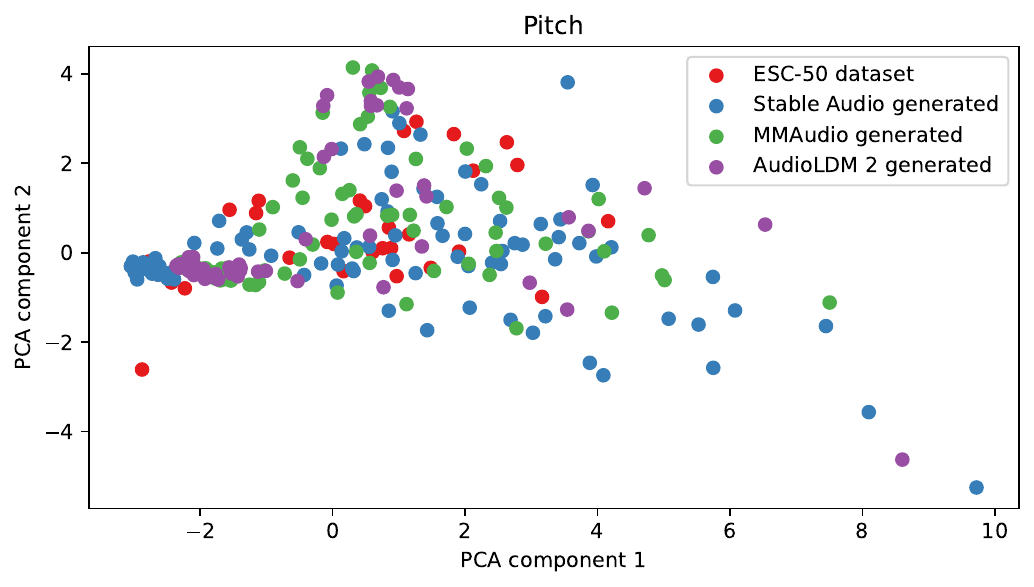}
\includegraphics[width=0.68\textwidth]{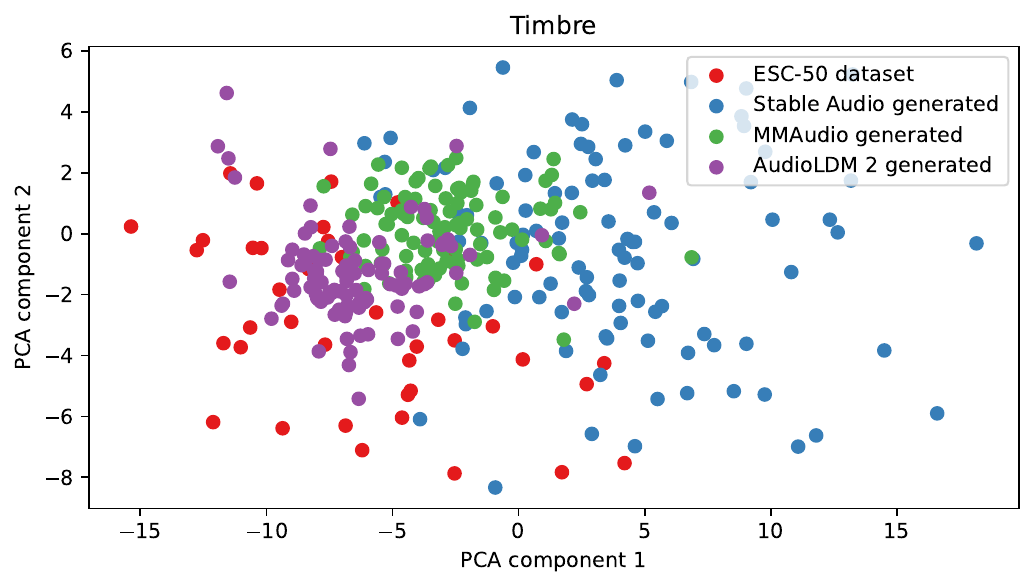}
\subsection{Prompt: Sound of brushing teeth}\centering
\includegraphics[width=0.68\textwidth]{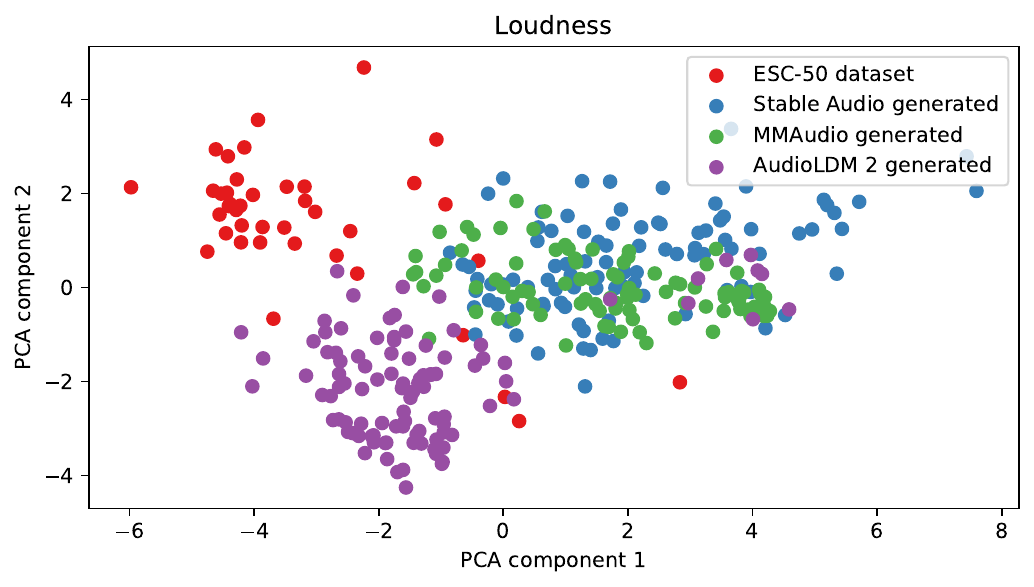}
\includegraphics[width=0.68\textwidth]{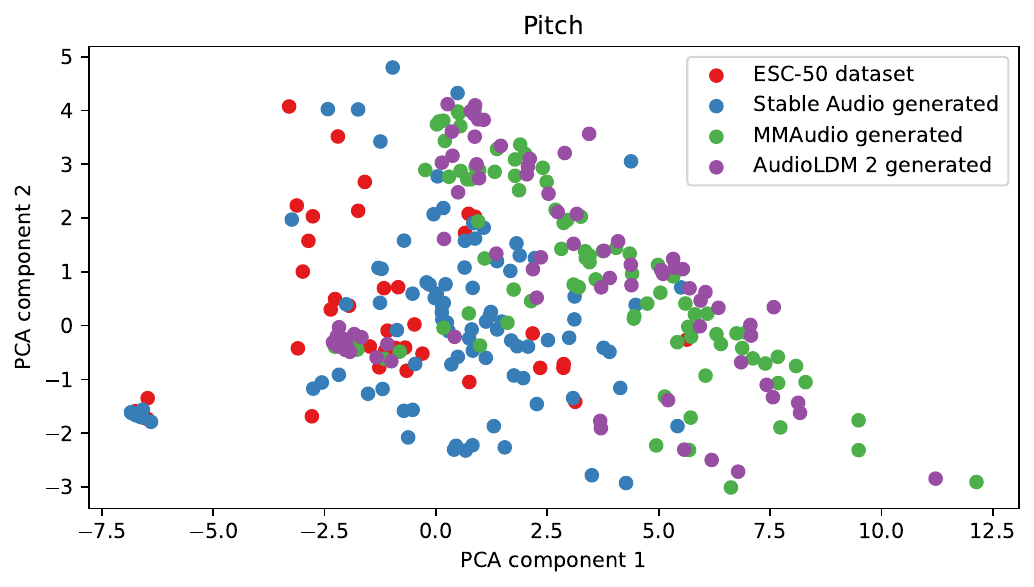}
\includegraphics[width=0.68\textwidth]{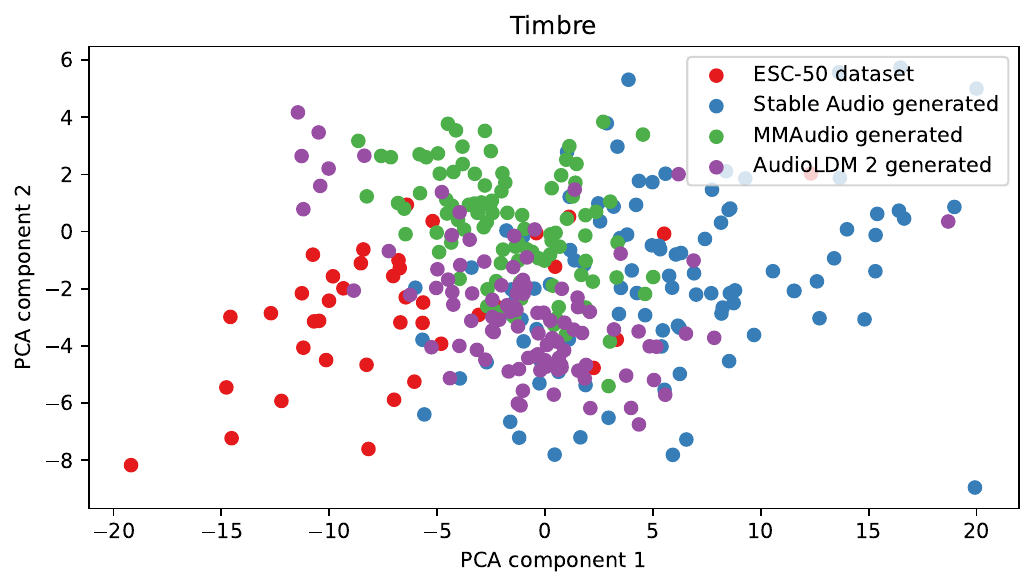}
\subsection{Prompt: Sound of can opening}\centering
\includegraphics[width=0.68\textwidth]{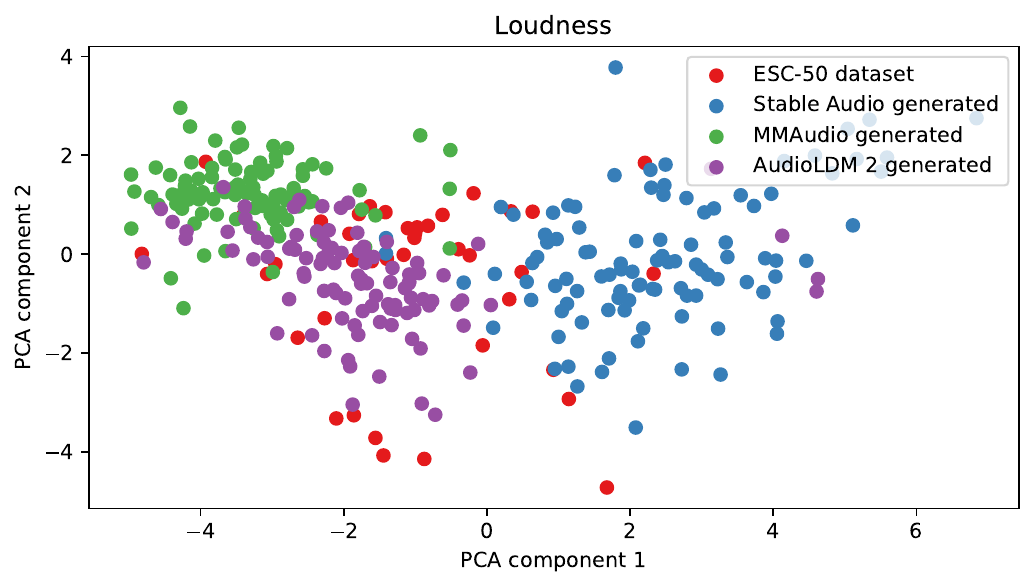}
\includegraphics[width=0.68\textwidth]{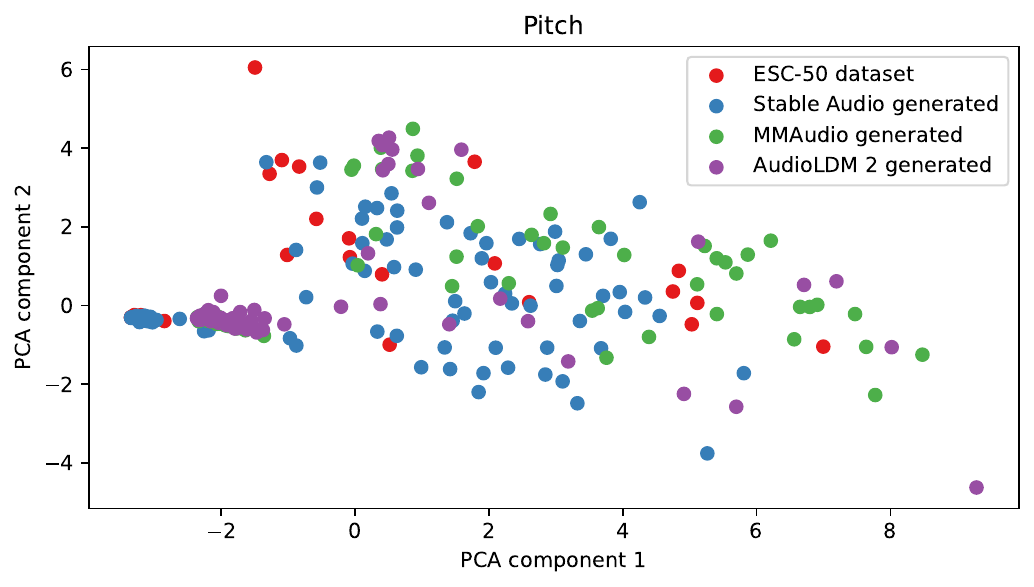}
\includegraphics[width=0.68\textwidth]{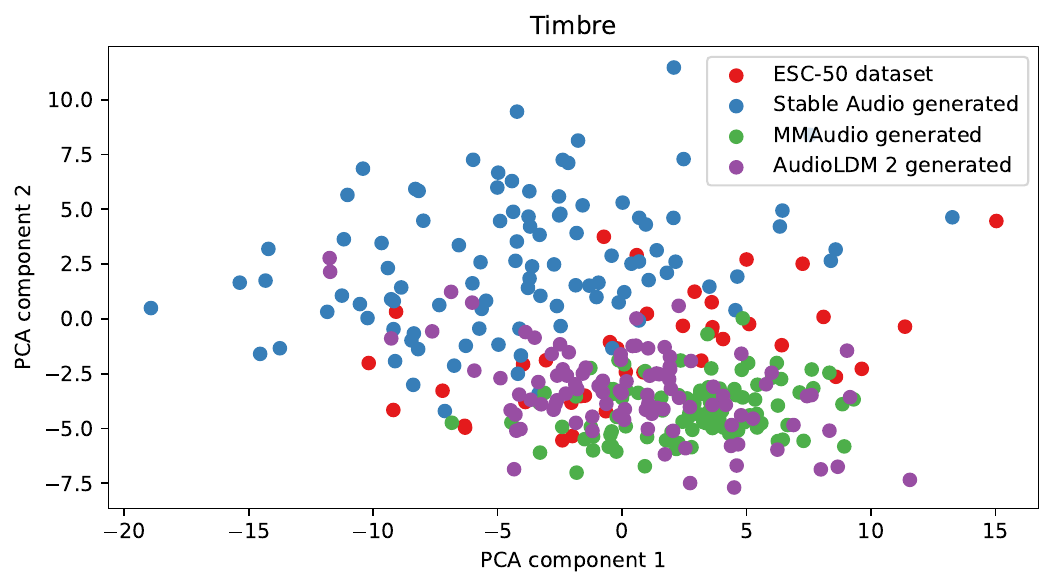}
\subsection{Prompt: Sound of car horn}\centering
\includegraphics[width=0.68\textwidth]{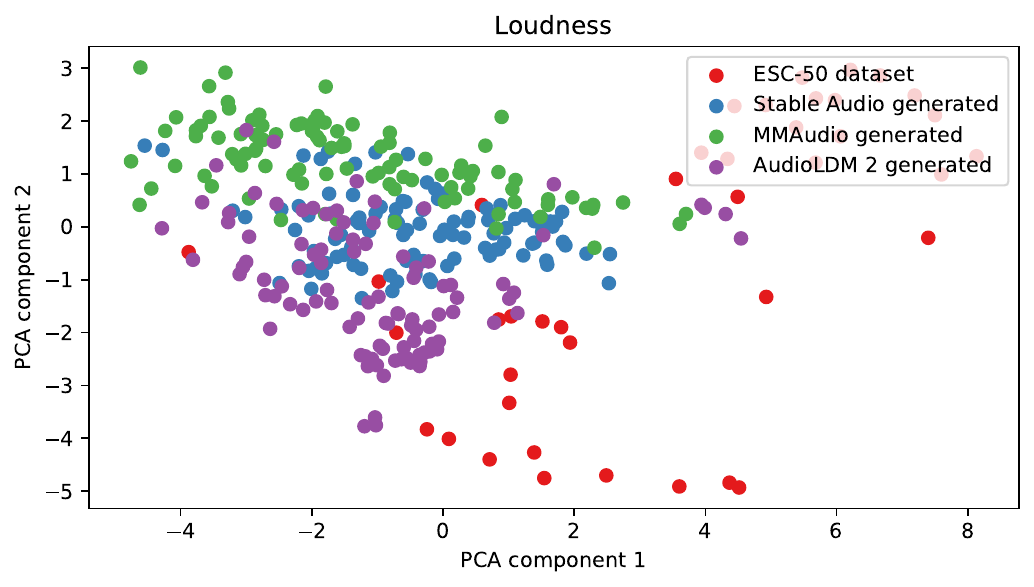}
\includegraphics[width=0.68\textwidth]{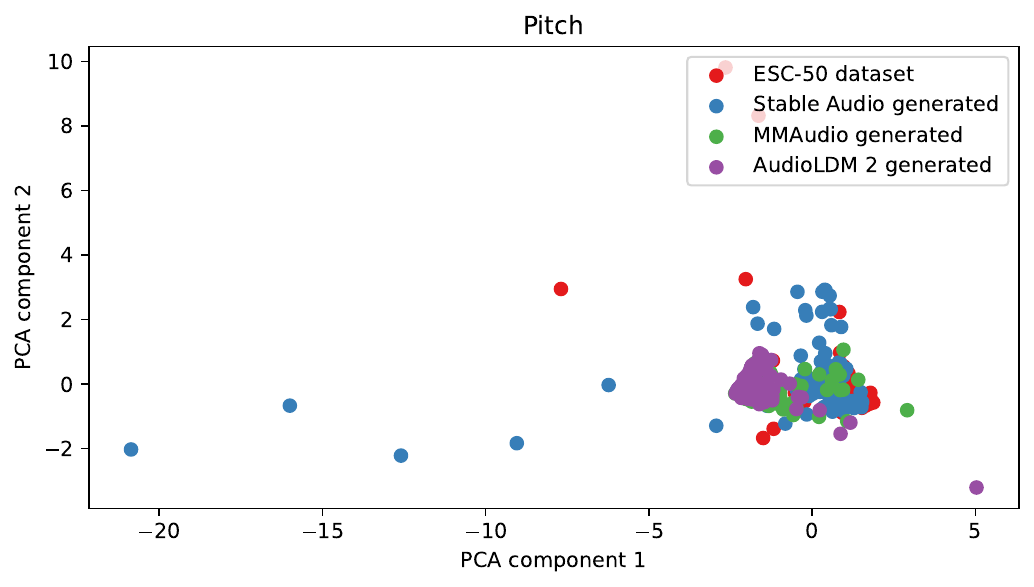}
\includegraphics[width=0.68\textwidth]{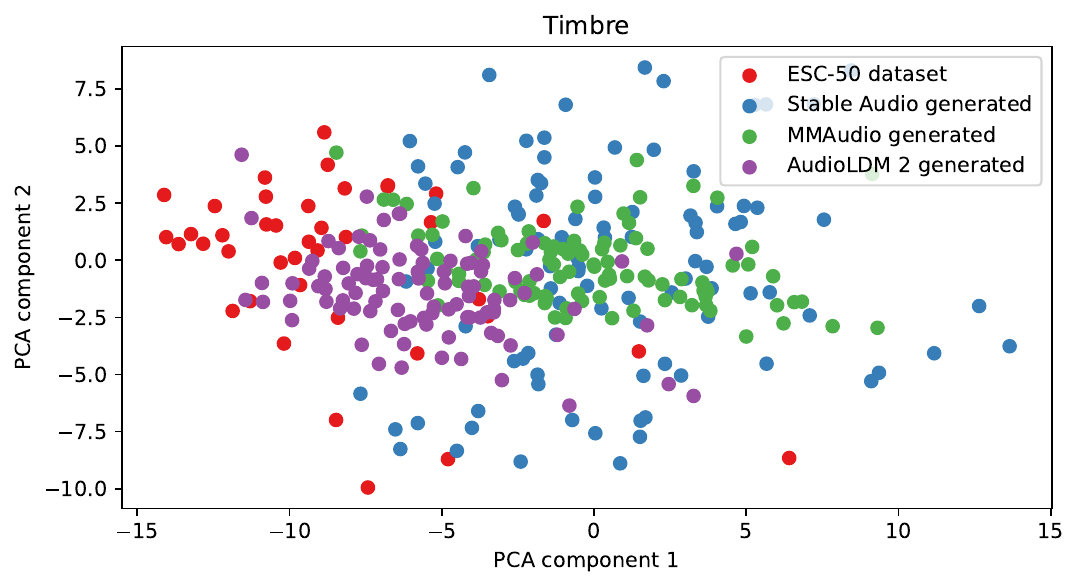}
\subsection{Prompt: Sound of cat}\centering
\includegraphics[width=0.68\textwidth]{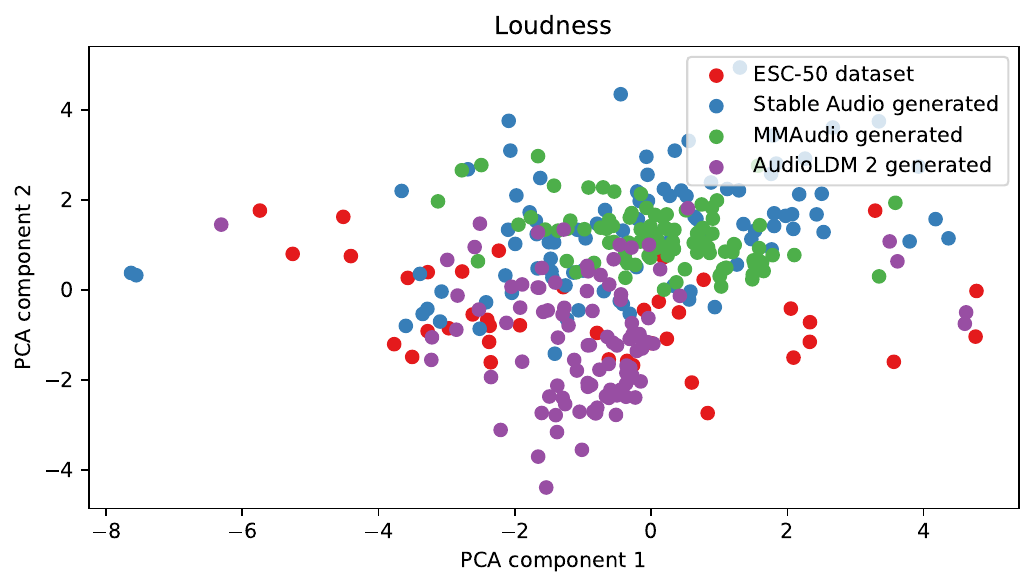}
\includegraphics[width=0.68\textwidth]{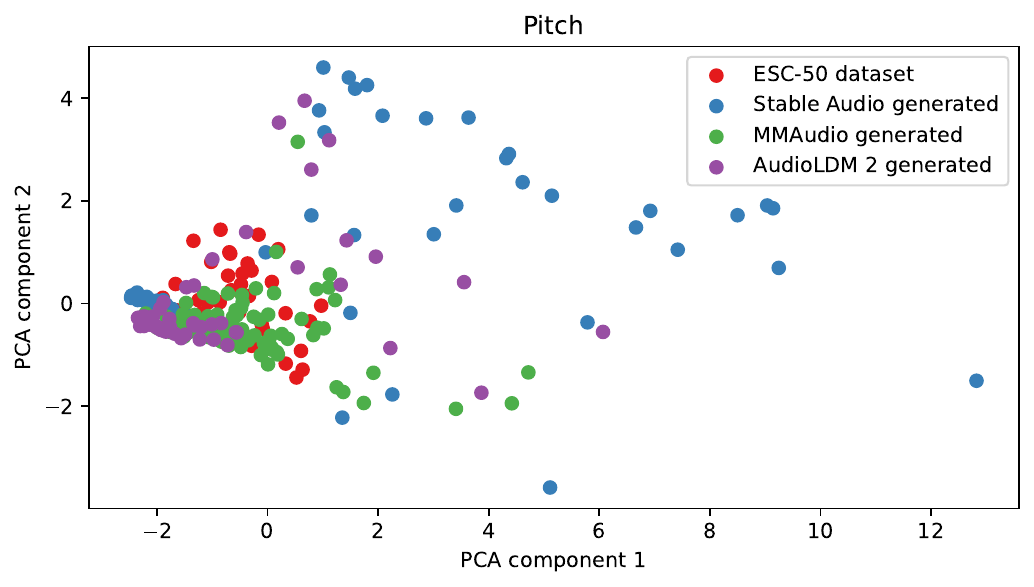}
\includegraphics[width=0.68\textwidth]{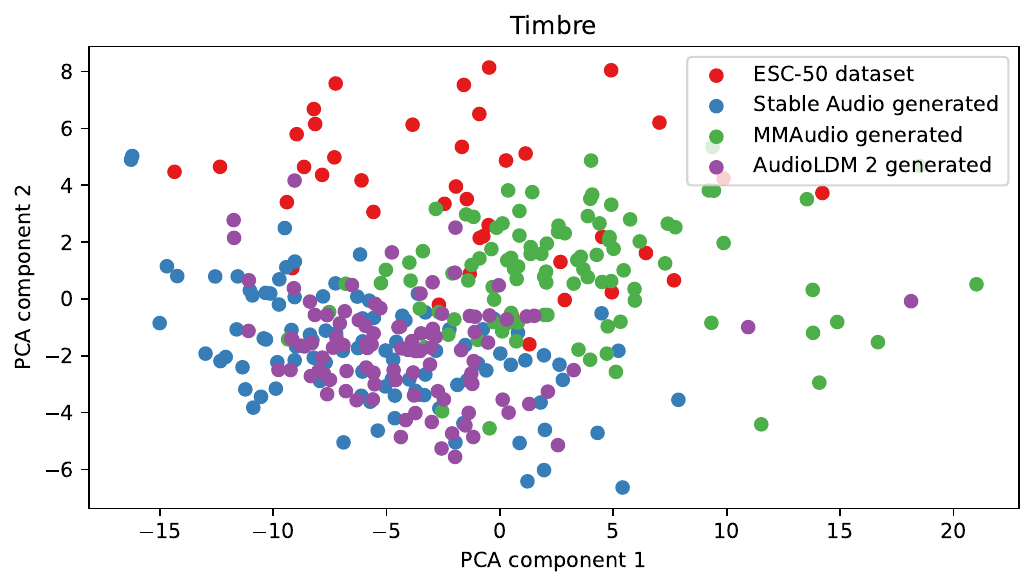}
\subsection{Prompt: Sound of chainsaw}\centering
\includegraphics[width=0.68\textwidth]{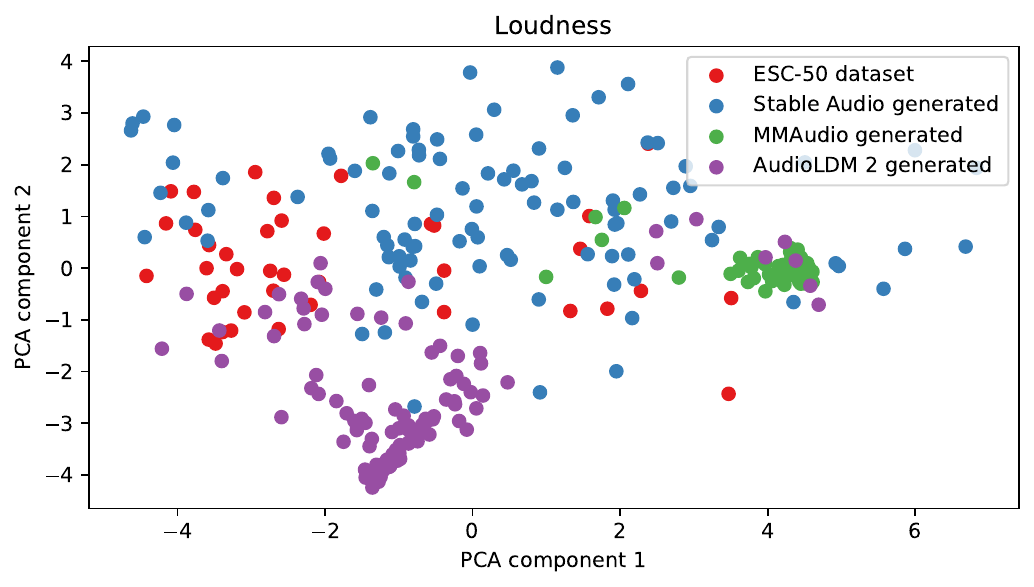}
\includegraphics[width=0.68\textwidth]{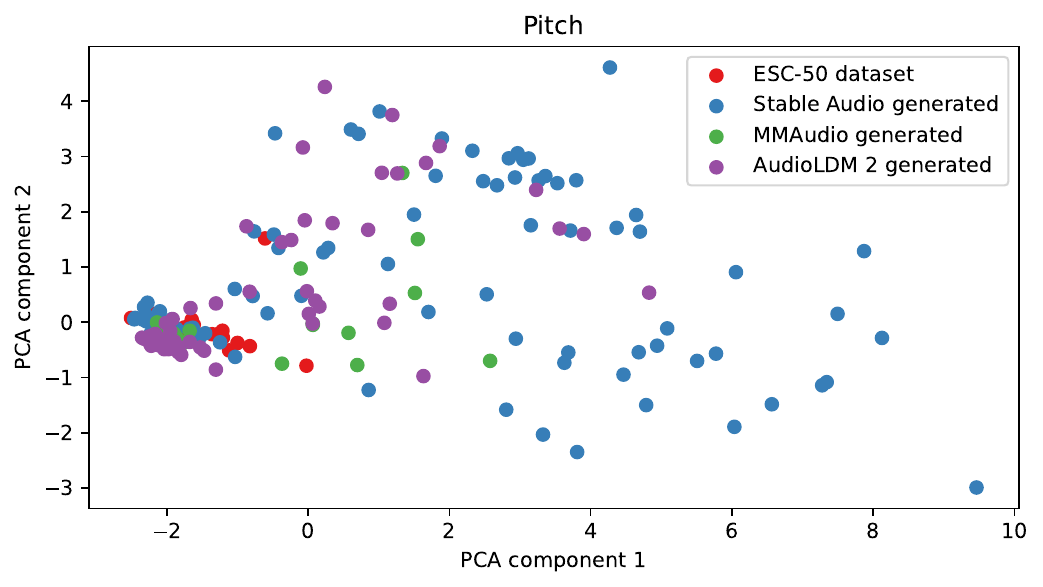}
\includegraphics[width=0.68\textwidth]{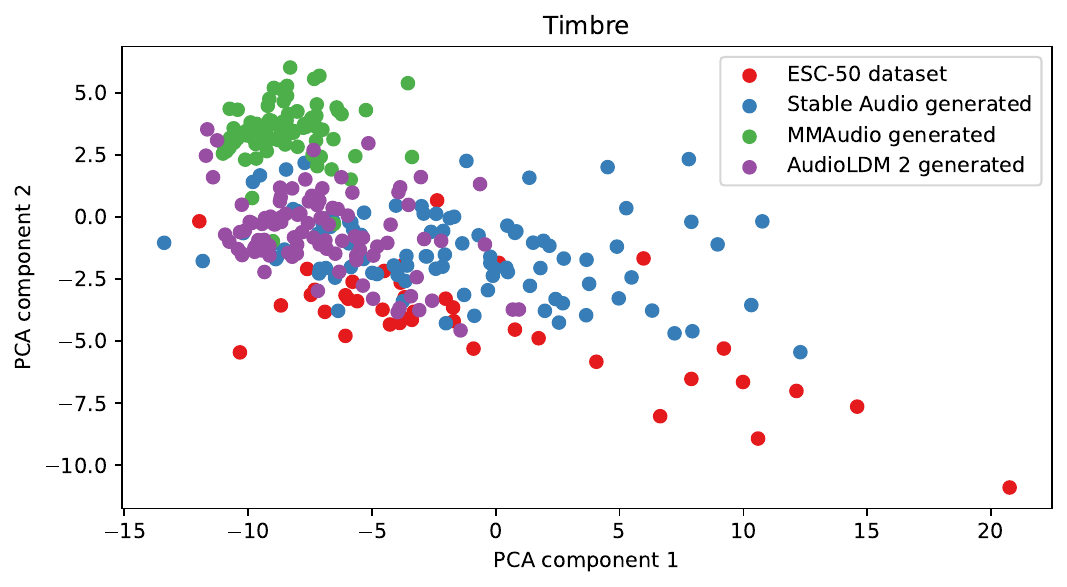}
\subsection{Prompt: Sound of chirping birds}\centering
\includegraphics[width=0.68\textwidth]{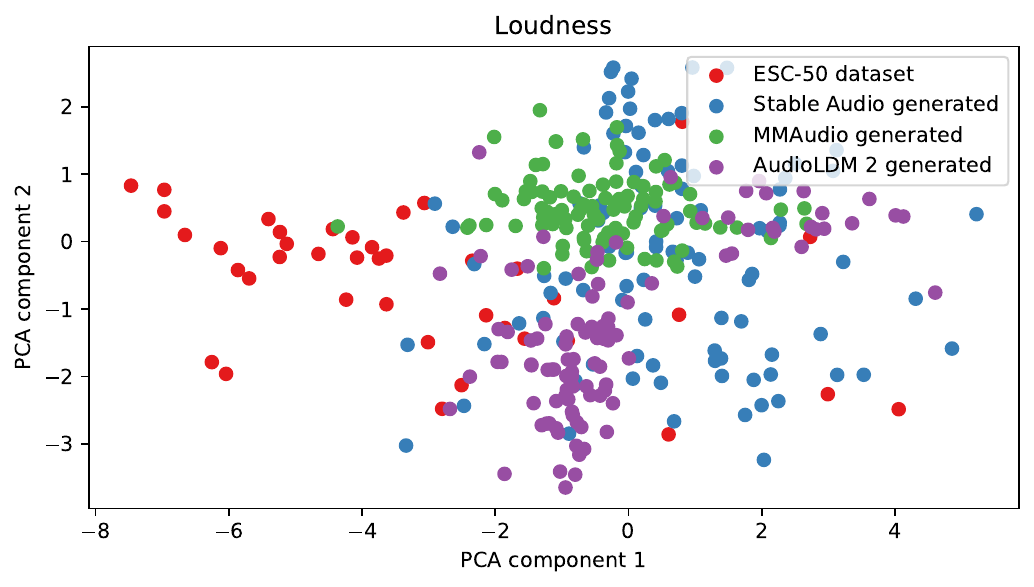}
\includegraphics[width=0.68\textwidth]{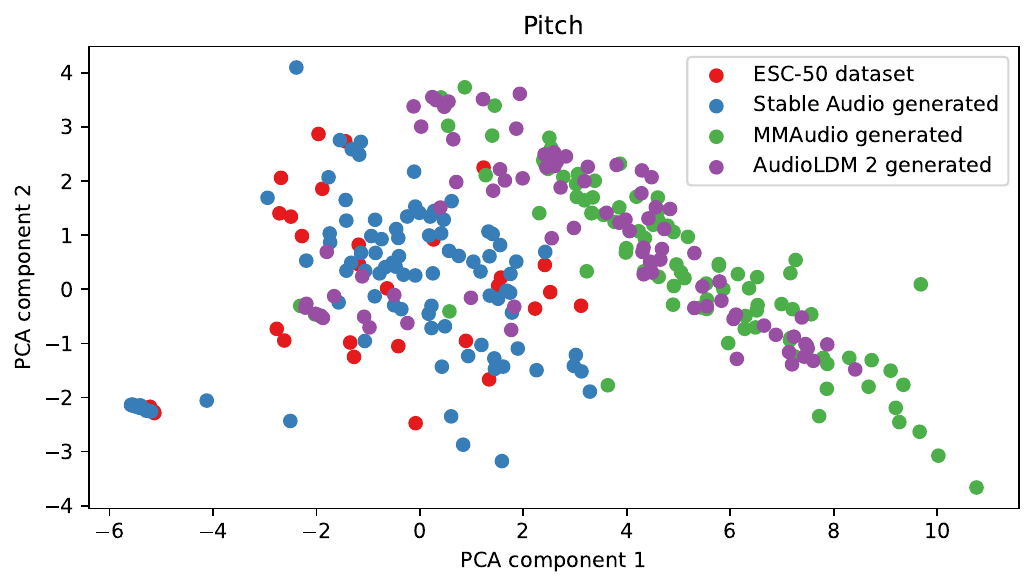}
\includegraphics[width=0.68\textwidth]{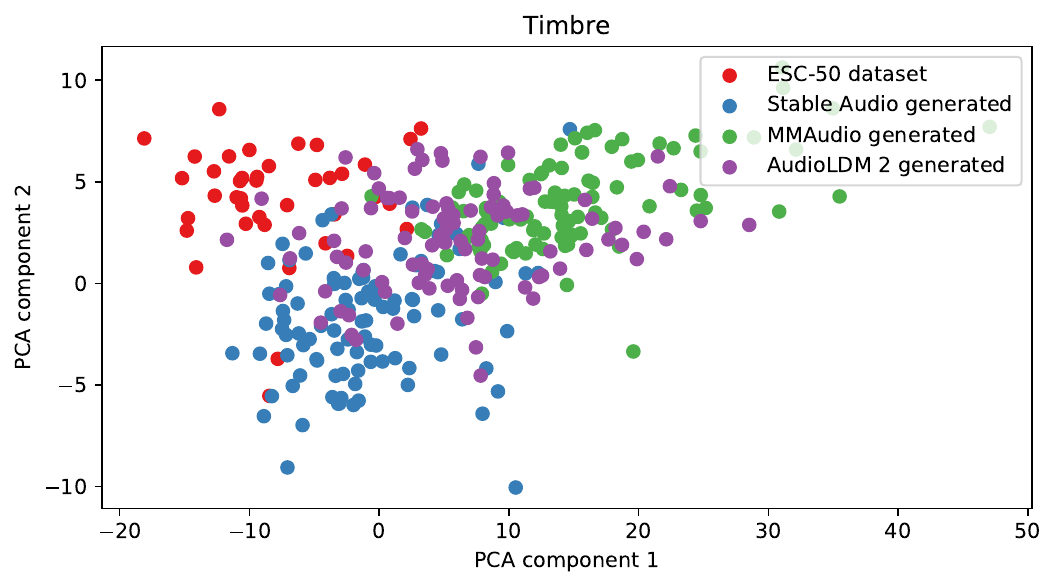}
\subsection{Prompt: Sound of church bells}\centering
\includegraphics[width=0.68\textwidth]{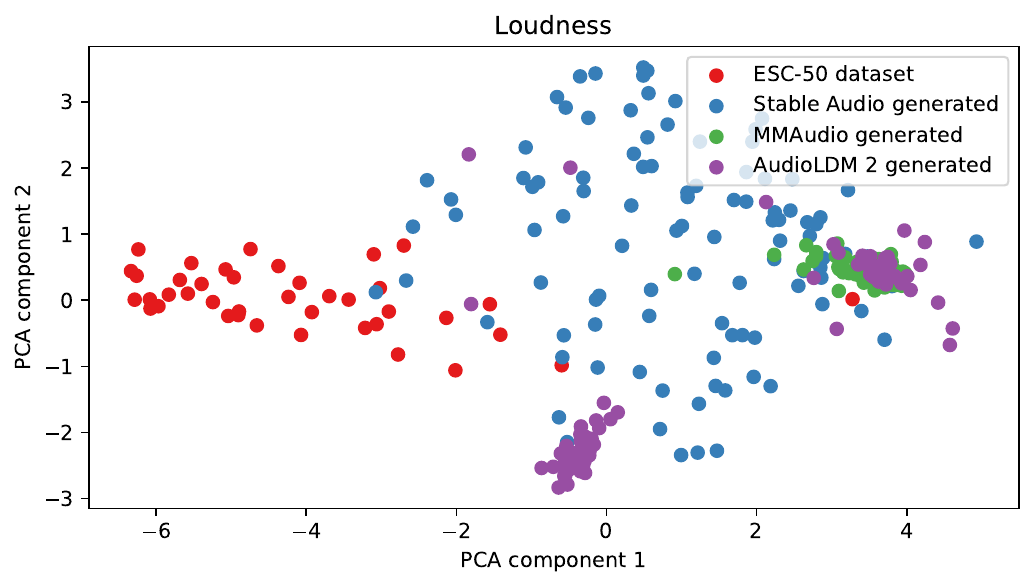}
\includegraphics[width=0.68\textwidth]{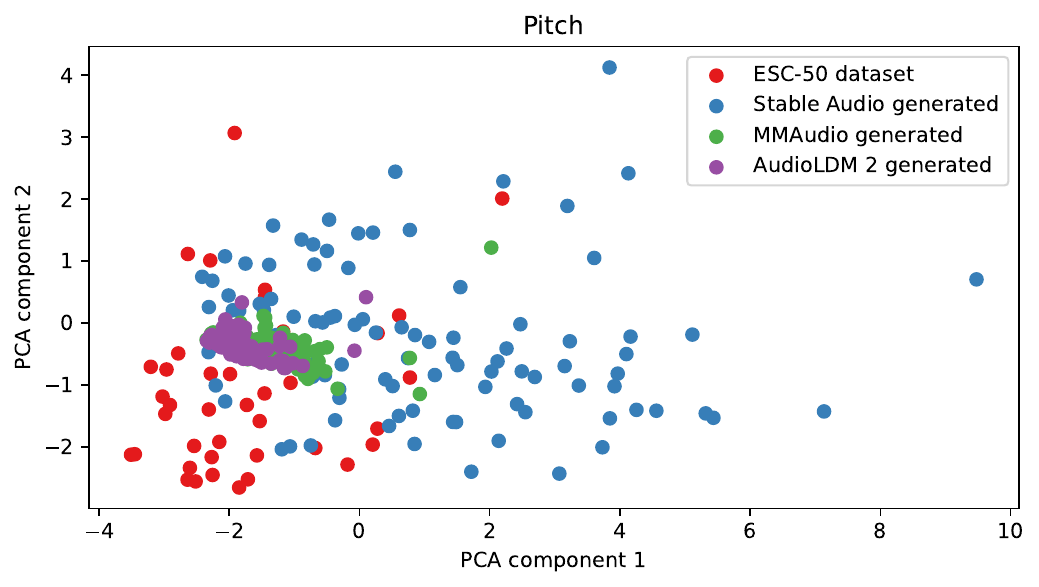}
\includegraphics[width=0.68\textwidth]{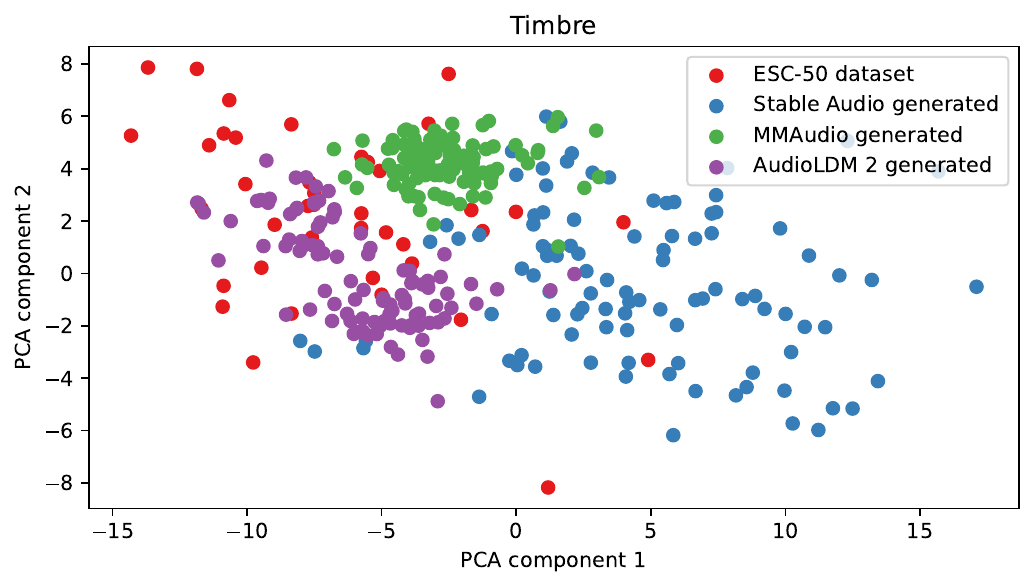}
\subsection{Prompt: Sound of clapping}\centering
\includegraphics[width=0.68\textwidth]{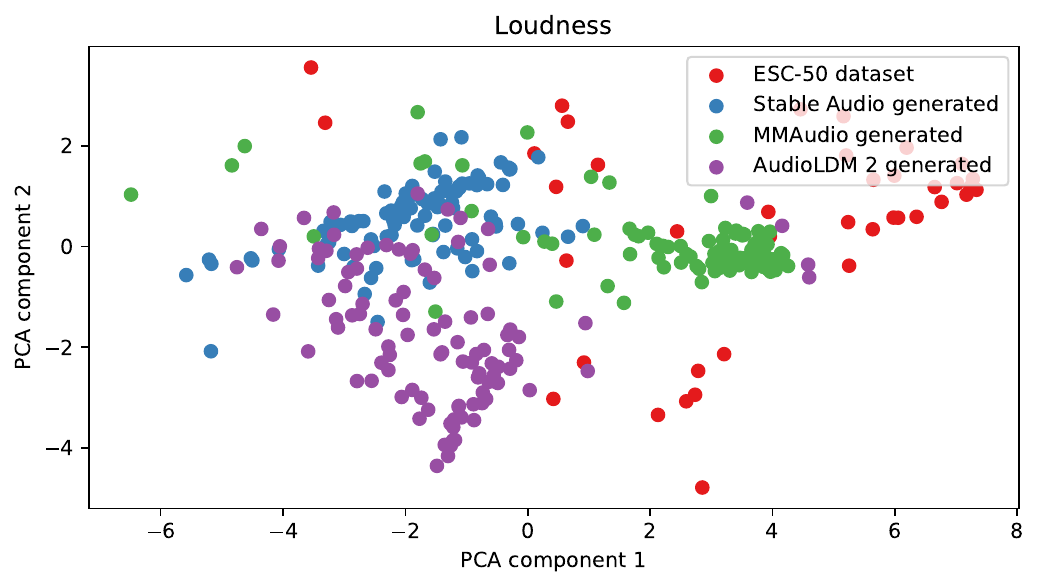}
\includegraphics[width=0.68\textwidth]{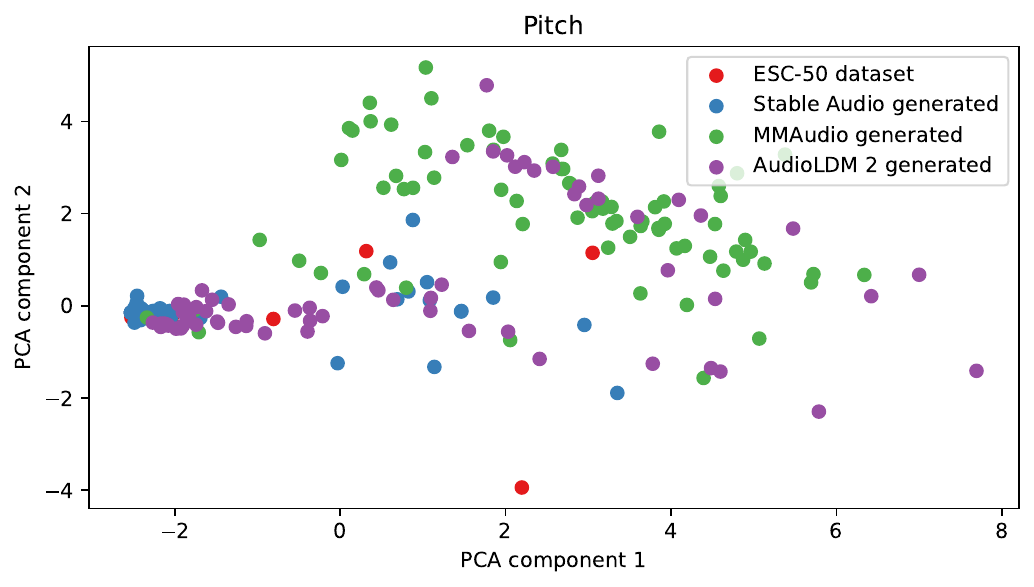}
\includegraphics[width=0.68\textwidth]{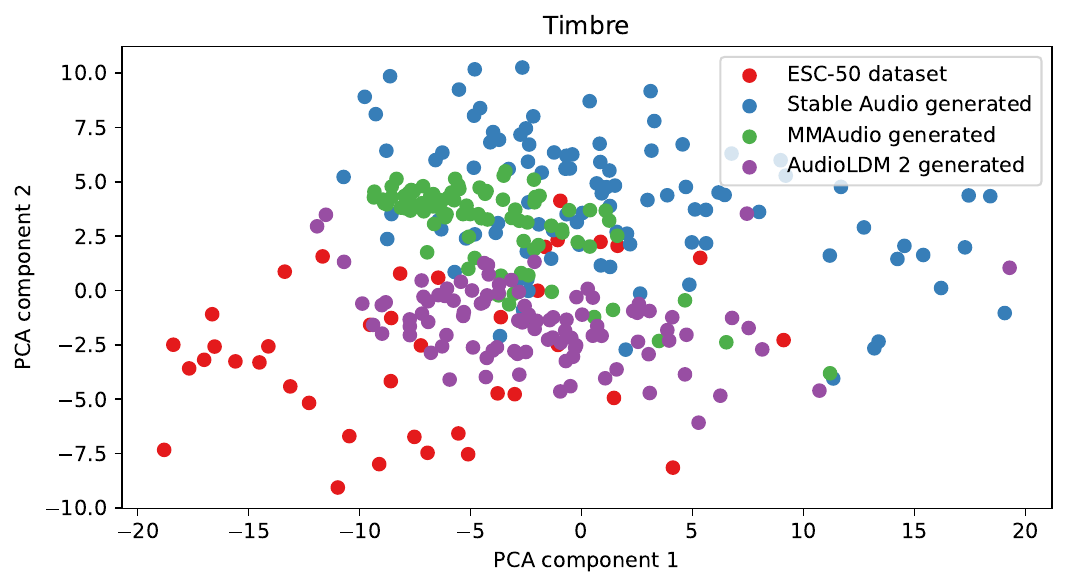}
\subsection{Prompt: Sound of clock alarm}\centering
\includegraphics[width=0.68\textwidth]{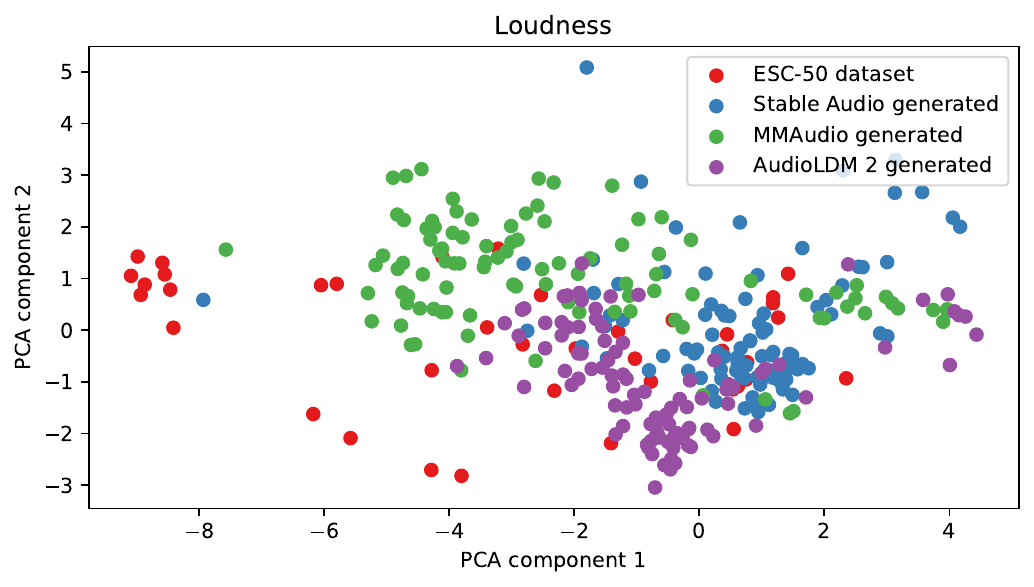}
\includegraphics[width=0.68\textwidth]{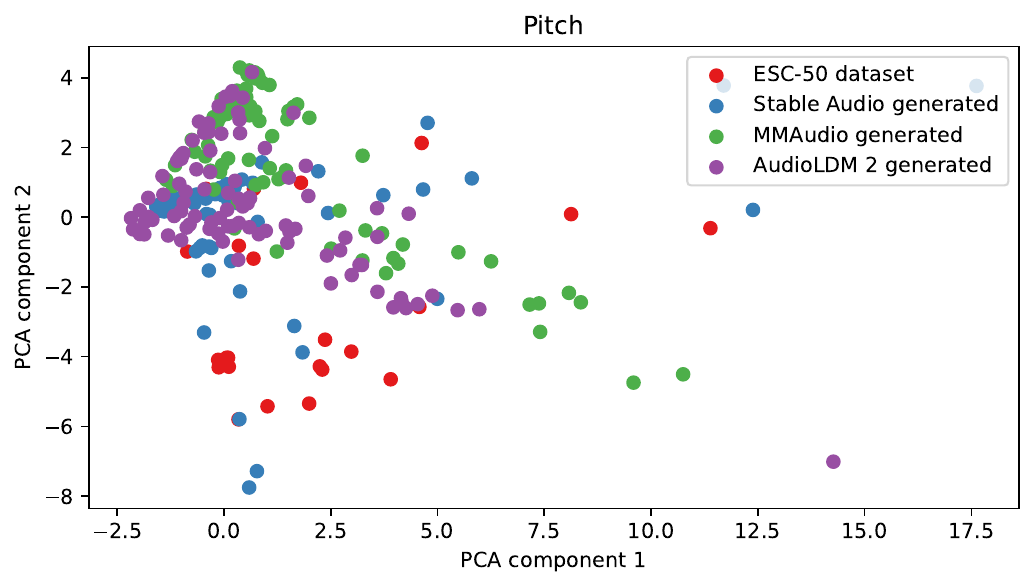}
\includegraphics[width=0.68\textwidth]{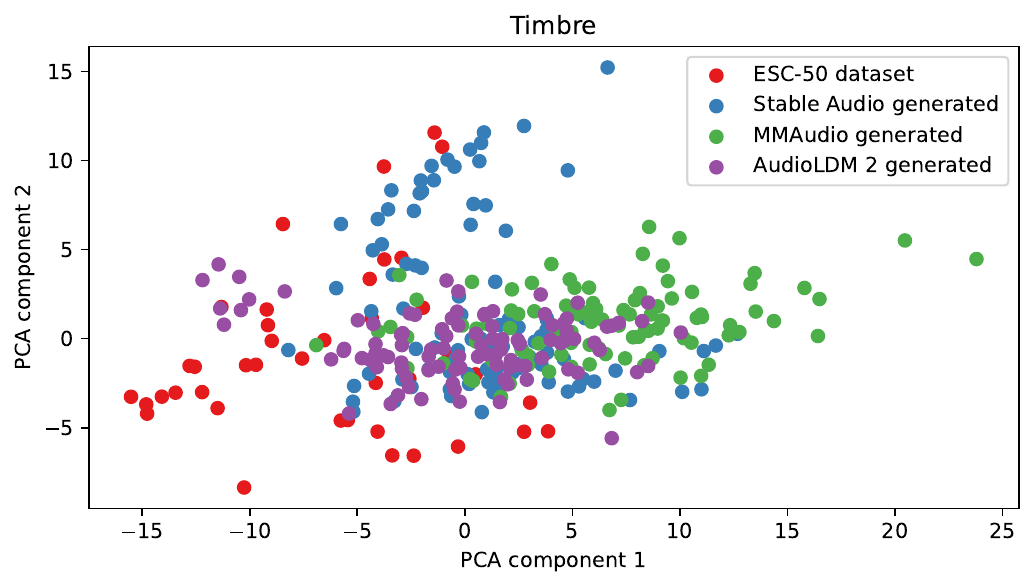}
\subsection{Prompt: Sound of clock tick}\centering
\includegraphics[width=0.68\textwidth]{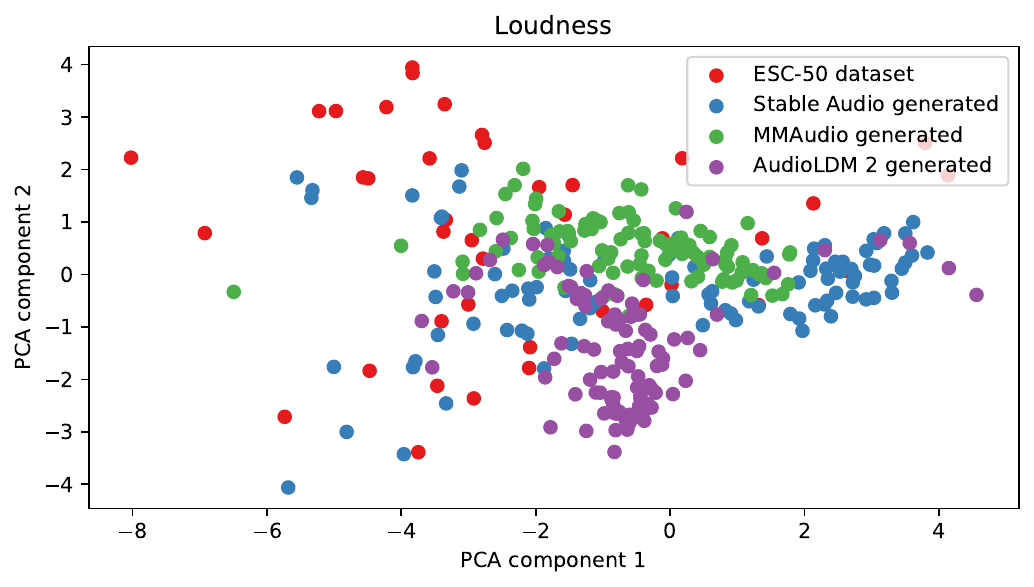}
\includegraphics[width=0.68\textwidth]{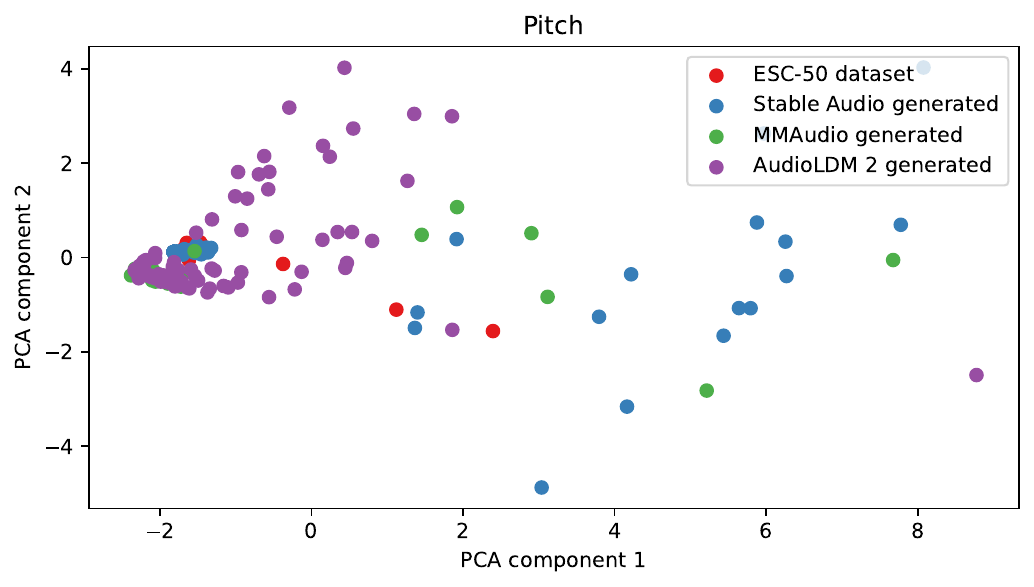}
\includegraphics[width=0.68\textwidth]{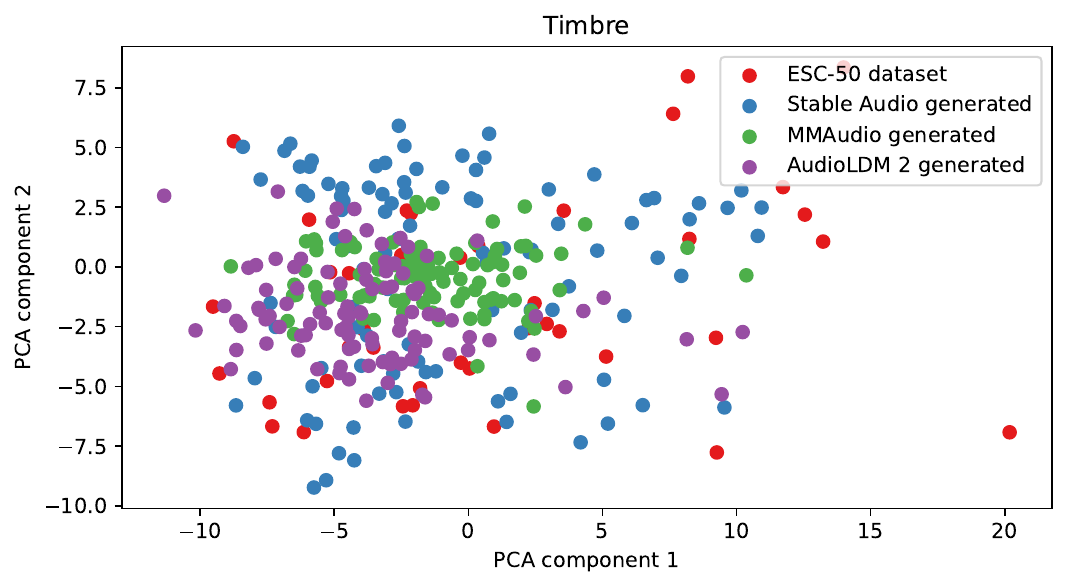}
\subsection{Prompt: Sound of coughing}\centering
\includegraphics[width=0.68\textwidth]{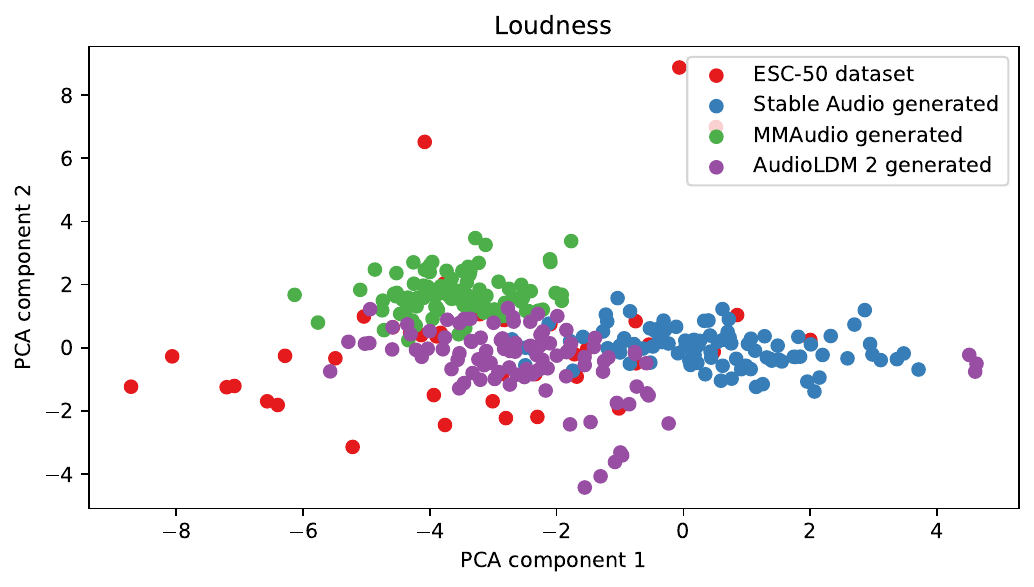}
\includegraphics[width=0.68\textwidth]{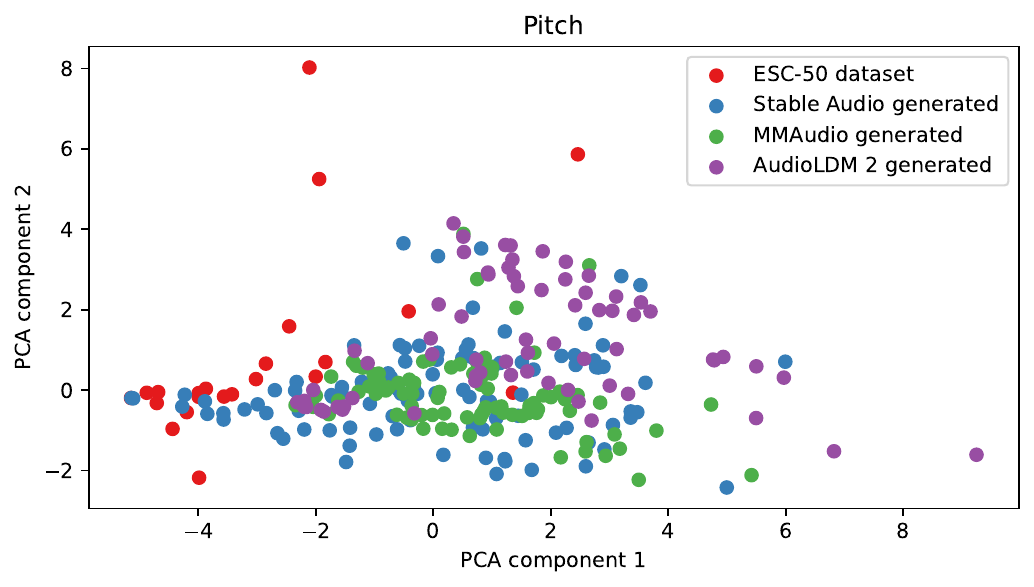}
\includegraphics[width=0.68\textwidth]{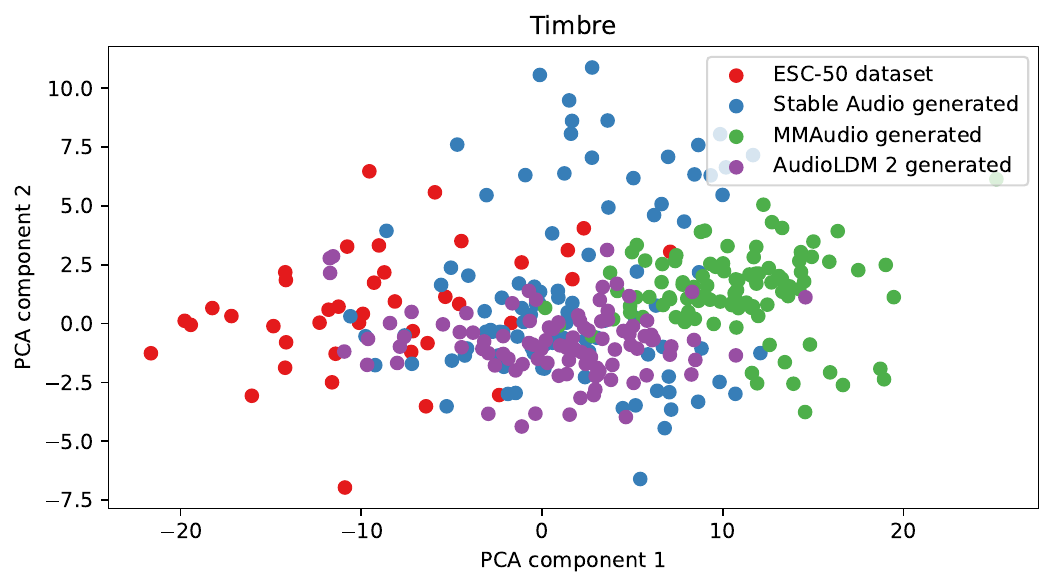}
\subsection{Prompt: Sound of cow}\centering
\includegraphics[width=0.68\textwidth]{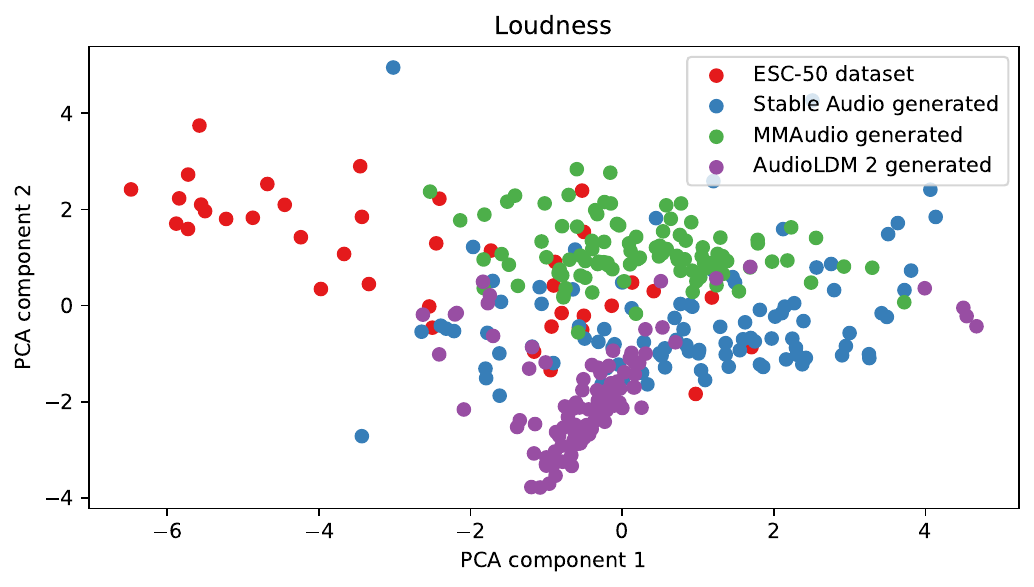}
\includegraphics[width=0.68\textwidth]{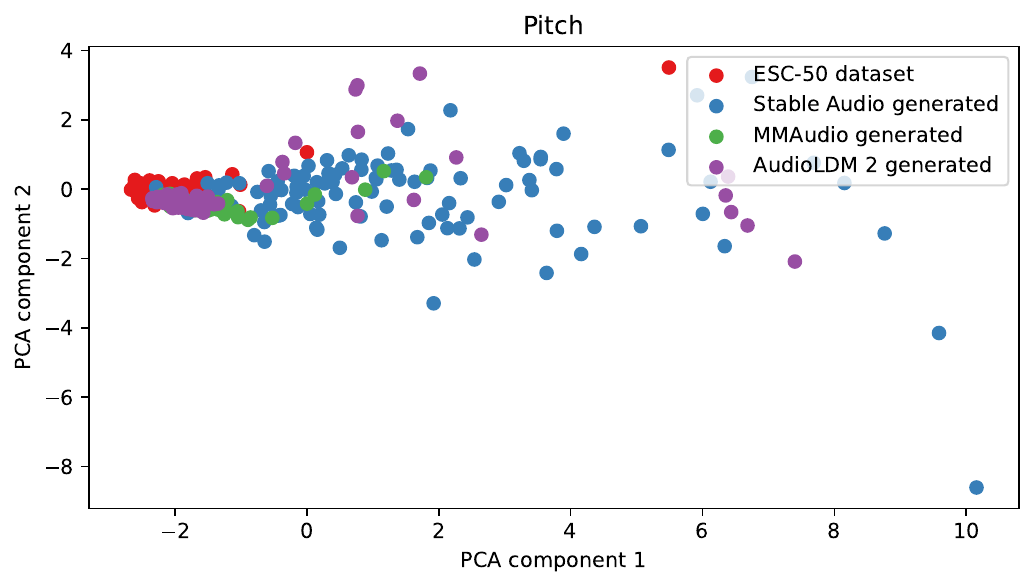}
\includegraphics[width=0.68\textwidth]{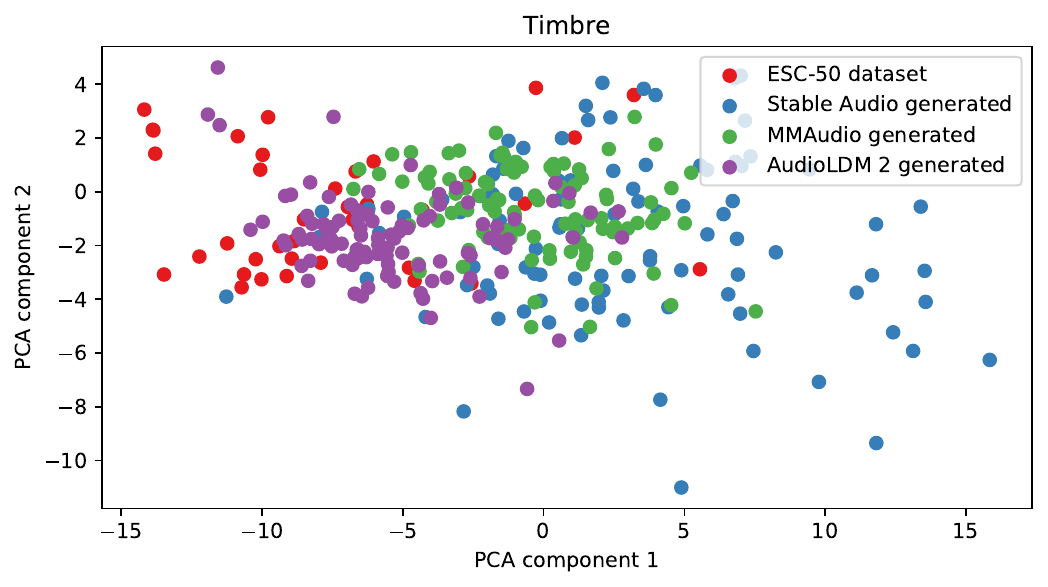}
\subsection{Prompt: Sound of crackling fire}\centering
\includegraphics[width=0.68\textwidth]{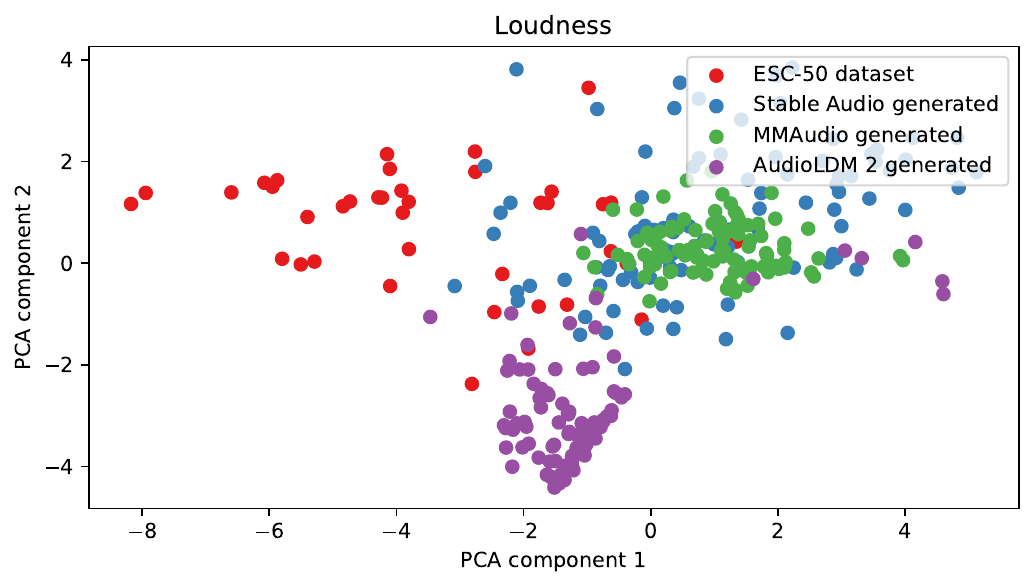}
\includegraphics[width=0.68\textwidth]{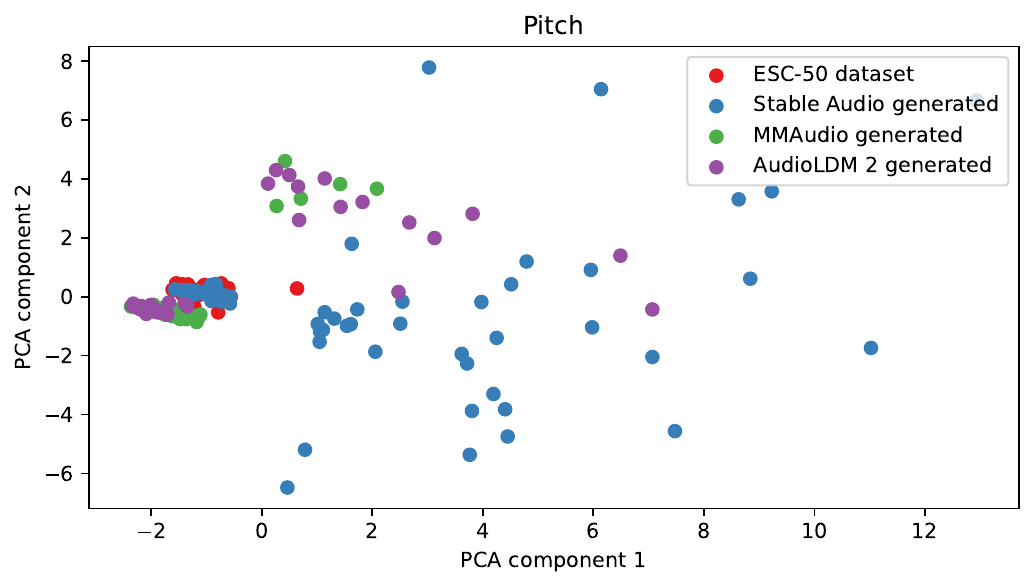}
\includegraphics[width=0.68\textwidth]{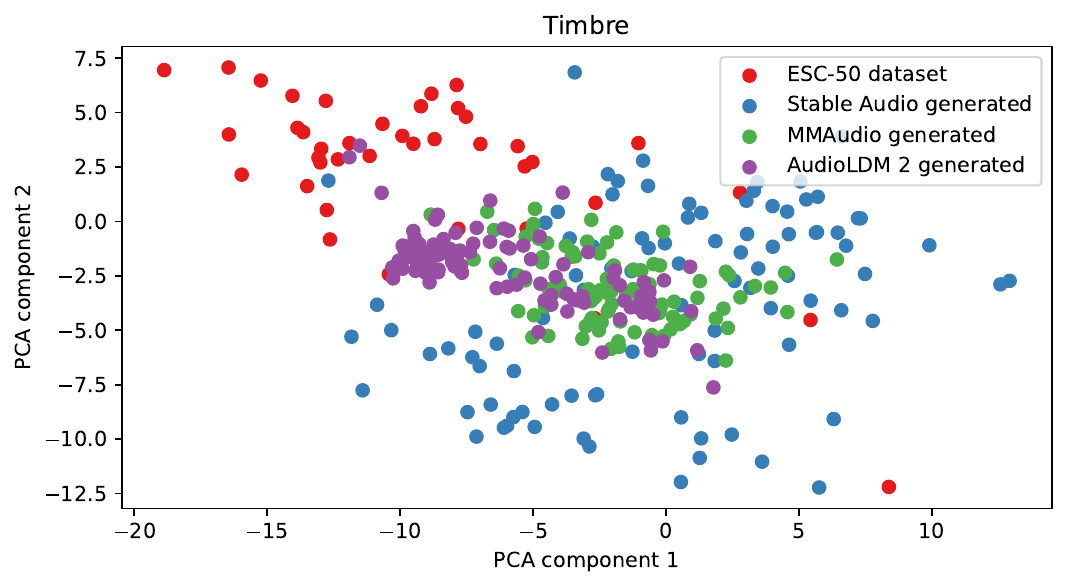}
\subsection{Prompt: Sound of crickets}\centering
\includegraphics[width=0.68\textwidth]{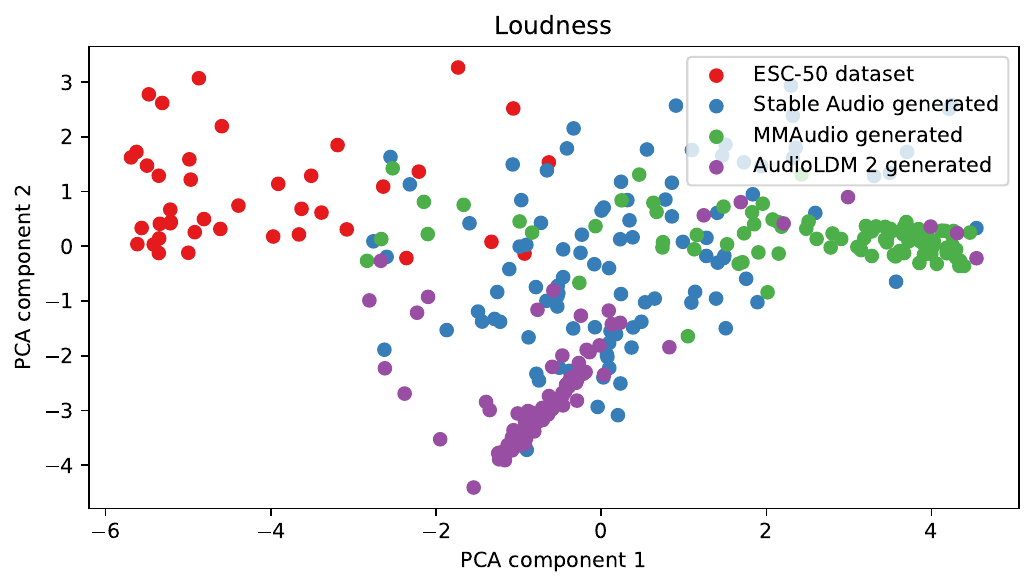}
\includegraphics[width=0.68\textwidth]{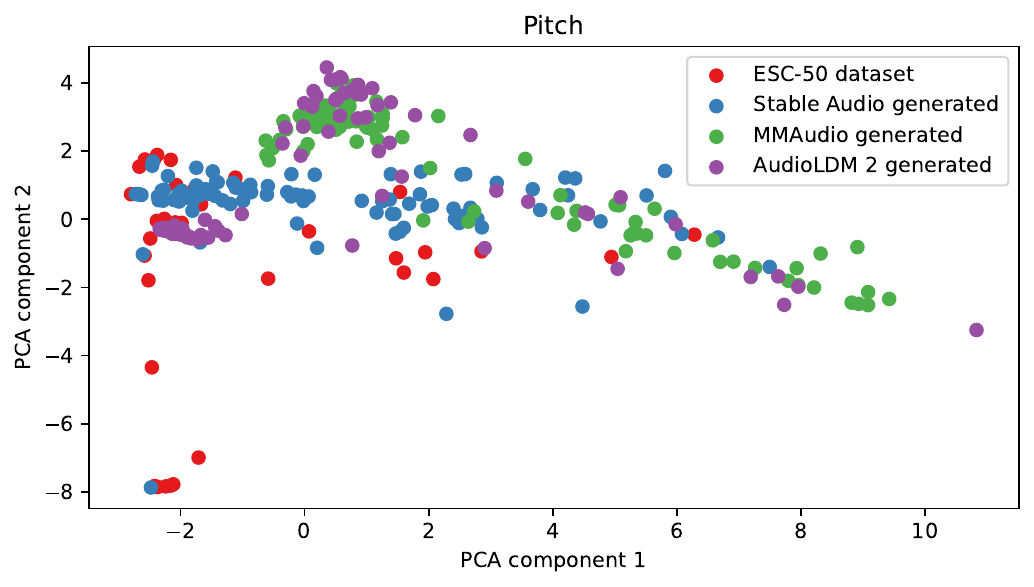}
\includegraphics[width=0.68\textwidth]{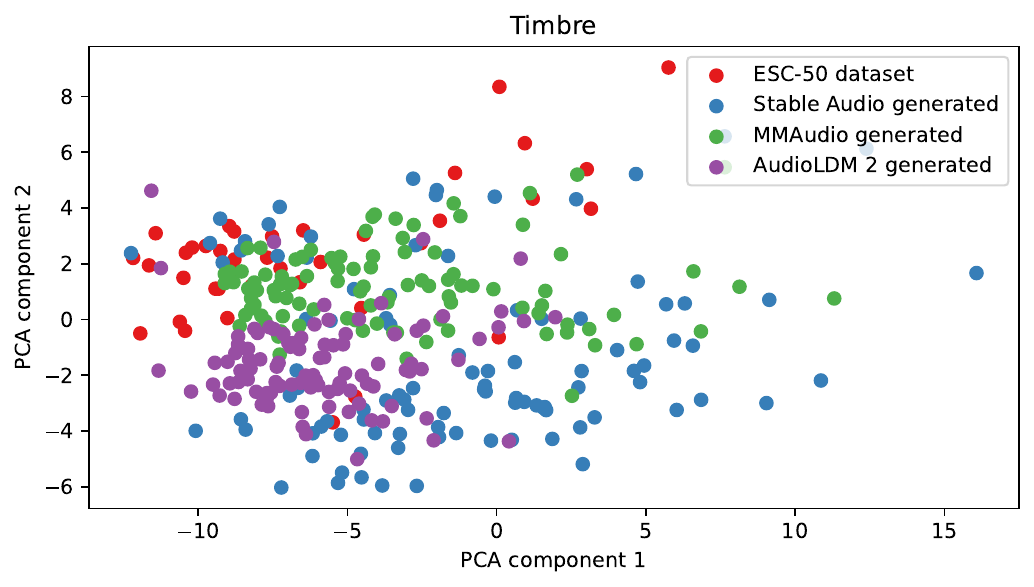}
\subsection{Prompt: Sound of crow}\centering
\includegraphics[width=0.68\textwidth]{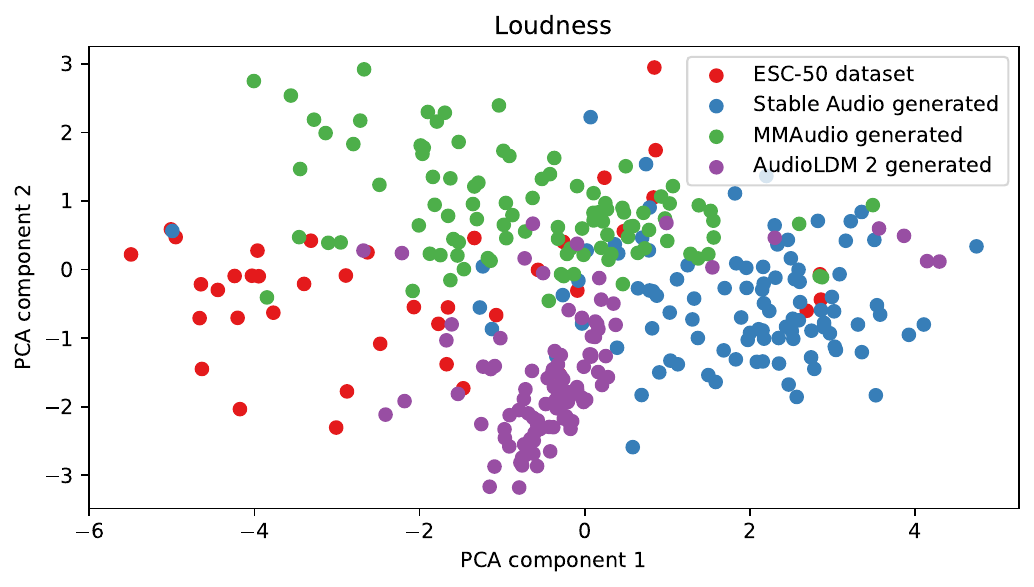}
\includegraphics[width=0.68\textwidth]{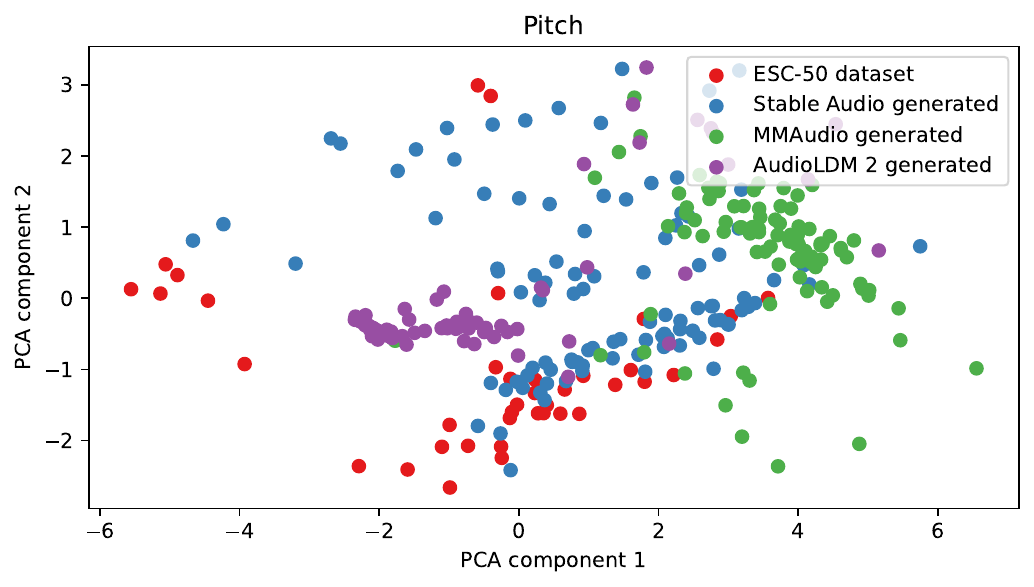}
\includegraphics[width=0.68\textwidth]{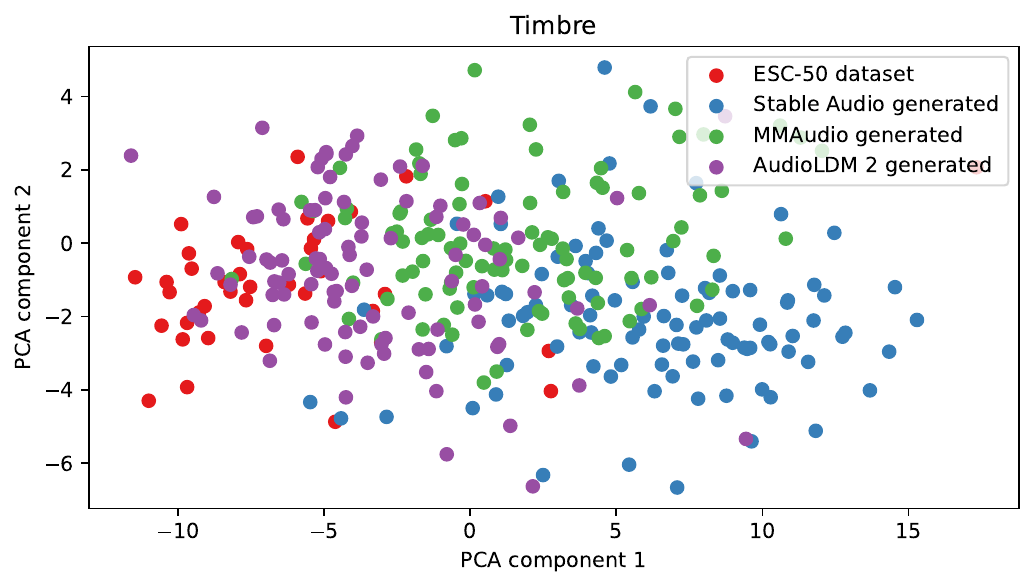}
\subsection{Prompt: Sound of crying baby}\centering
\includegraphics[width=0.68\textwidth]{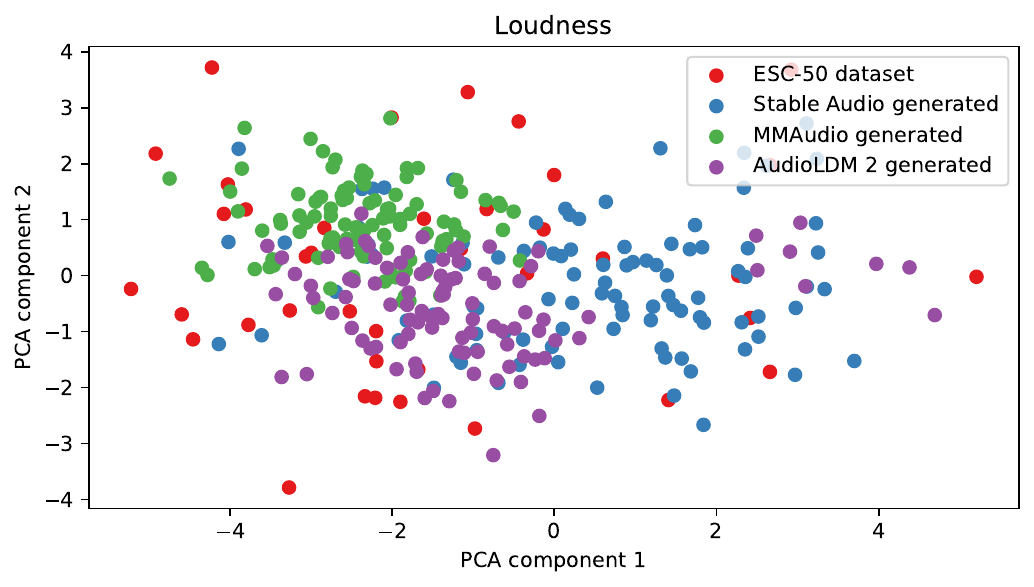}
\includegraphics[width=0.68\textwidth]{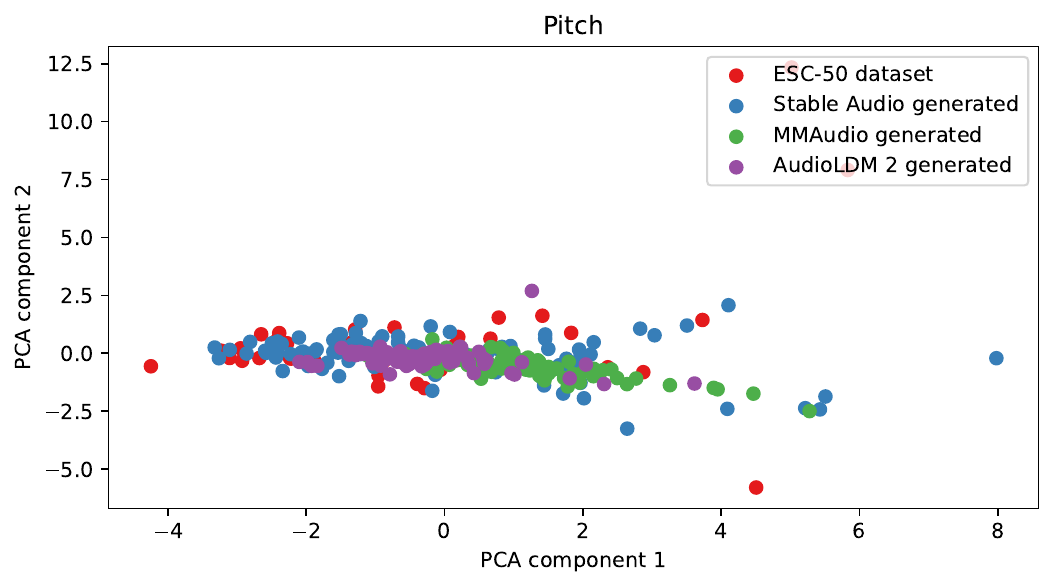}
\includegraphics[width=0.68\textwidth]{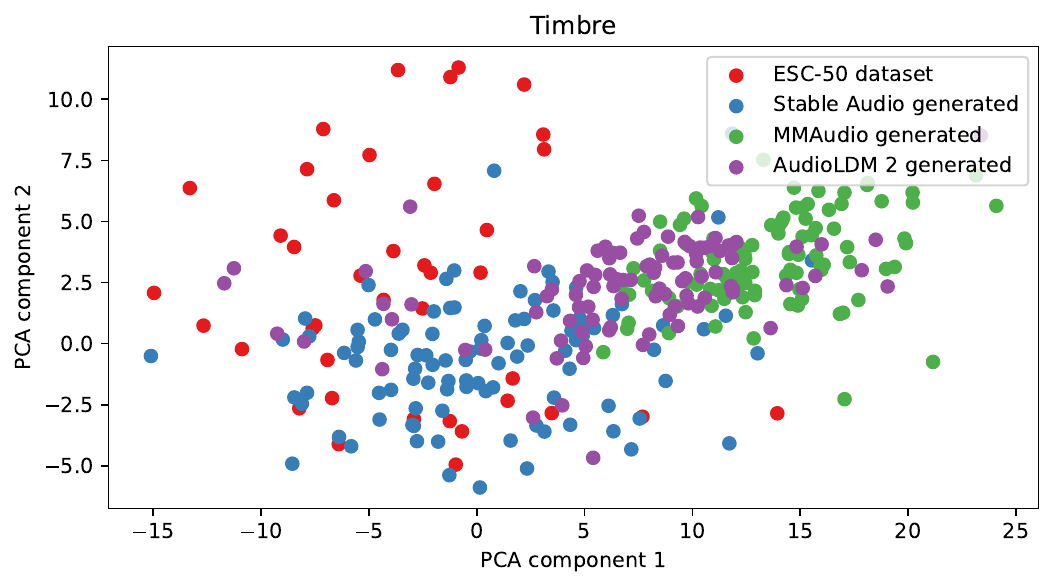}
\subsection{Prompt: Sound of dog}\centering
\includegraphics[width=0.68\textwidth]{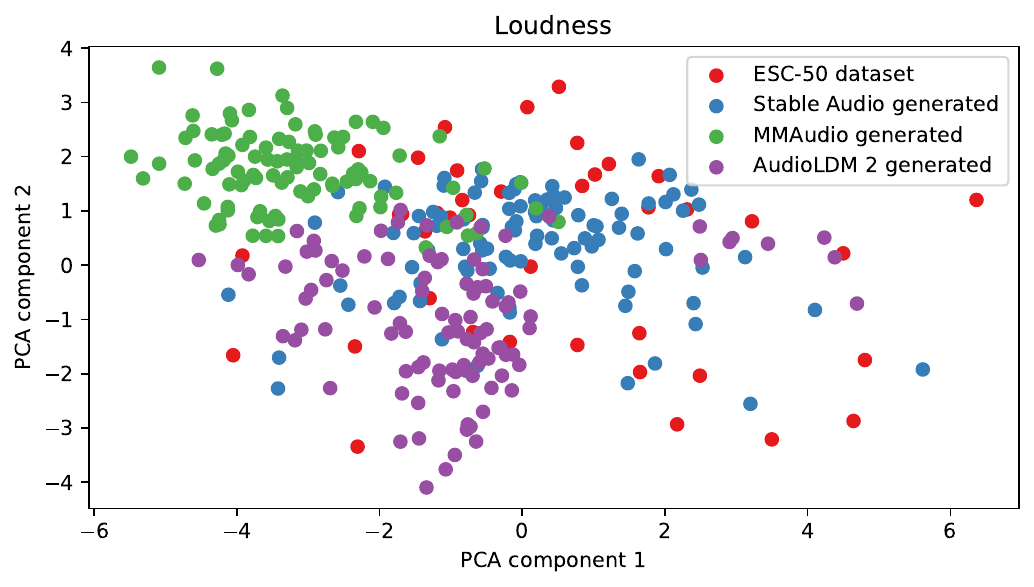}
\includegraphics[width=0.68\textwidth]{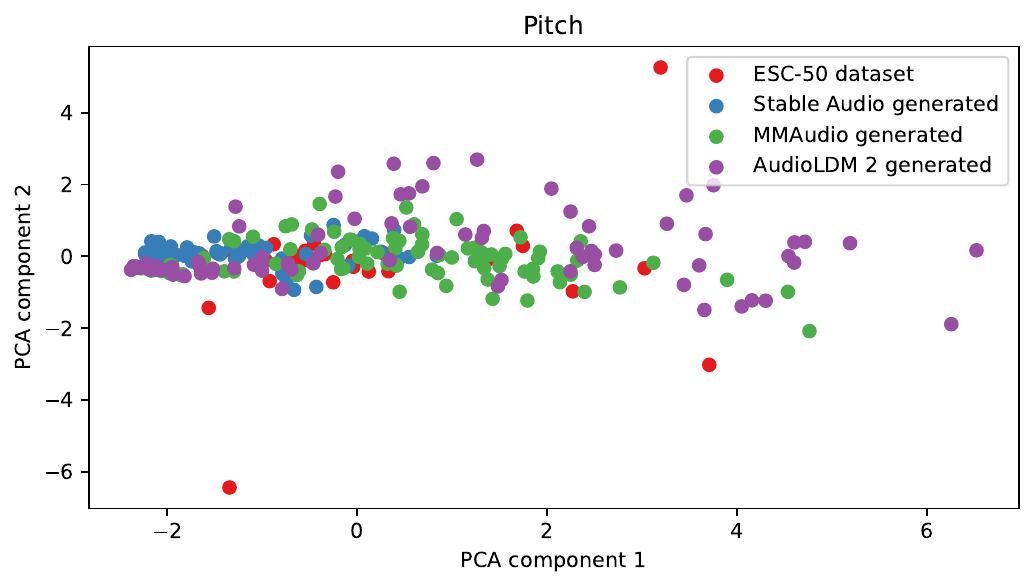}
\includegraphics[width=0.68\textwidth]{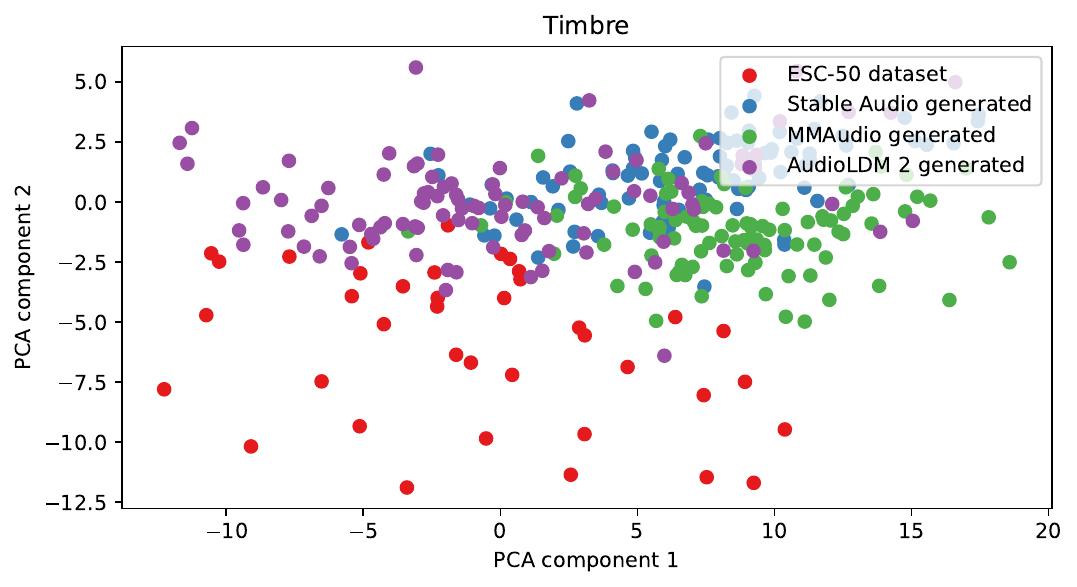}
\subsection{Prompt: Sound of door, wood creaks}\centering
\includegraphics[width=0.68\textwidth]{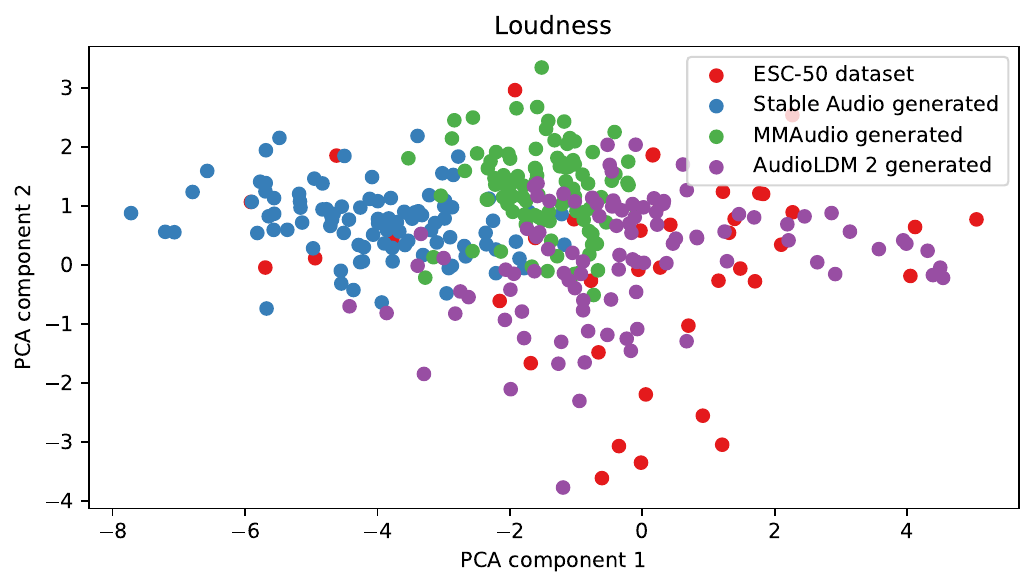}
\includegraphics[width=0.68\textwidth]{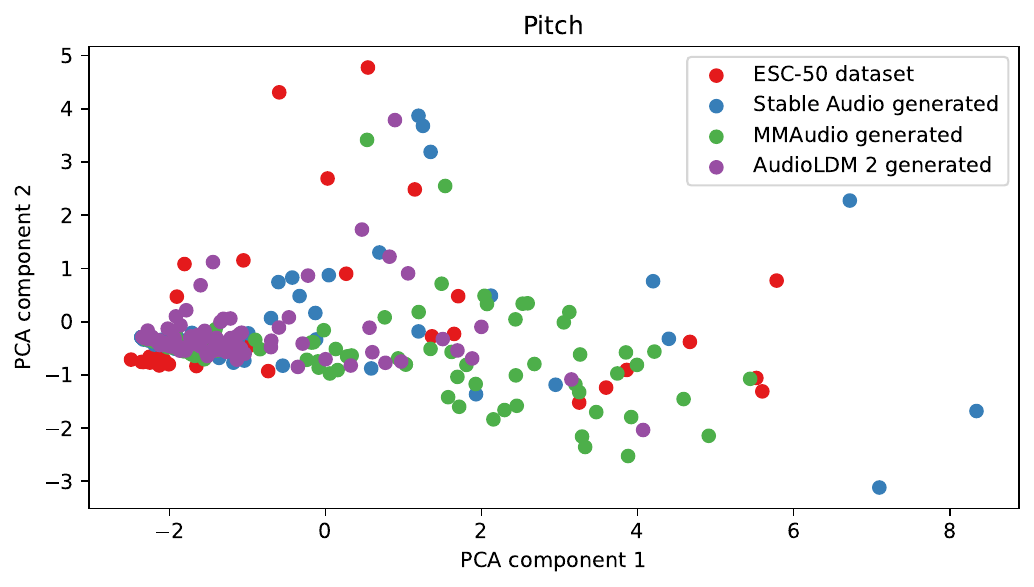}
\includegraphics[width=0.68\textwidth]{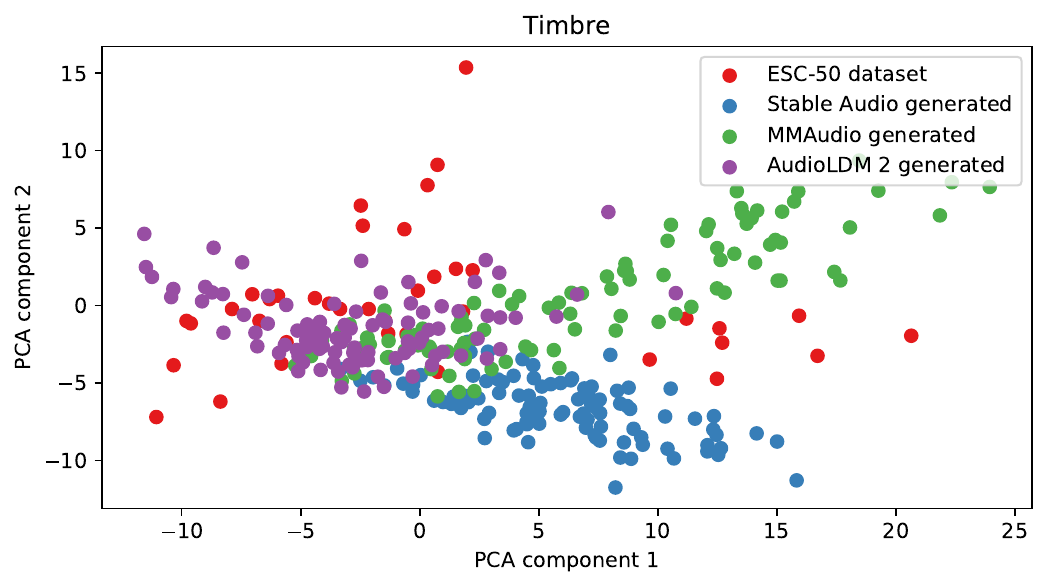}
\subsection{Prompt: Sound of door knock}\centering
\includegraphics[width=0.68\textwidth]{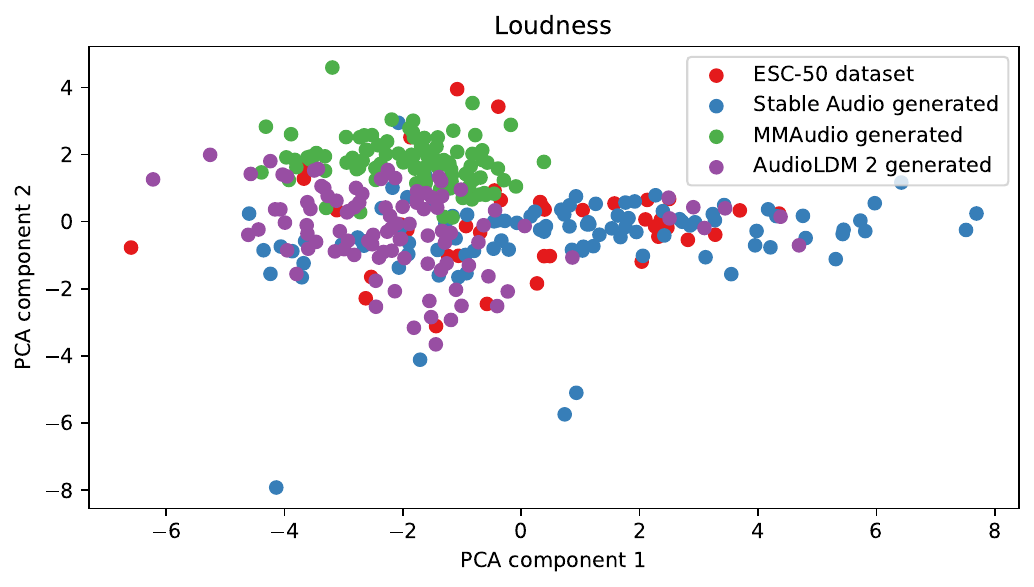}
\includegraphics[width=0.68\textwidth]{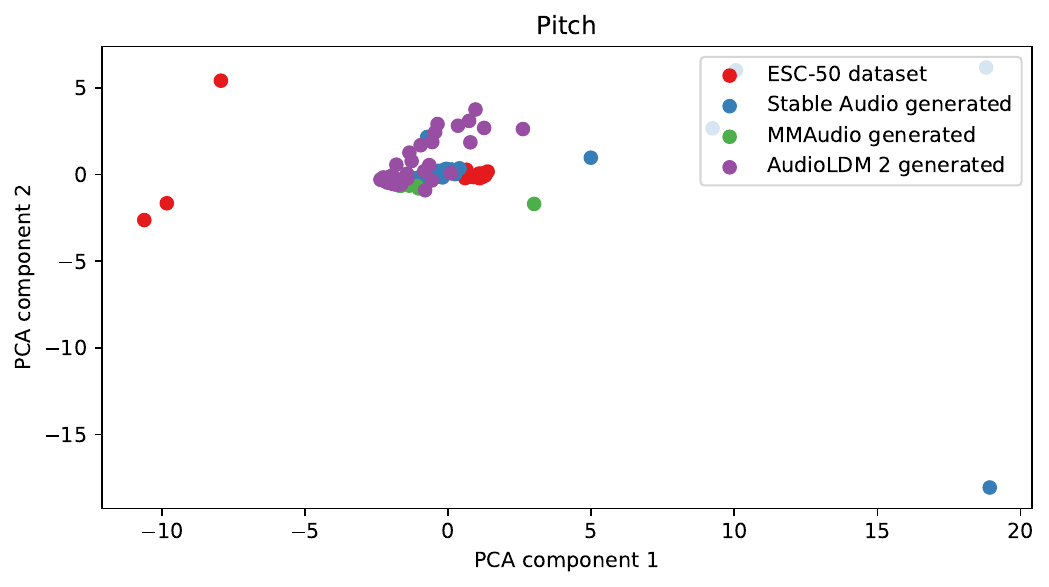}
\includegraphics[width=0.68\textwidth]{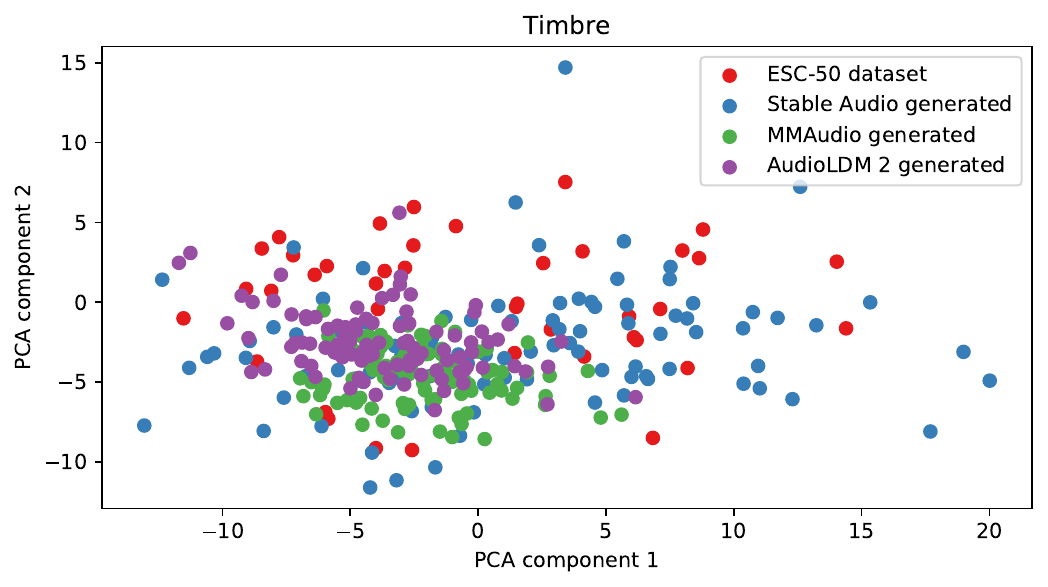}
\subsection{Prompt: Sound of drinking, sipping}\centering
\includegraphics[width=0.68\textwidth]{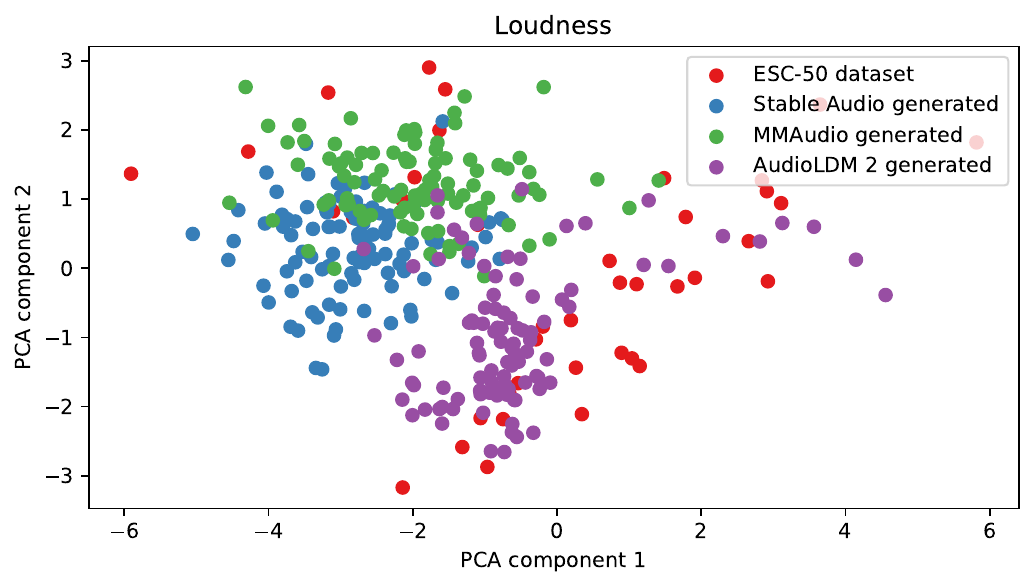}
\includegraphics[width=0.68\textwidth]{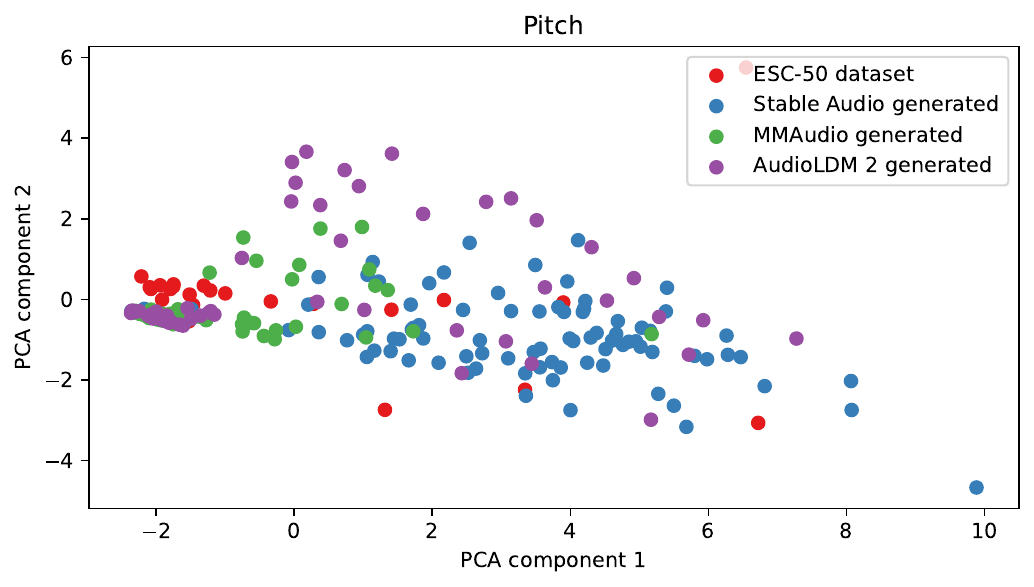}
\includegraphics[width=0.68\textwidth]{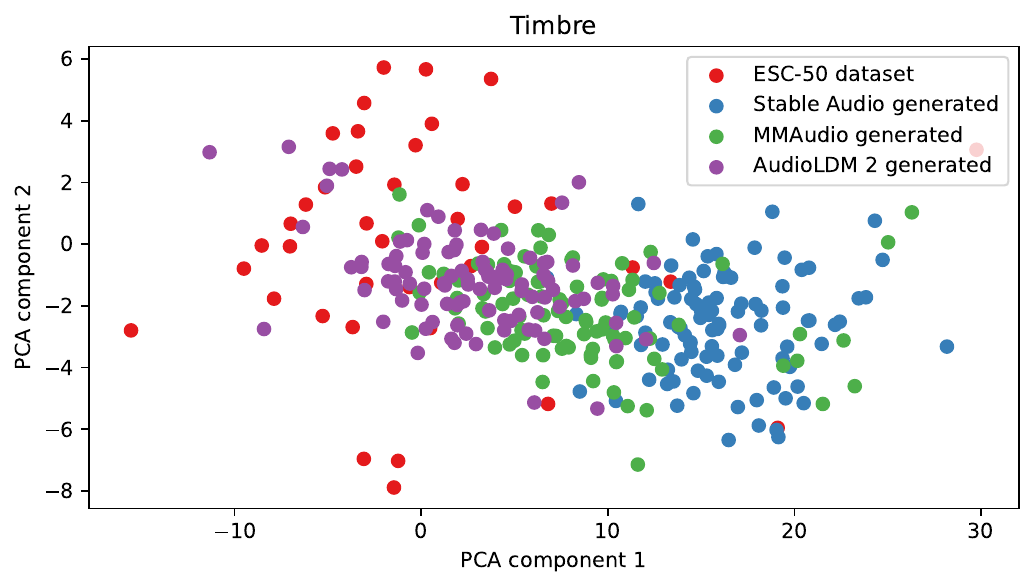}
\subsection{Prompt: Sound of engine}\centering
\includegraphics[width=0.68\textwidth]{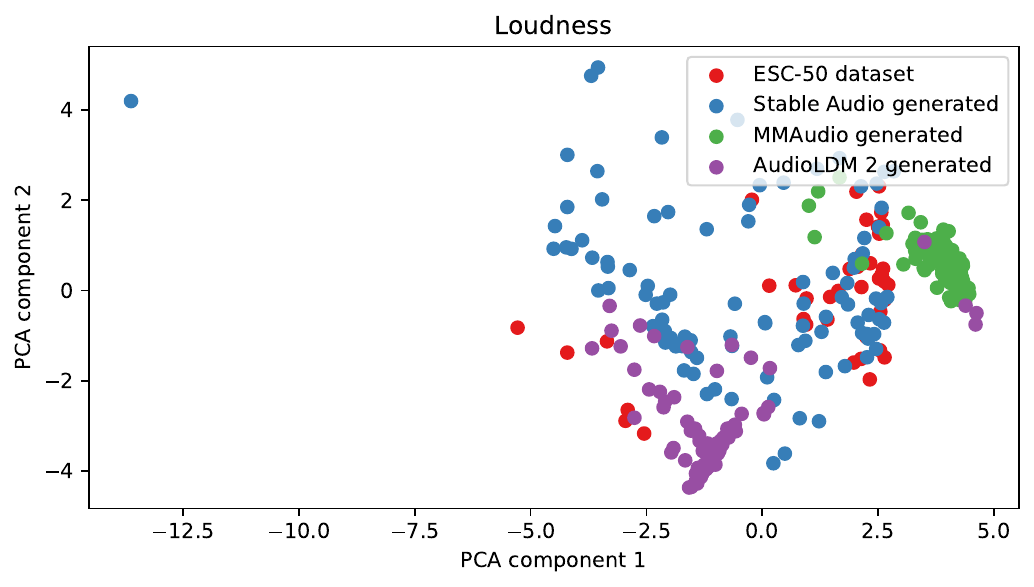}
\includegraphics[width=0.68\textwidth]{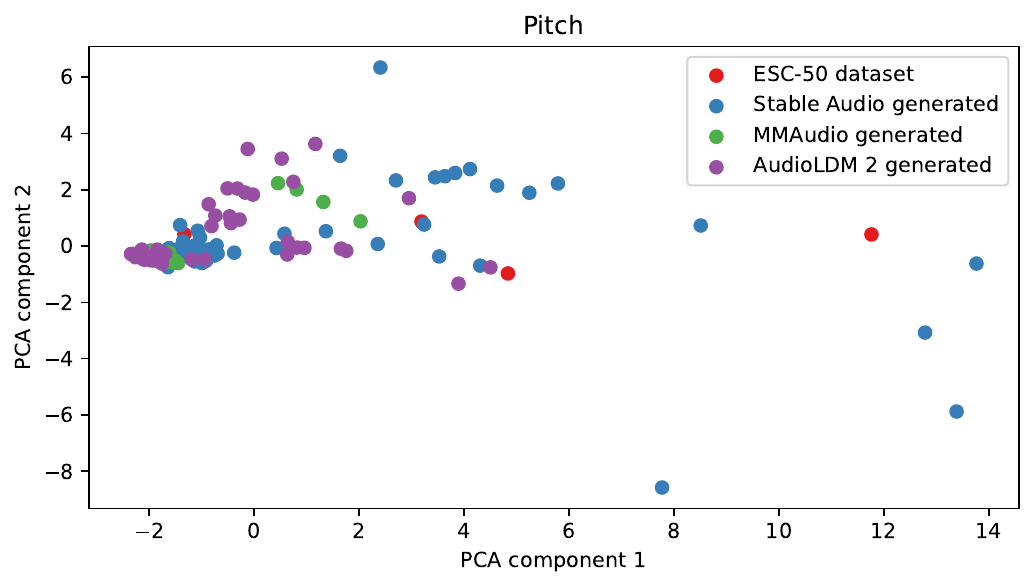}
\includegraphics[width=0.68\textwidth]{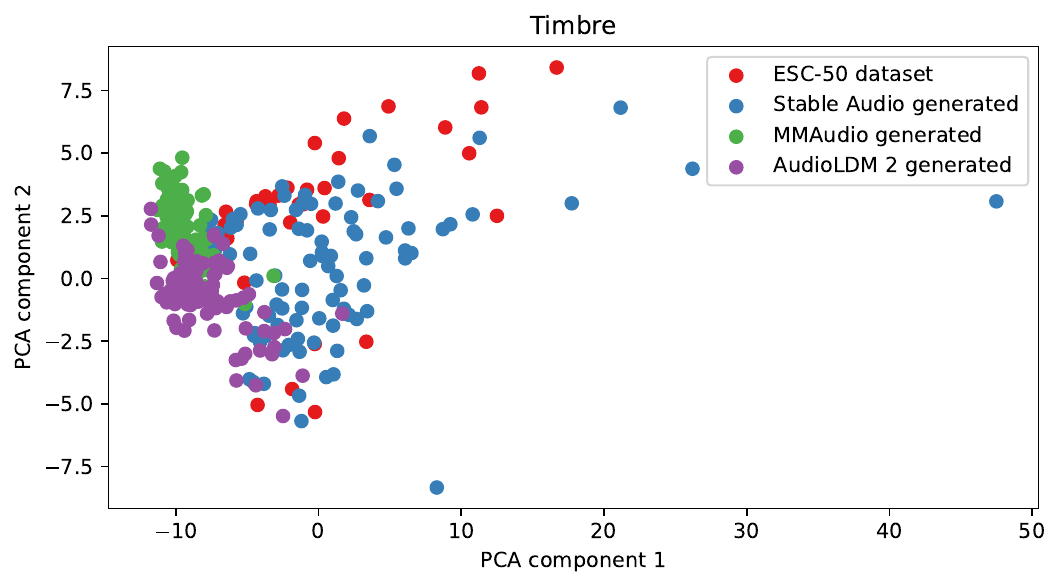}
\subsection{Prompt: Sound of fireworks}\centering
\includegraphics[width=0.68\textwidth]{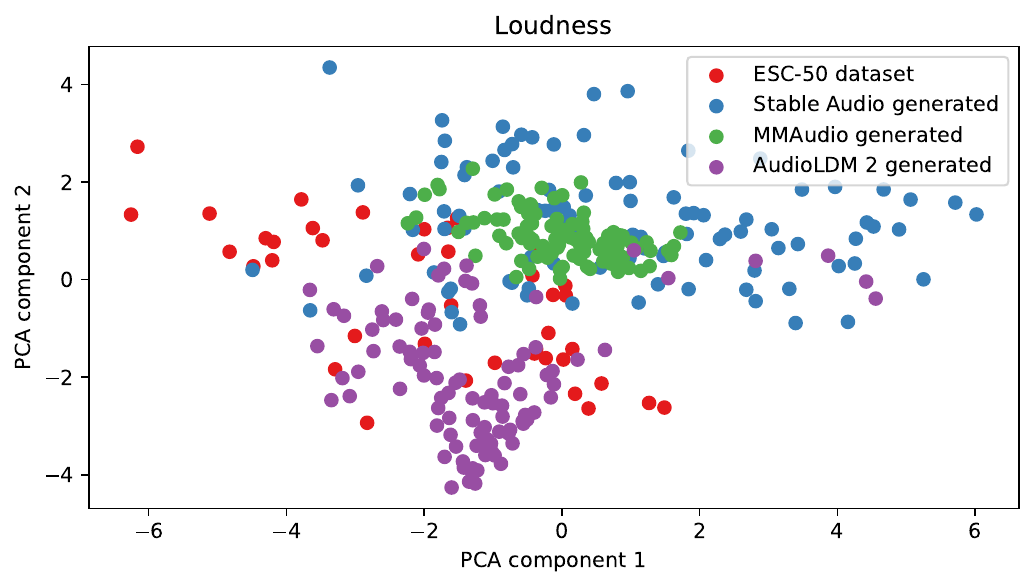}
\includegraphics[width=0.68\textwidth]{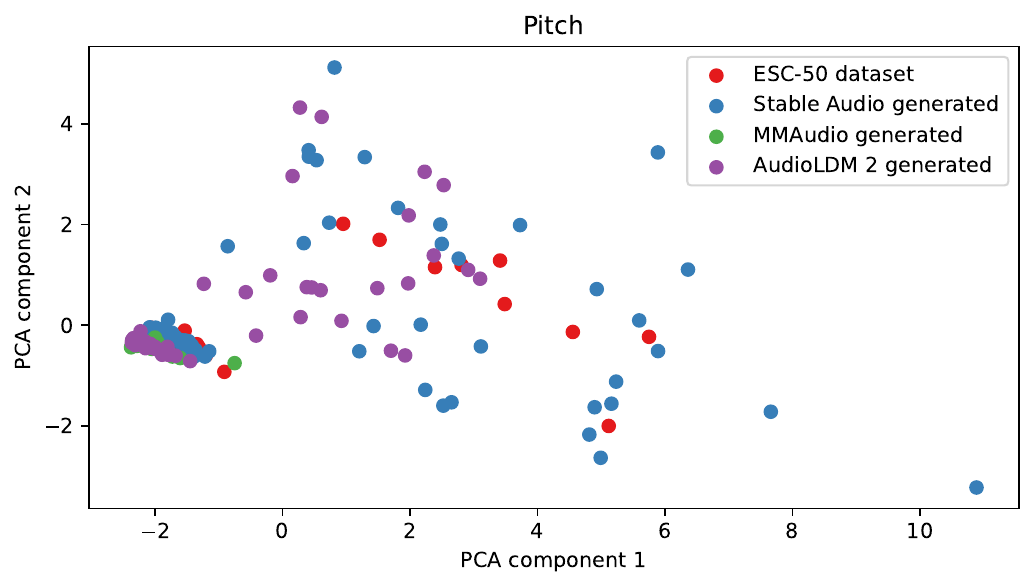}
\includegraphics[width=0.68\textwidth]{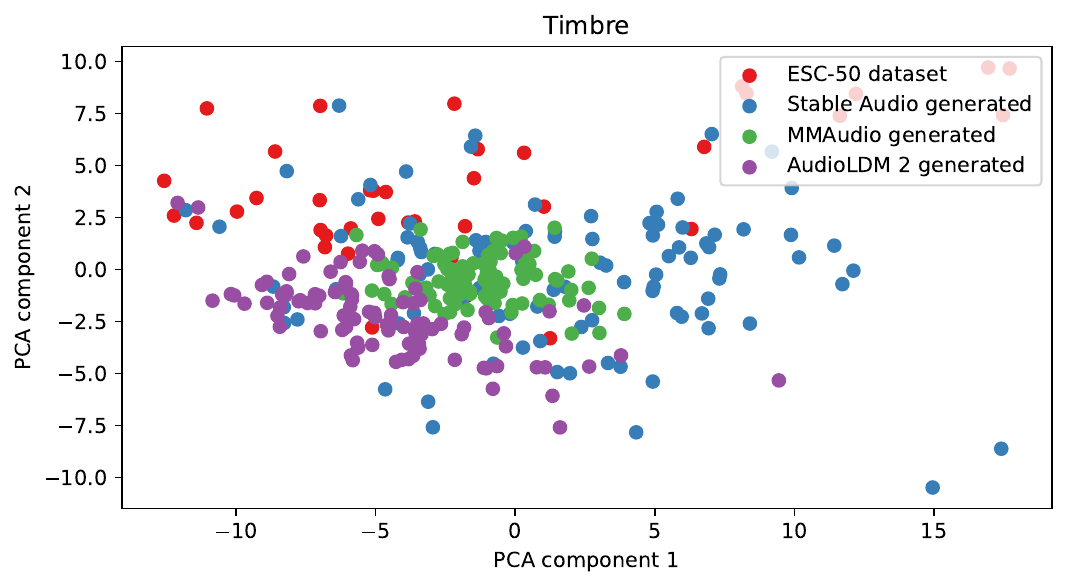}
\subsection{Prompt: Sound of footsteps}\centering
\includegraphics[width=0.68\textwidth]{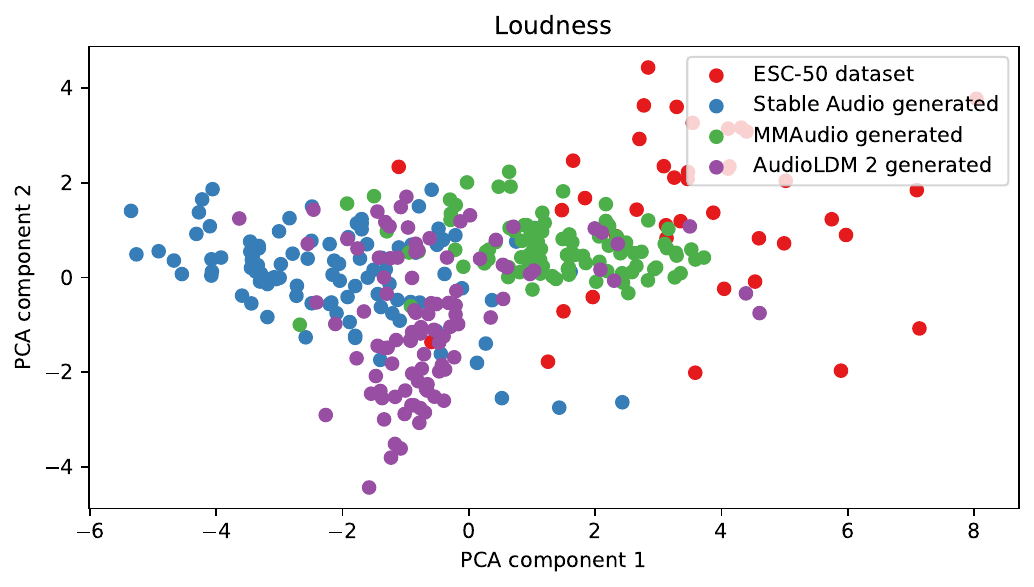}
\includegraphics[width=0.68\textwidth]{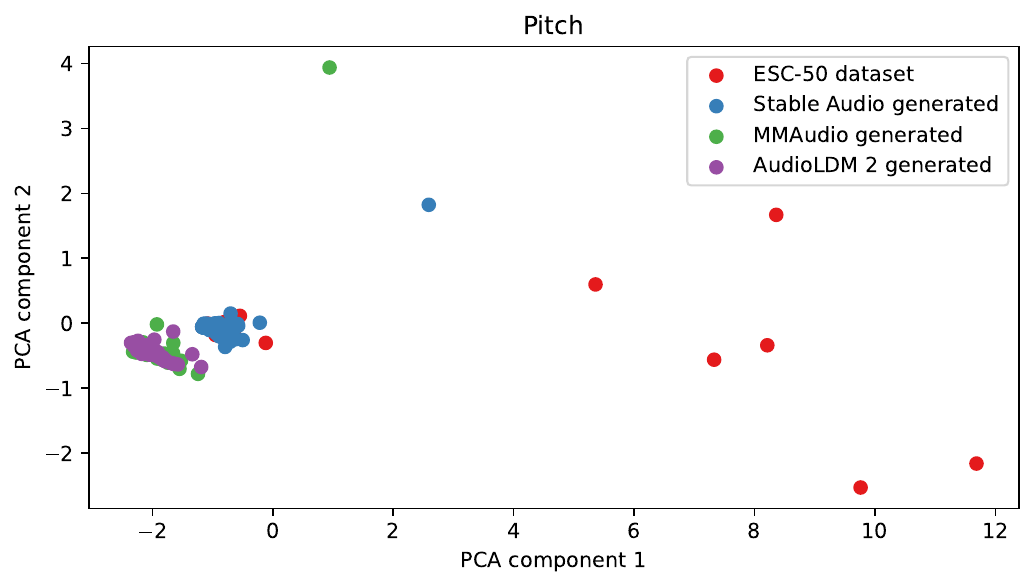}
\includegraphics[width=0.68\textwidth]{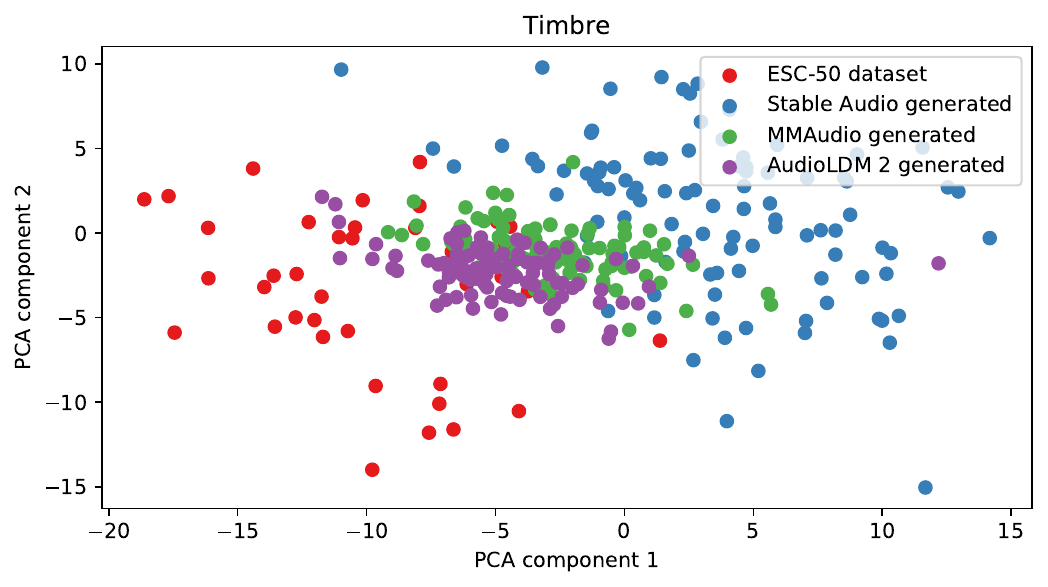}
\subsection{Prompt: Sound of frog}\centering
\includegraphics[width=0.68\textwidth]{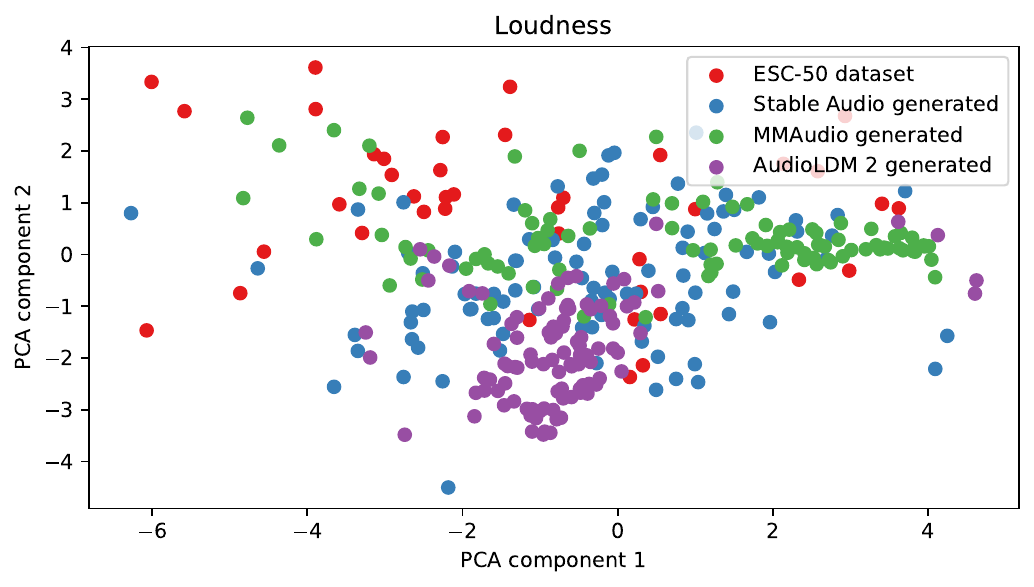}
\includegraphics[width=0.68\textwidth]{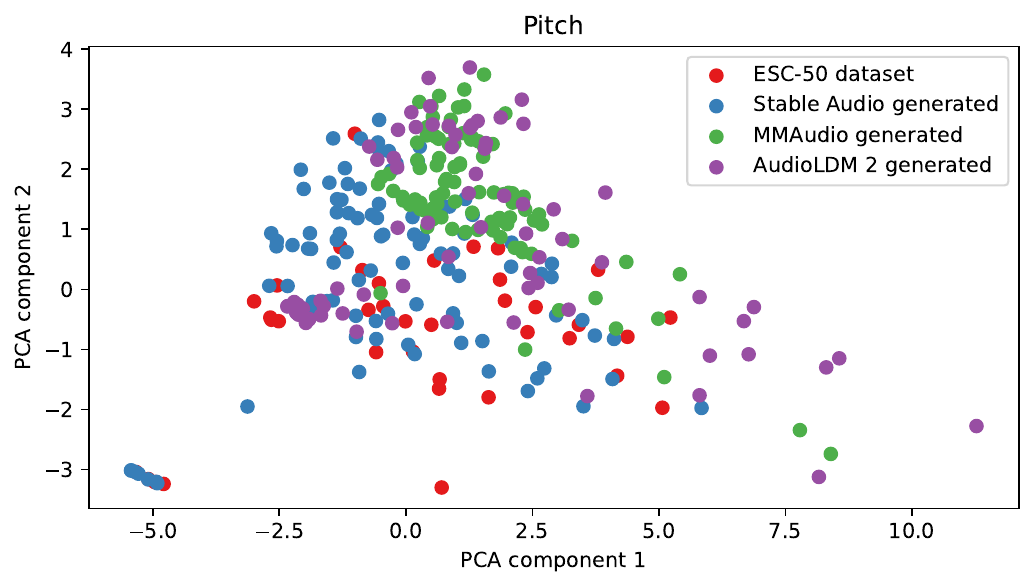}
\includegraphics[width=0.68\textwidth]{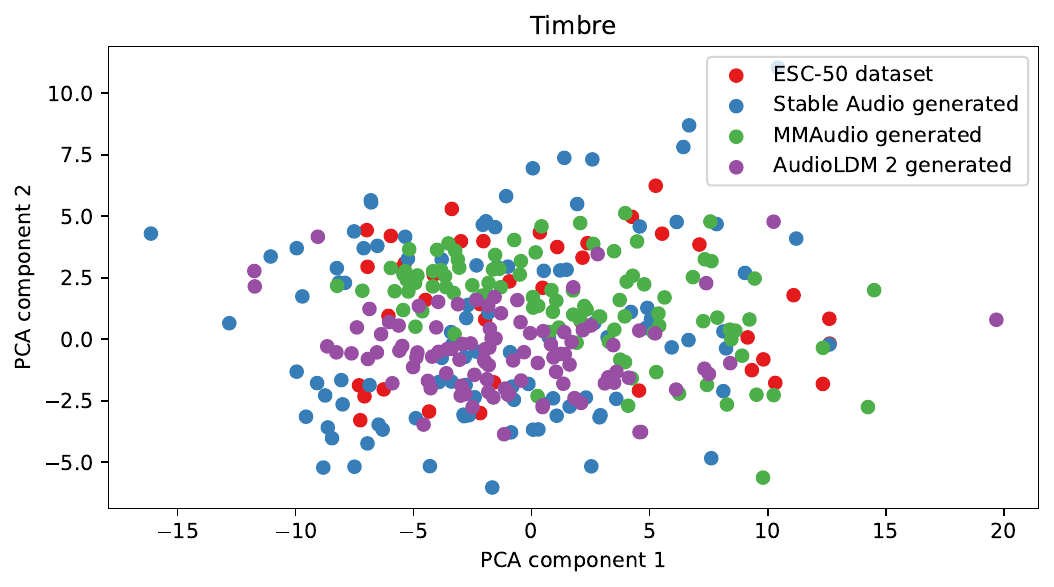}
\subsection{Prompt: Sound of glass breaking}\centering
\includegraphics[width=0.68\textwidth]{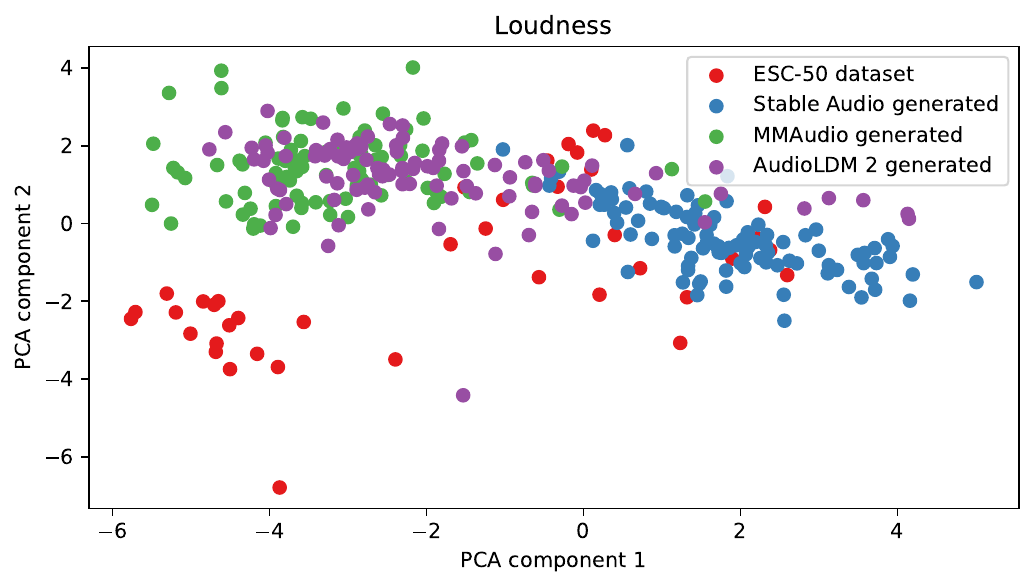}
\includegraphics[width=0.68\textwidth]{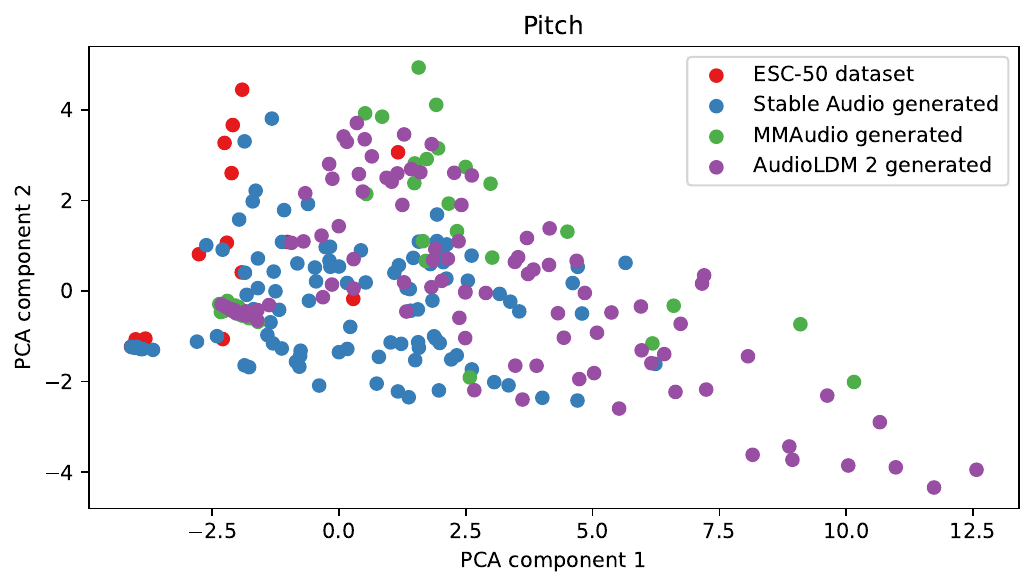}
\includegraphics[width=0.68\textwidth]{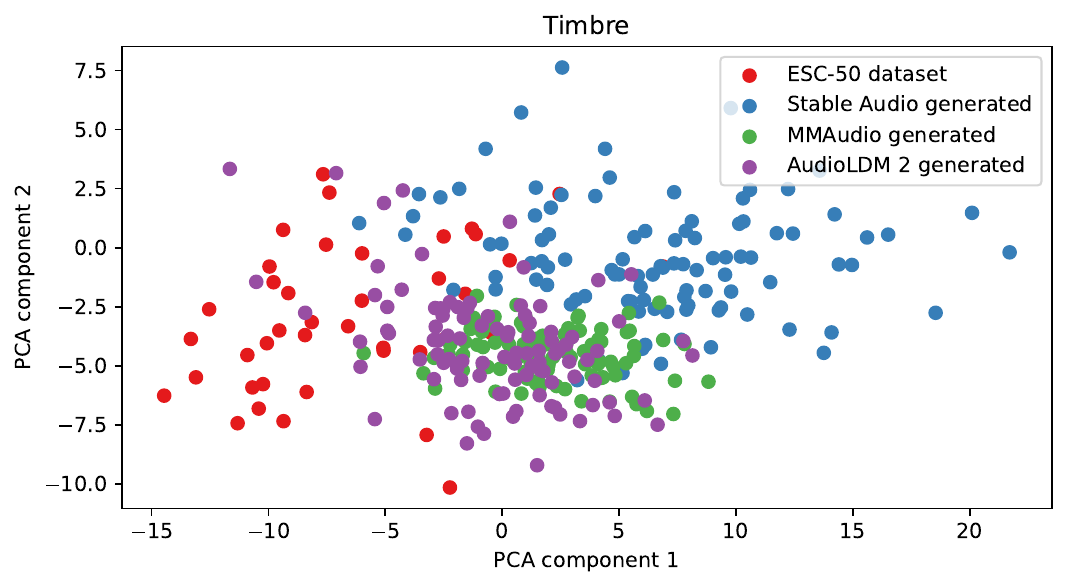}
\subsection{Prompt: Sound of hand saw}\centering
\includegraphics[width=0.68\textwidth]{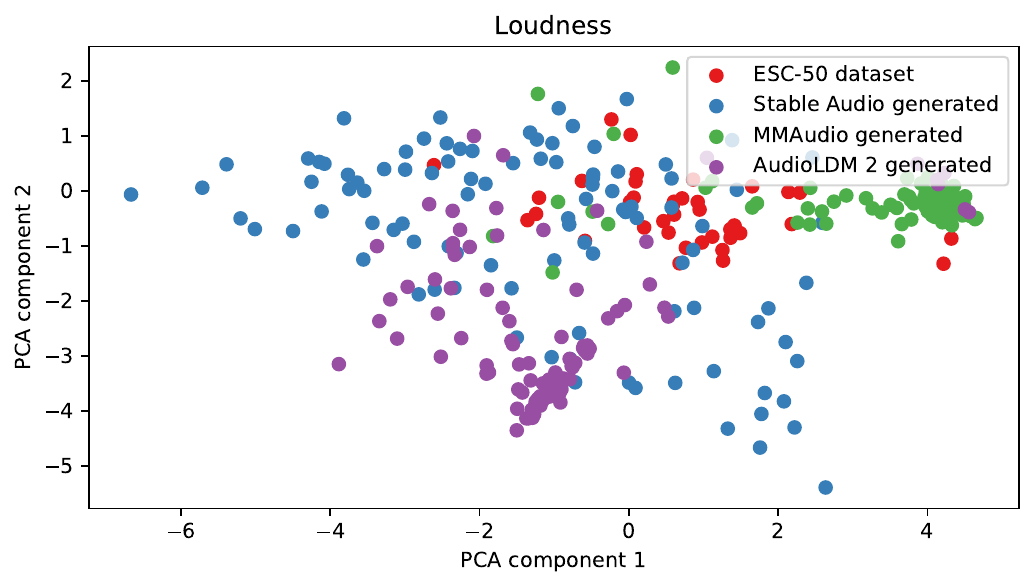}
\includegraphics[width=0.68\textwidth]{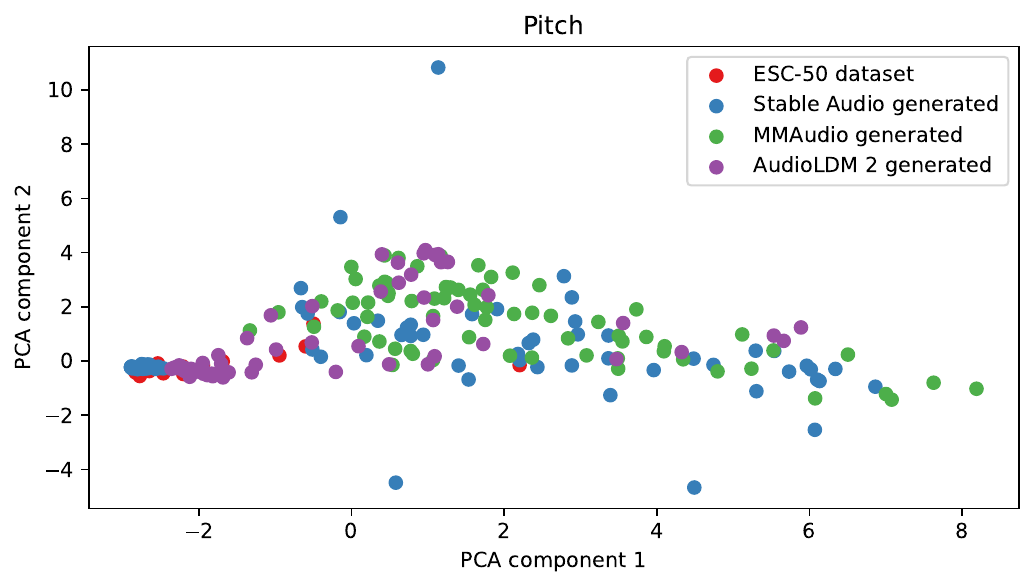}
\includegraphics[width=0.68\textwidth]{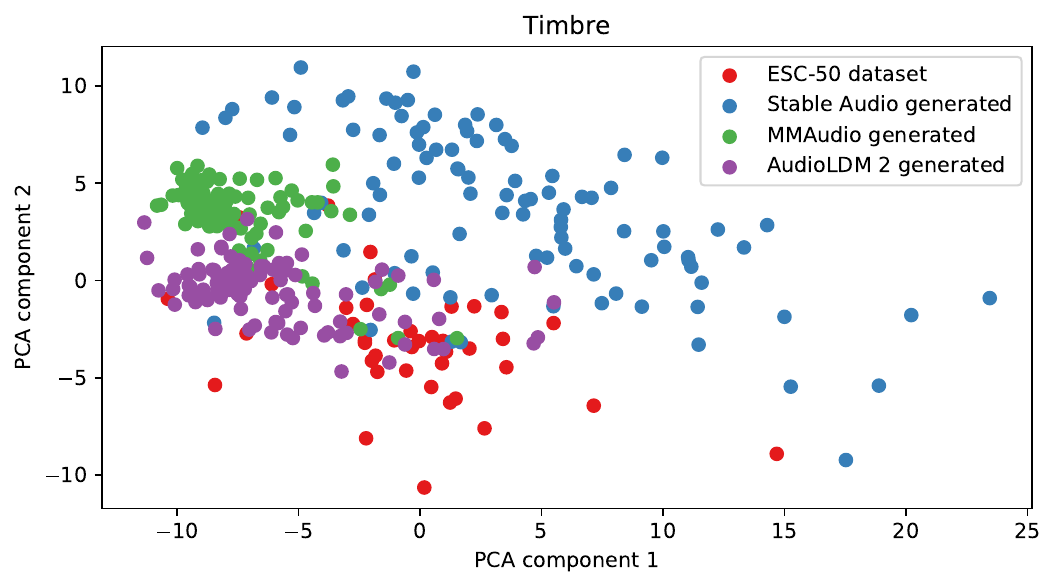}
\subsection{Prompt: Sound of helicopter}\centering
\includegraphics[width=0.68\textwidth]{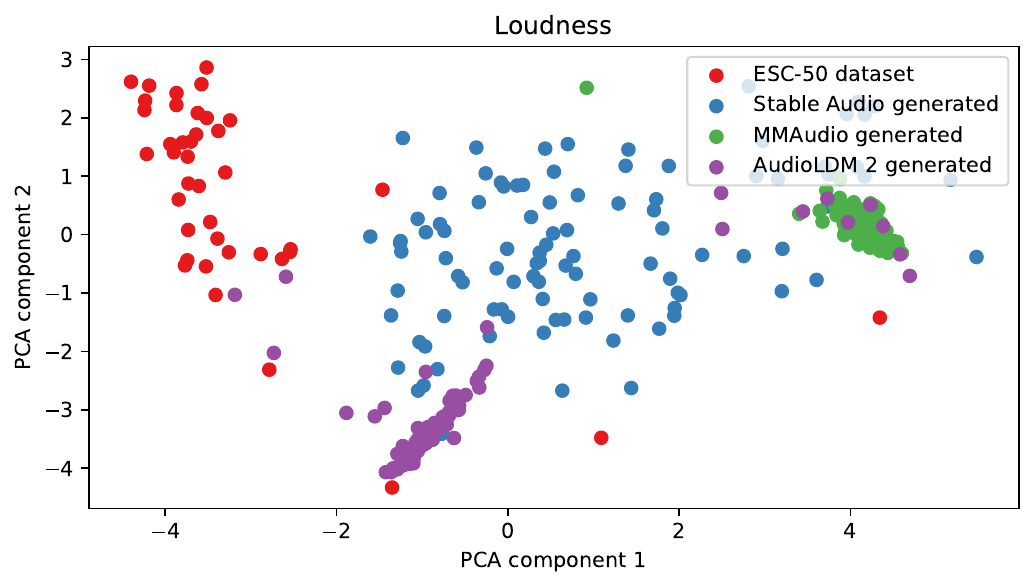}
\includegraphics[width=0.68\textwidth]{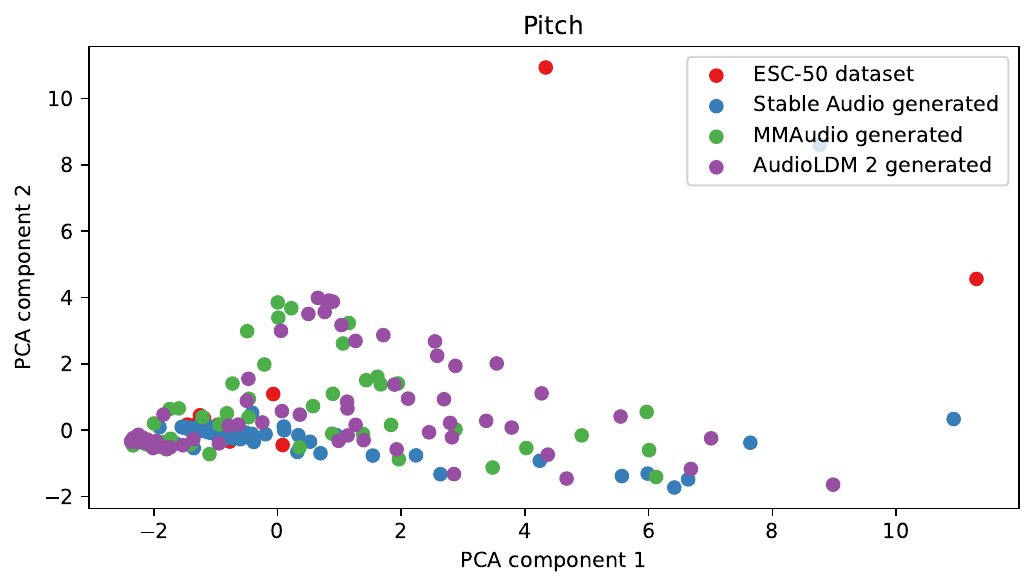}
\includegraphics[width=0.68\textwidth]{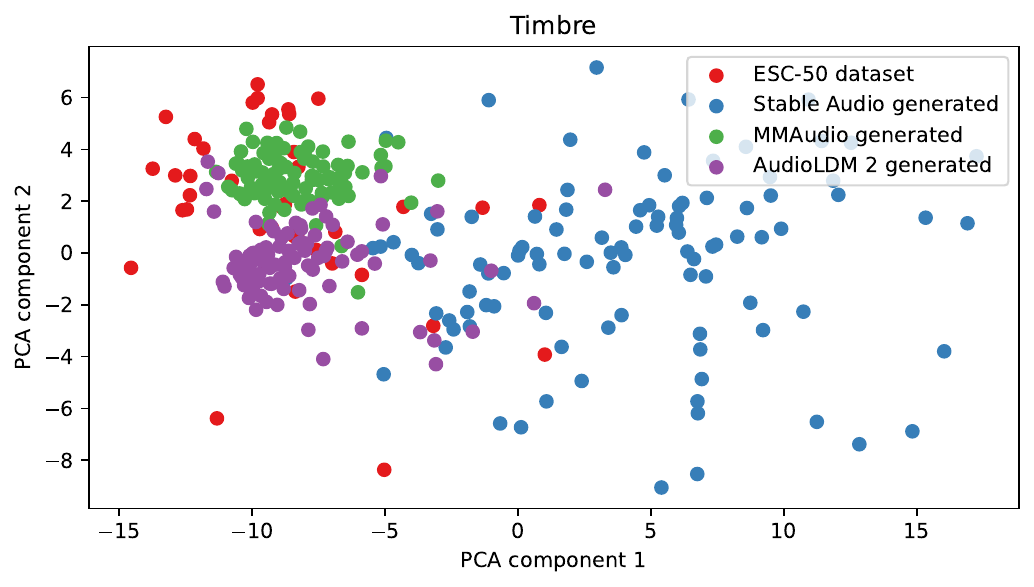}
\subsection{Prompt: Sound of hen}\centering
\includegraphics[width=0.68\textwidth]{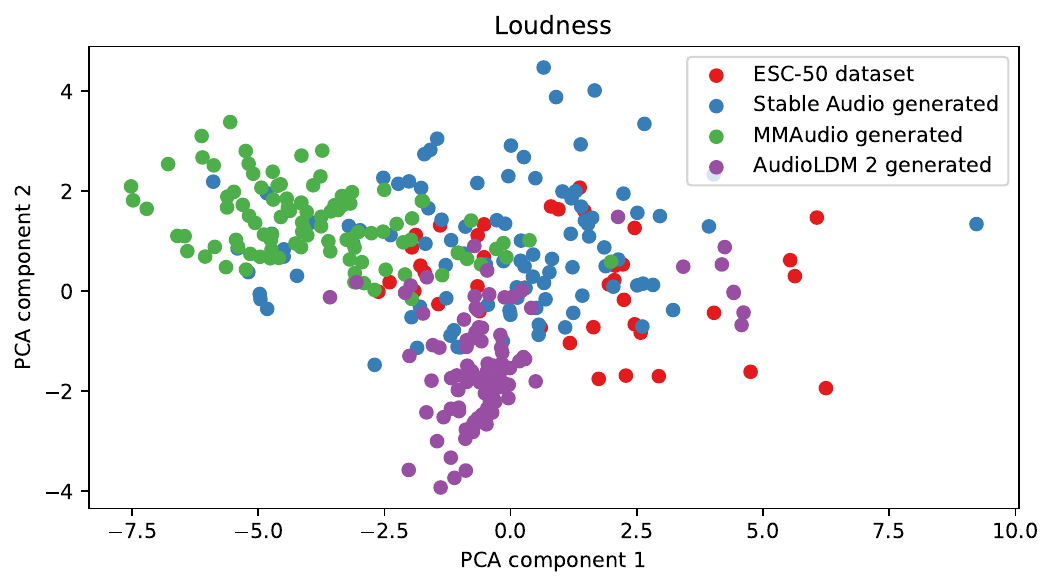}
\includegraphics[width=0.68\textwidth]{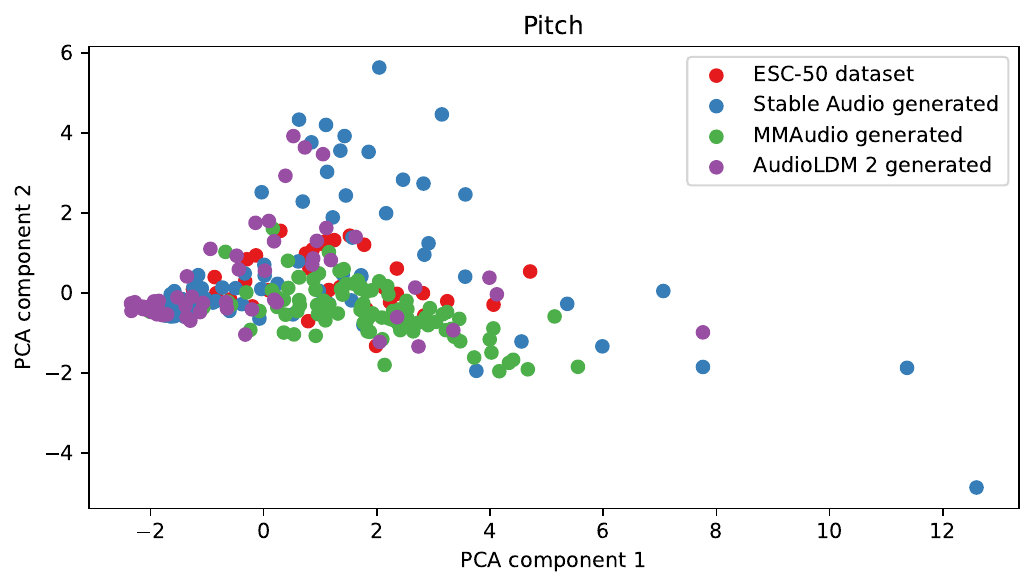}
\includegraphics[width=0.68\textwidth]{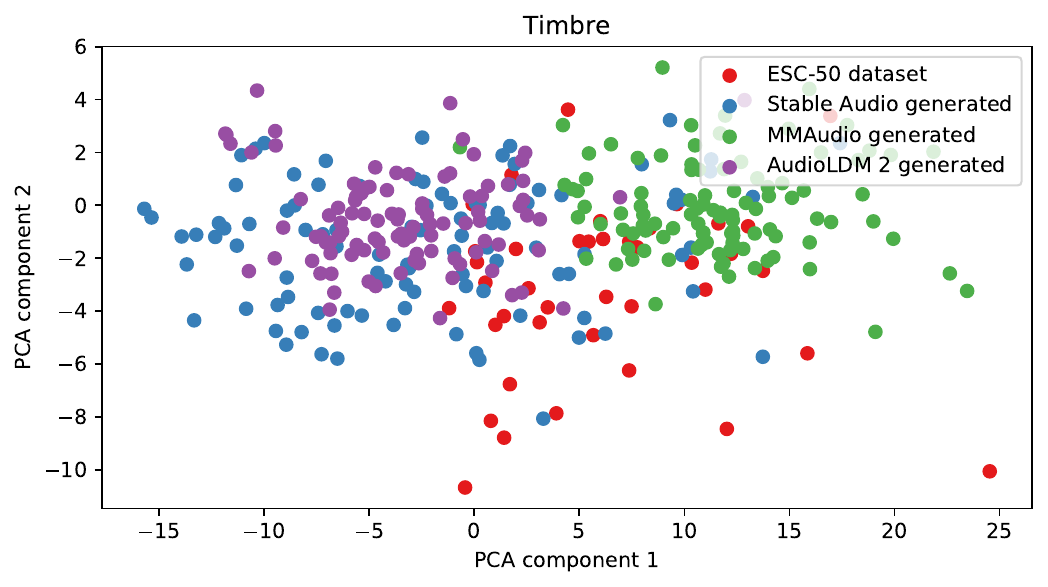}
\subsection{Prompt: Sound of insects (flying)}\centering
\includegraphics[width=0.68\textwidth]{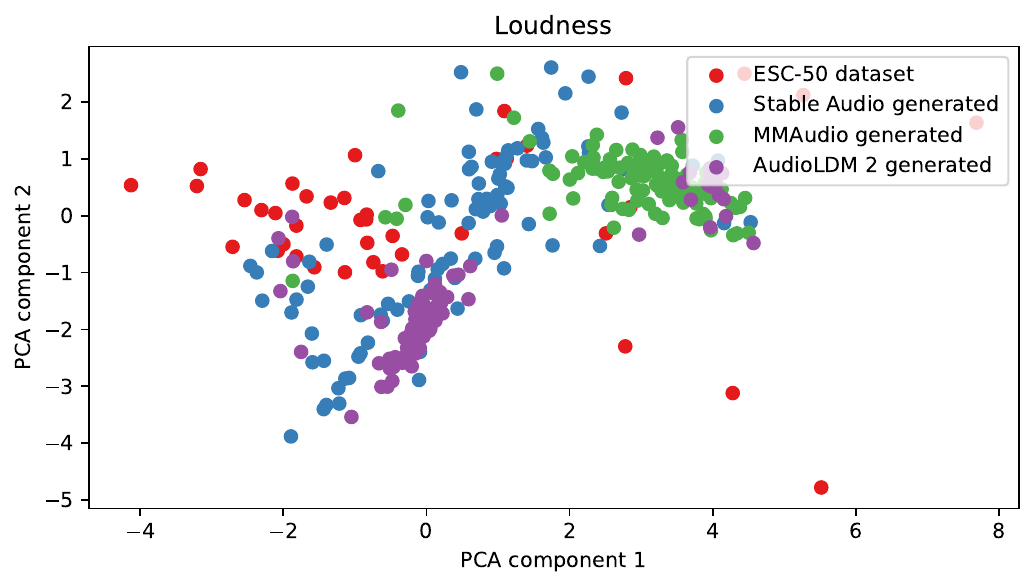}
\includegraphics[width=0.68\textwidth]{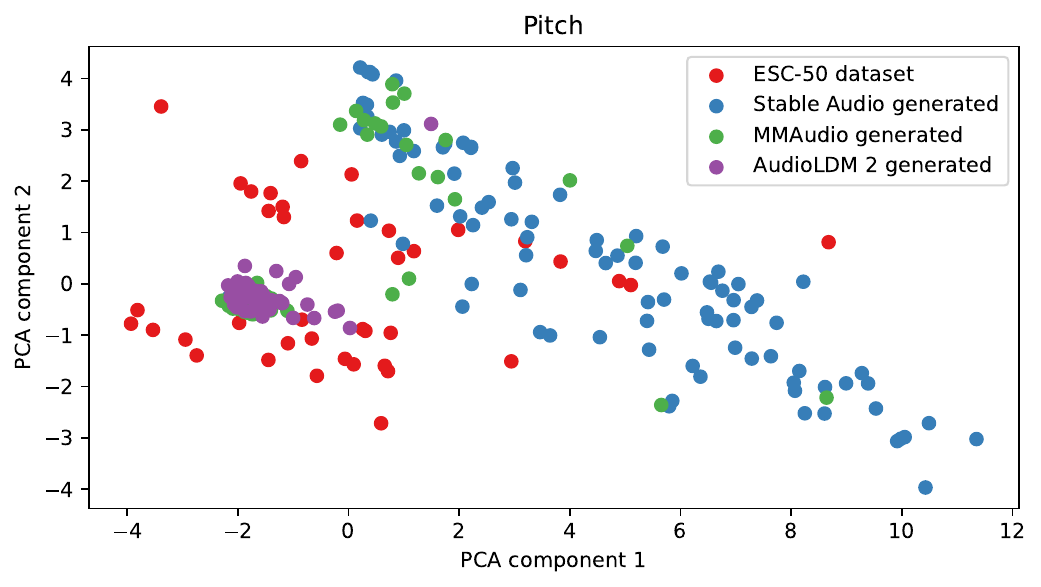}
\includegraphics[width=0.68\textwidth]{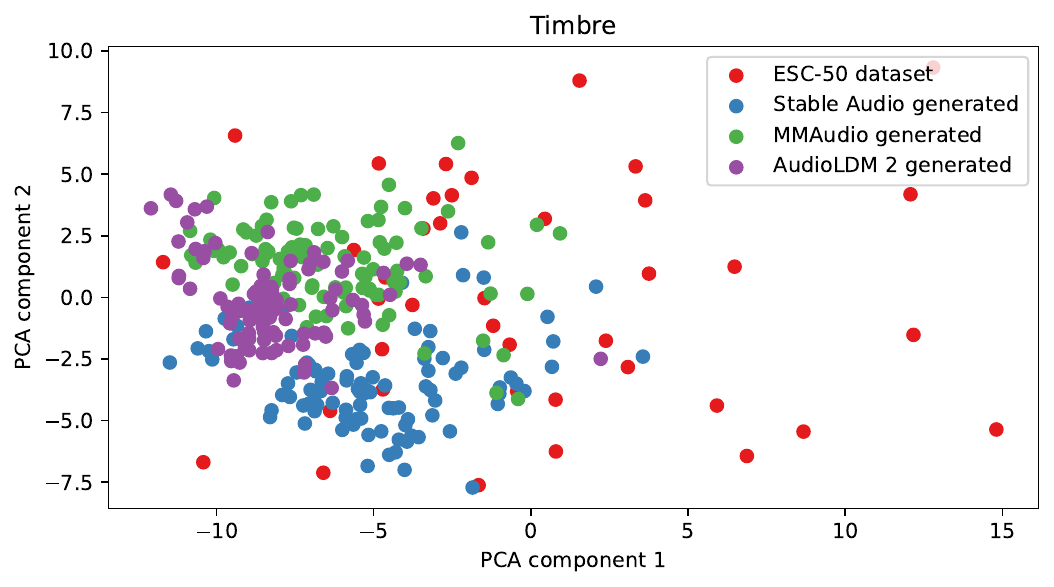}
\subsection{Prompt: Sound of keyboard typing}\centering
\includegraphics[width=0.68\textwidth]{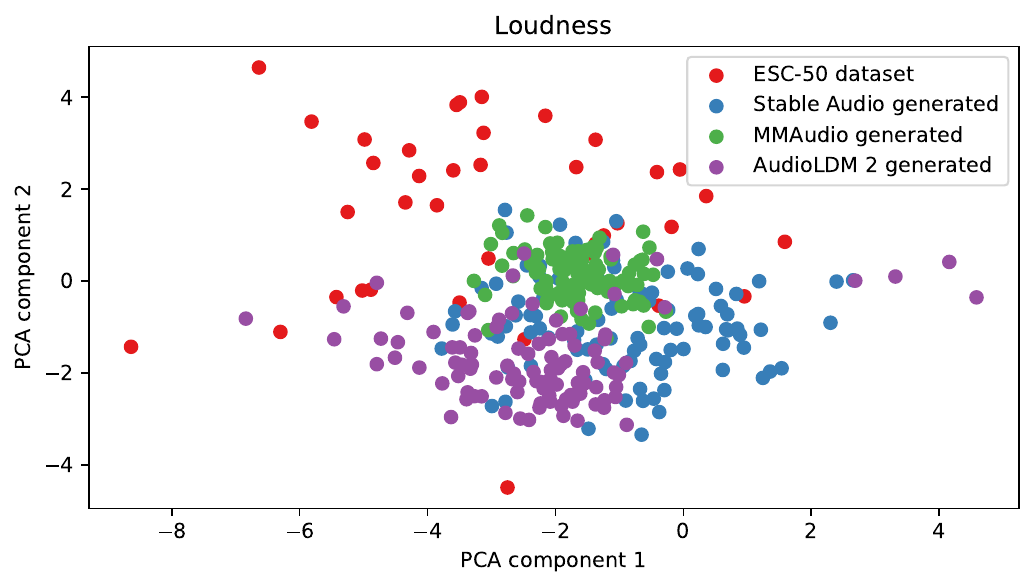}
\includegraphics[width=0.68\textwidth]{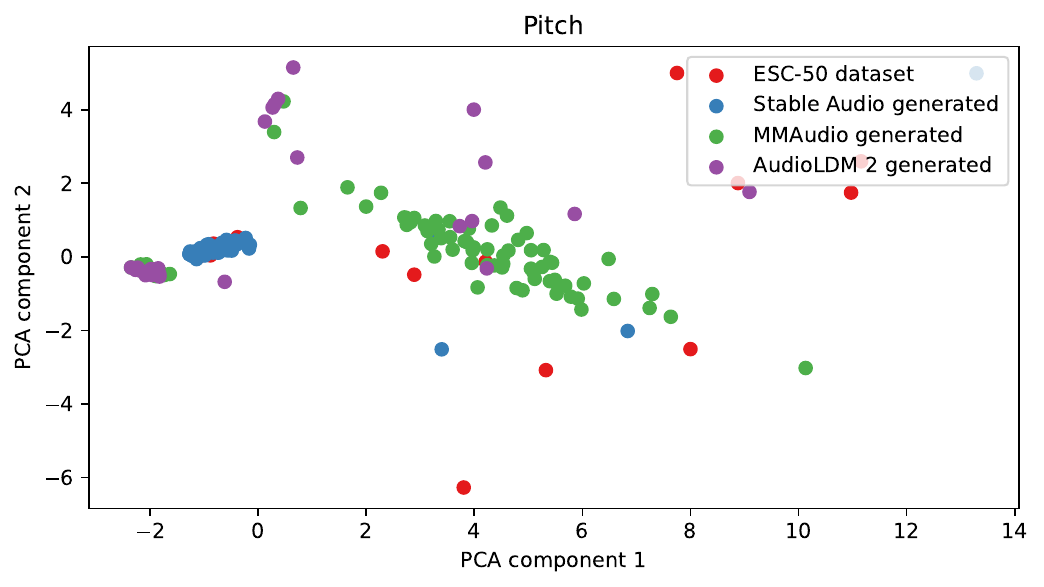}
\includegraphics[width=0.68\textwidth]{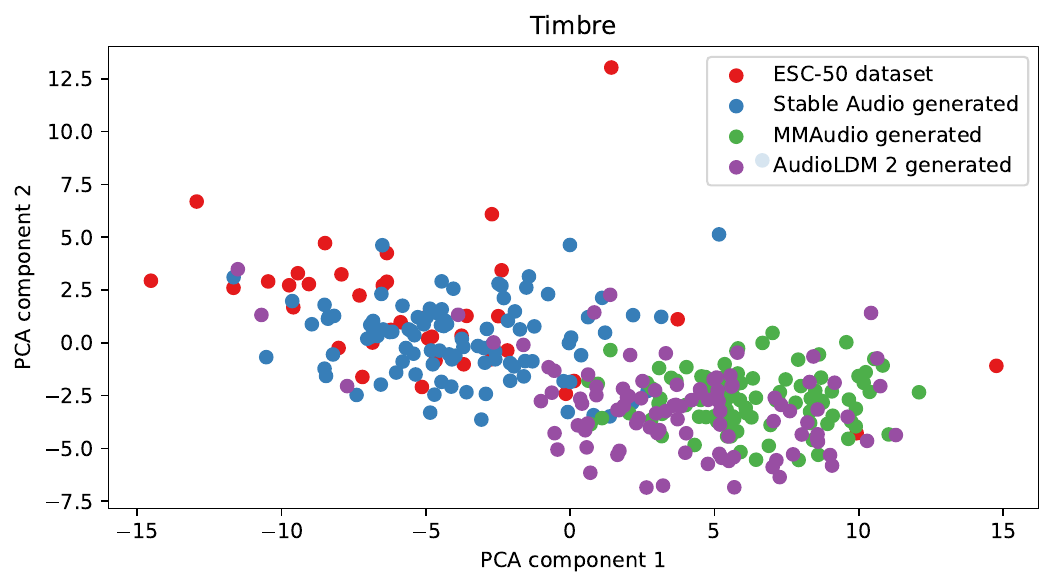}
\subsection{Prompt: Sound of laughing}\centering
\includegraphics[width=0.68\textwidth]{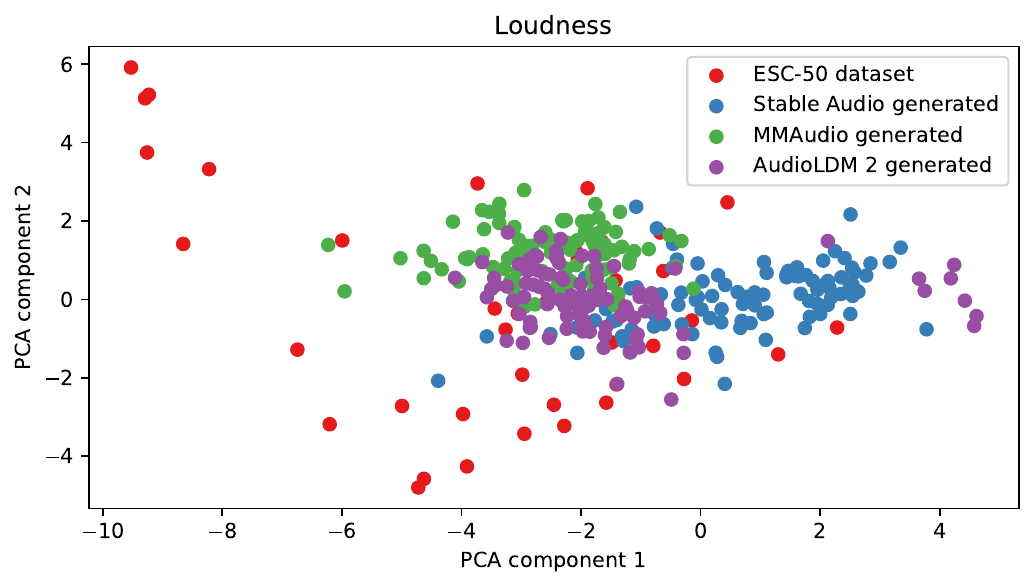}
\includegraphics[width=0.68\textwidth]{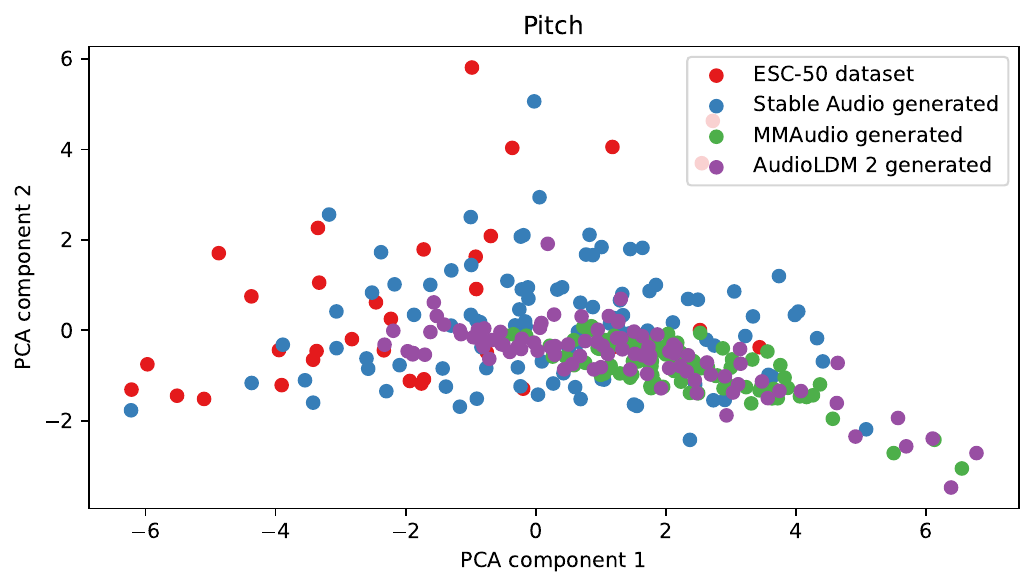}
\includegraphics[width=0.68\textwidth]{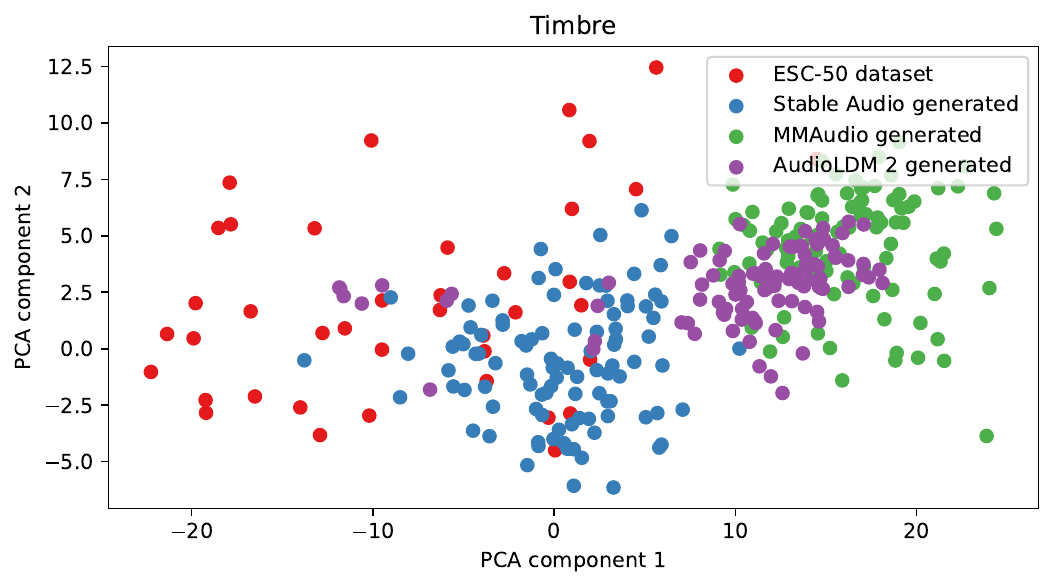}
\subsection{Prompt: Sound of mouse click}\centering
\includegraphics[width=0.68\textwidth]{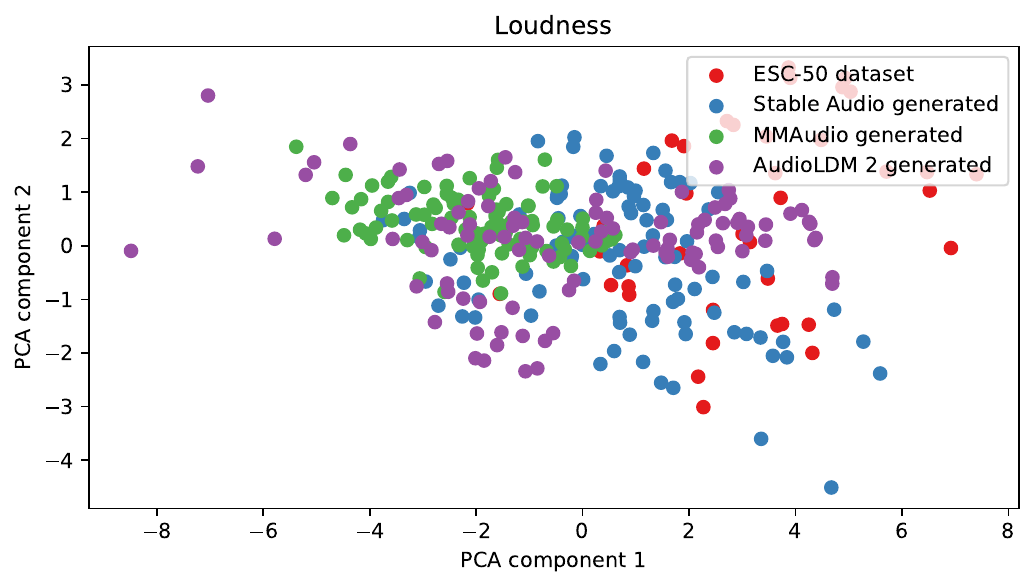}
\includegraphics[width=0.68\textwidth]{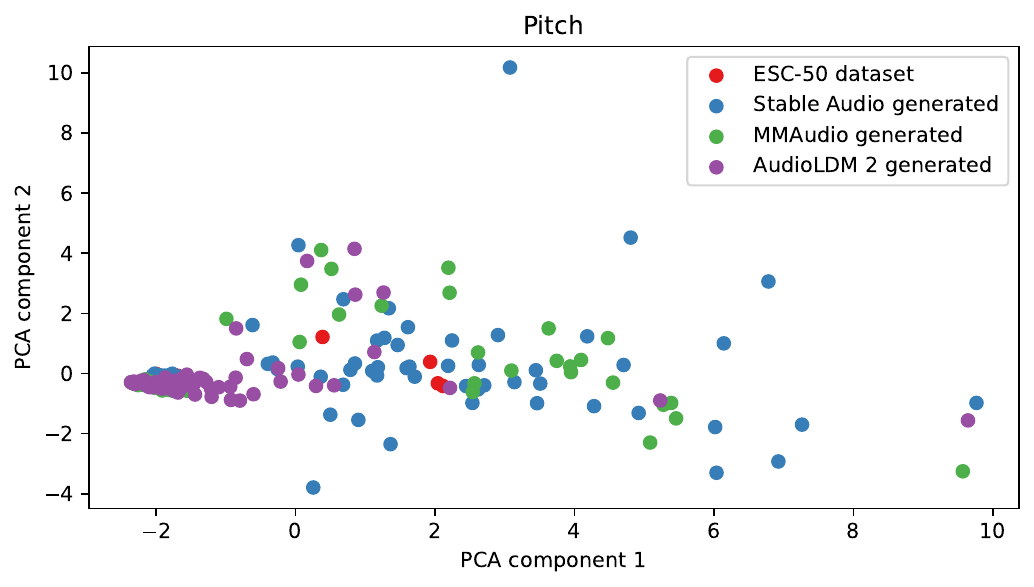}
\includegraphics[width=0.68\textwidth]{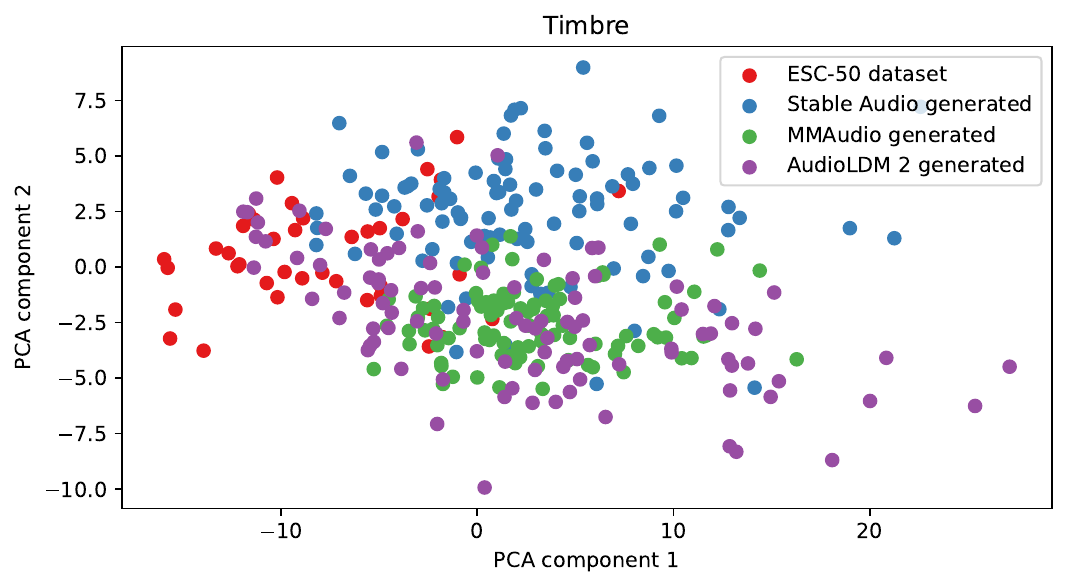}
\subsection{Prompt: Sound of pig}\centering
\includegraphics[width=0.68\textwidth]{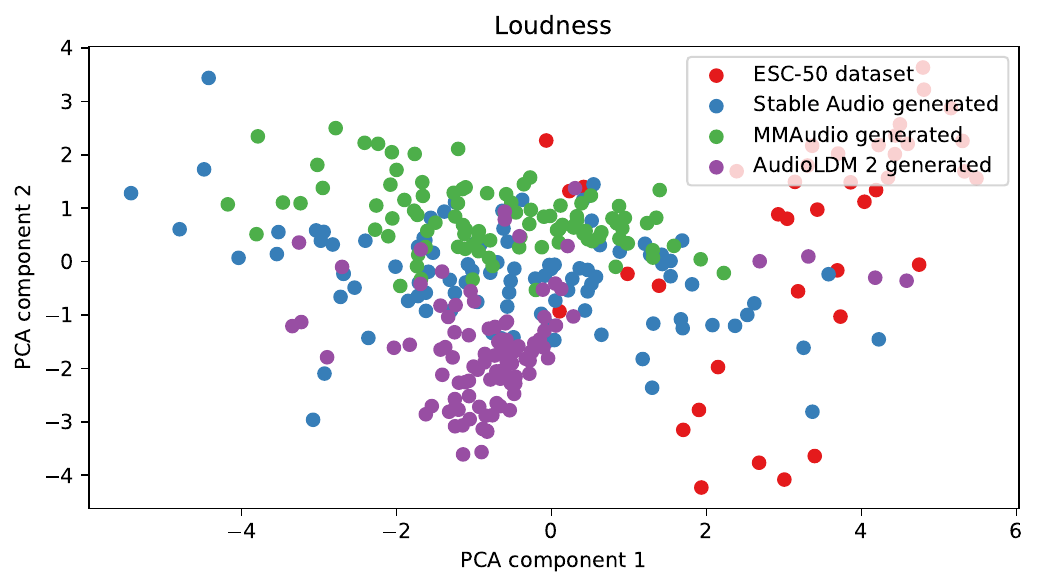}
\includegraphics[width=0.68\textwidth]{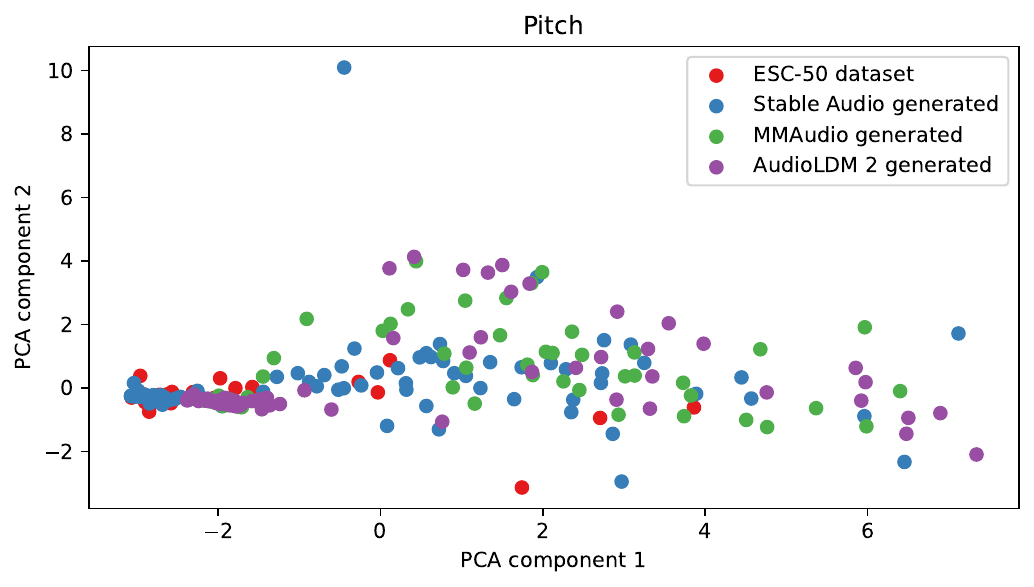}
\includegraphics[width=0.68\textwidth]{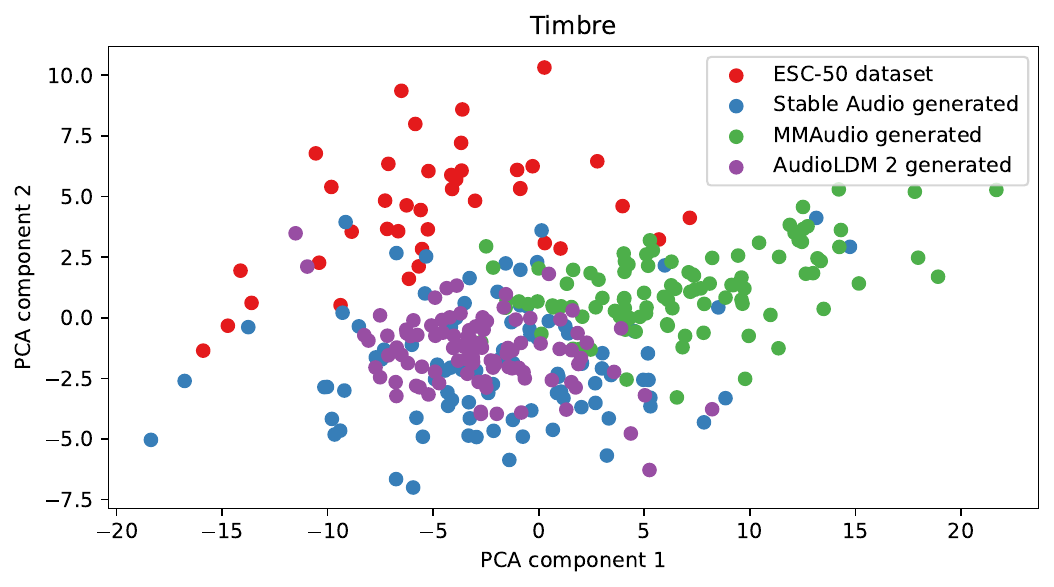}
\subsection{Prompt: Sound of pouring water}\centering
\includegraphics[width=0.68\textwidth]{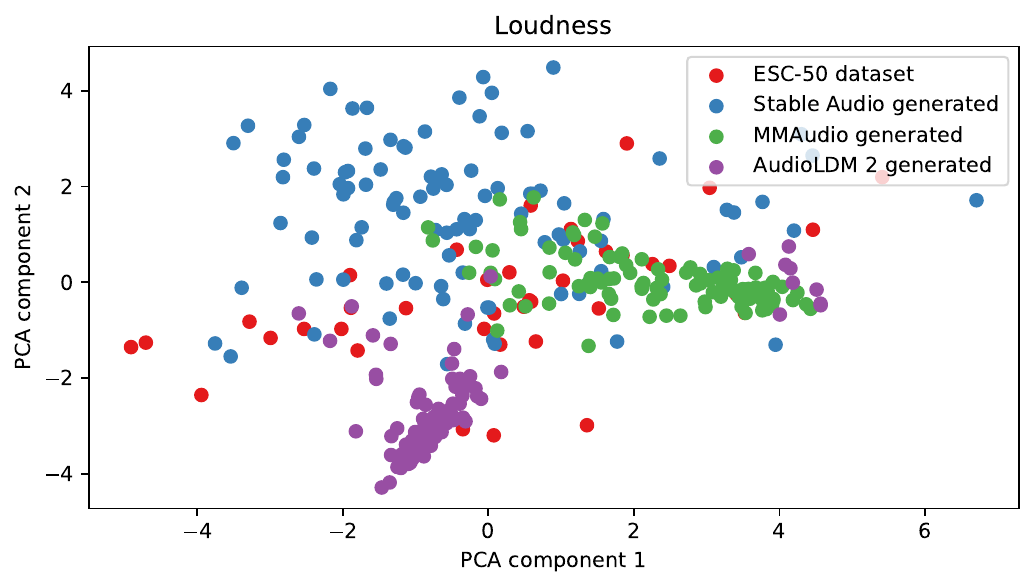}
\includegraphics[width=0.68\textwidth]{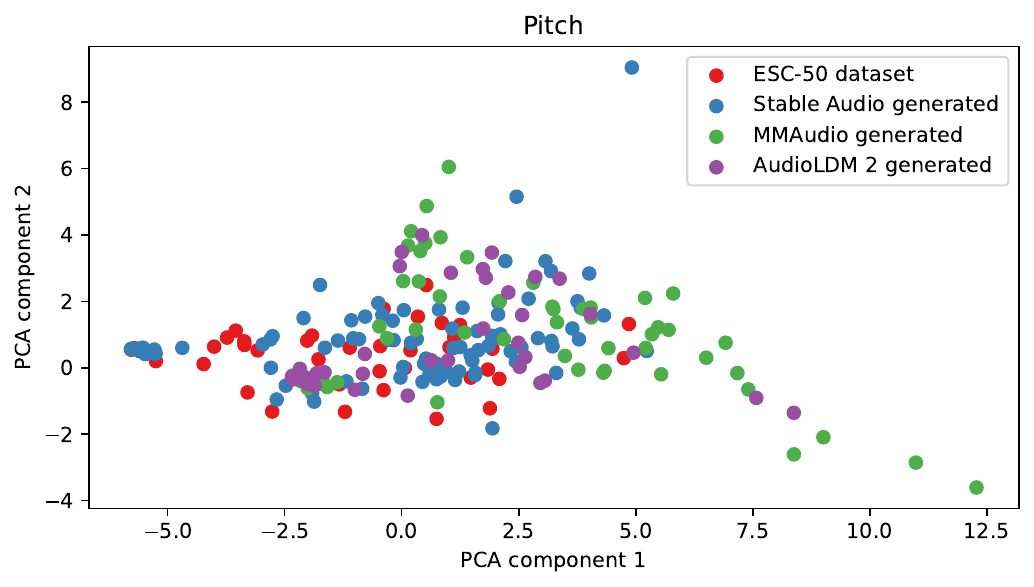}
\includegraphics[width=0.68\textwidth]{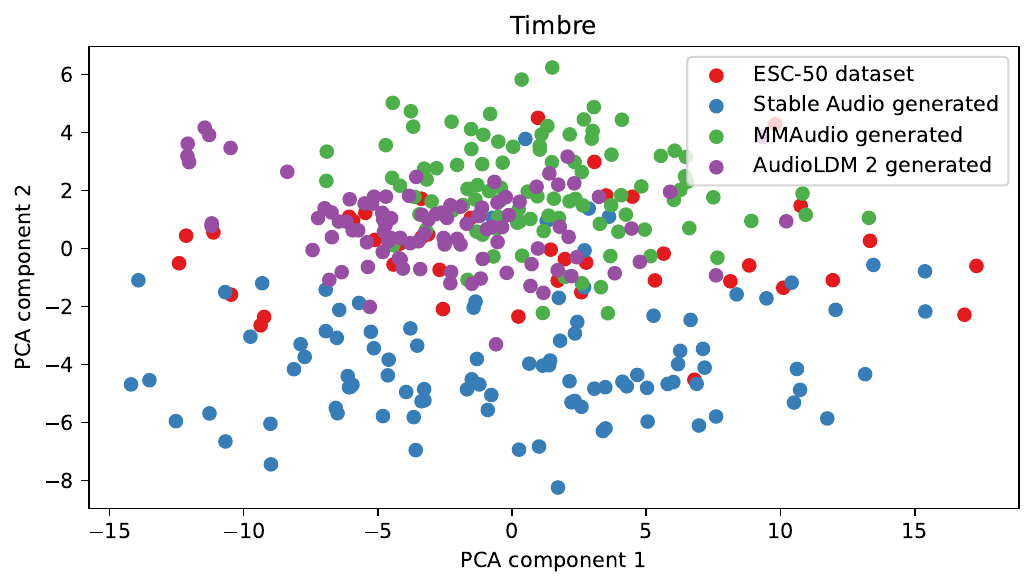}
\subsection{Prompt: Sound of rain}\centering
\includegraphics[width=0.68\textwidth]{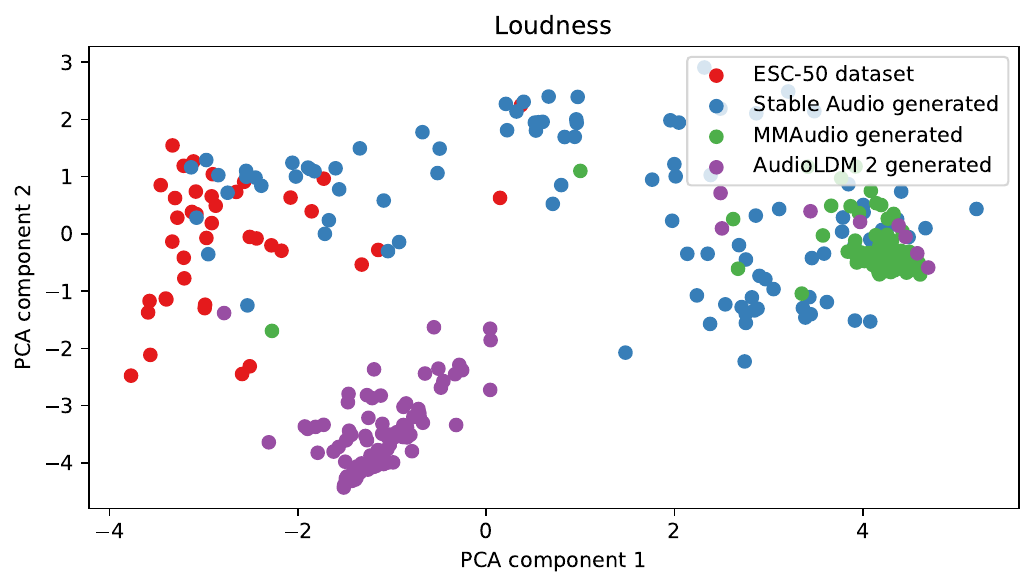}
\includegraphics[width=0.68\textwidth]{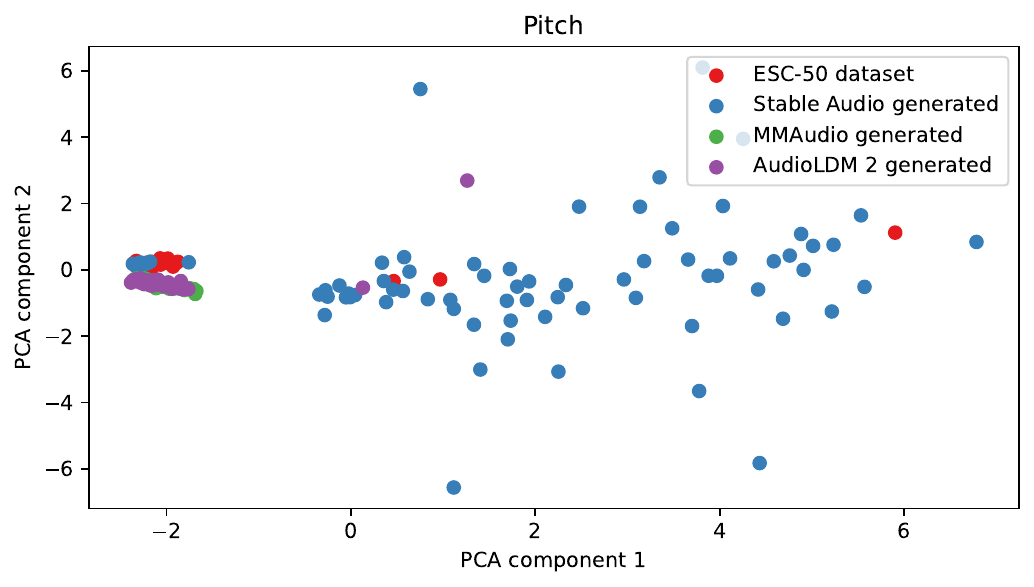}
\includegraphics[width=0.68\textwidth]{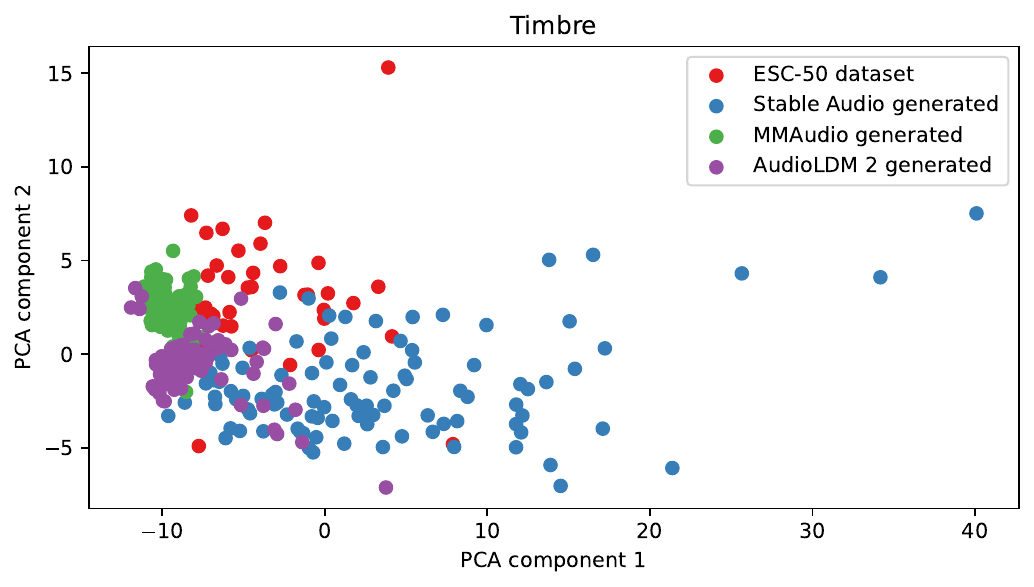}
\subsection{Prompt: Sound of rooster}\centering
\includegraphics[width=0.68\textwidth]{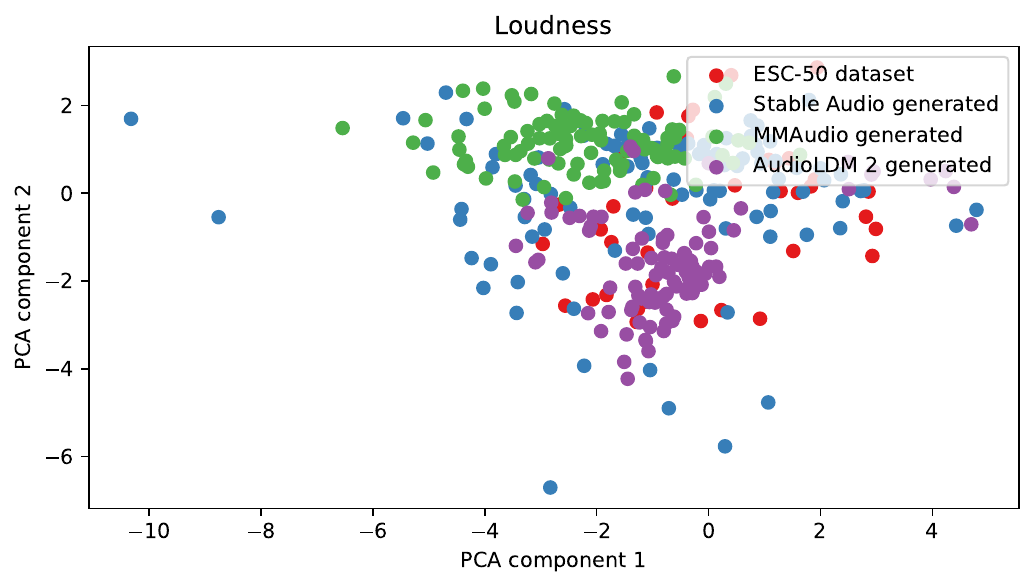}
\includegraphics[width=0.68\textwidth]{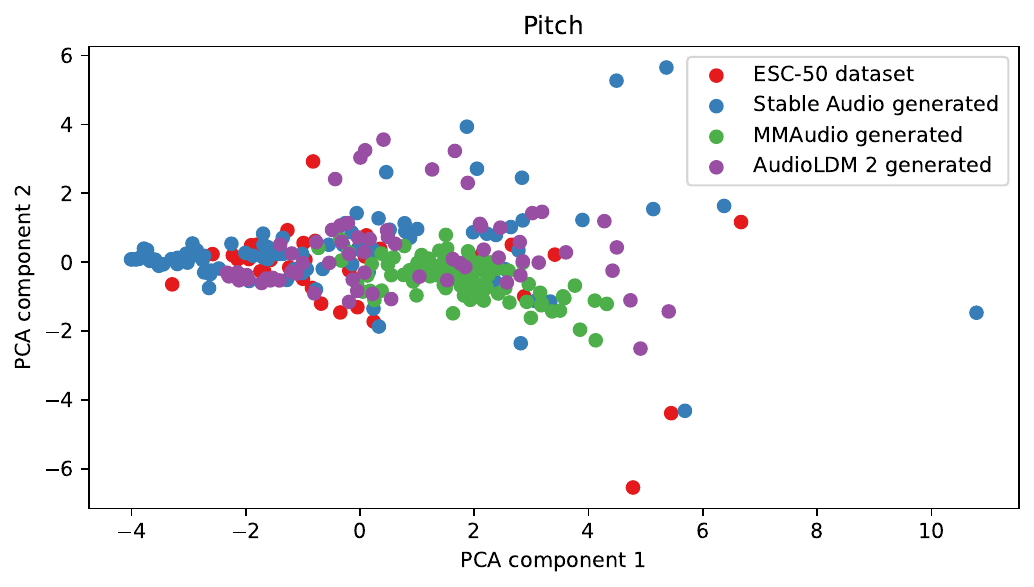}
\includegraphics[width=0.68\textwidth]{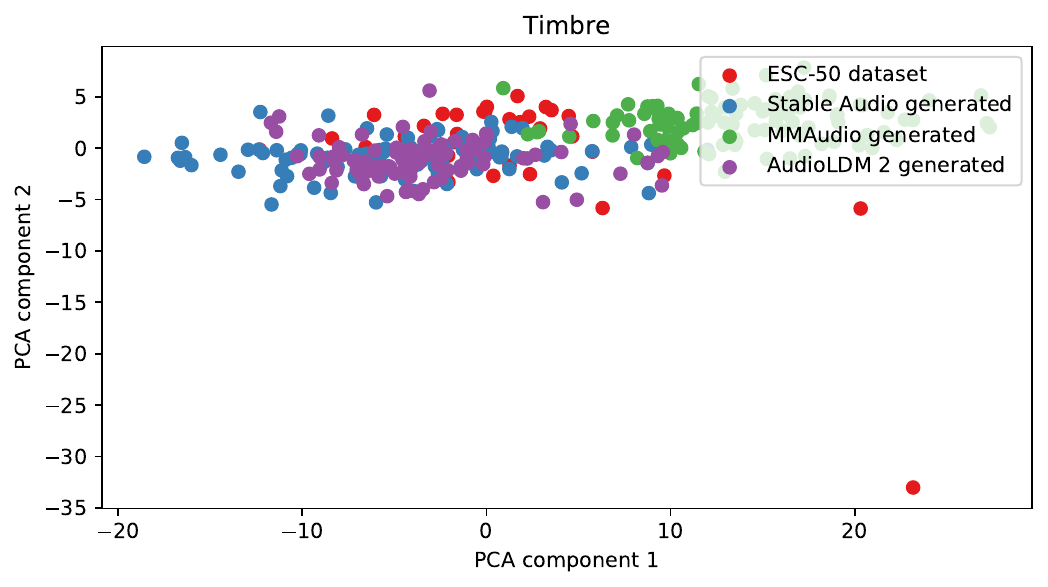}
\subsection{Prompt: Sound of sea waves}\centering
\includegraphics[width=0.68\textwidth]{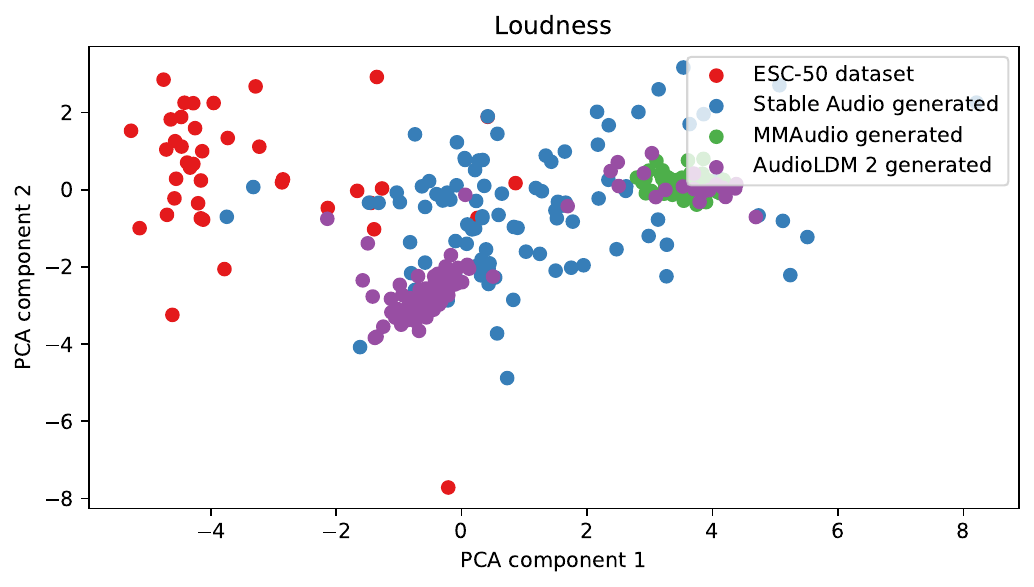}
\includegraphics[width=0.68\textwidth]{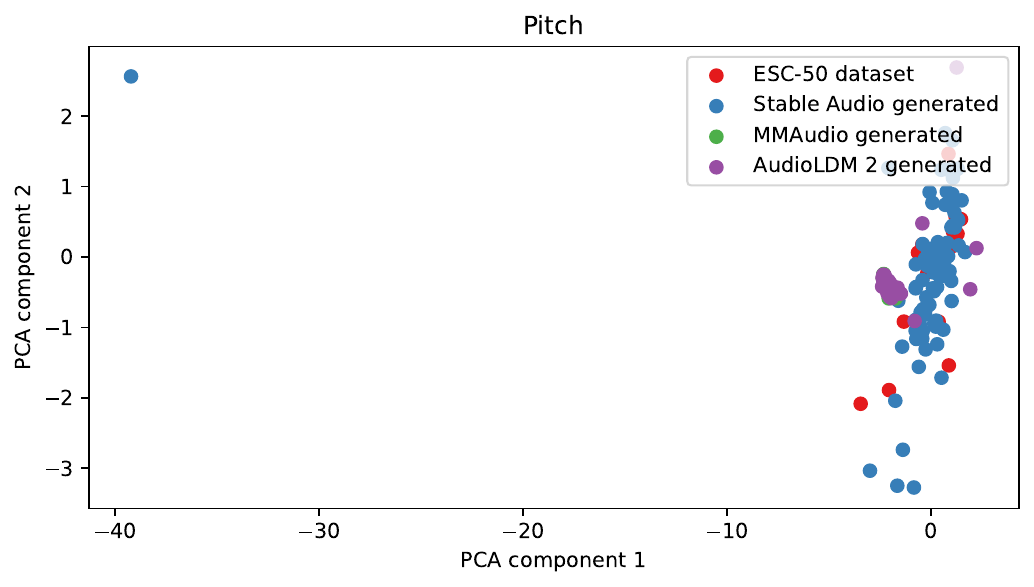}
\includegraphics[width=0.68\textwidth]{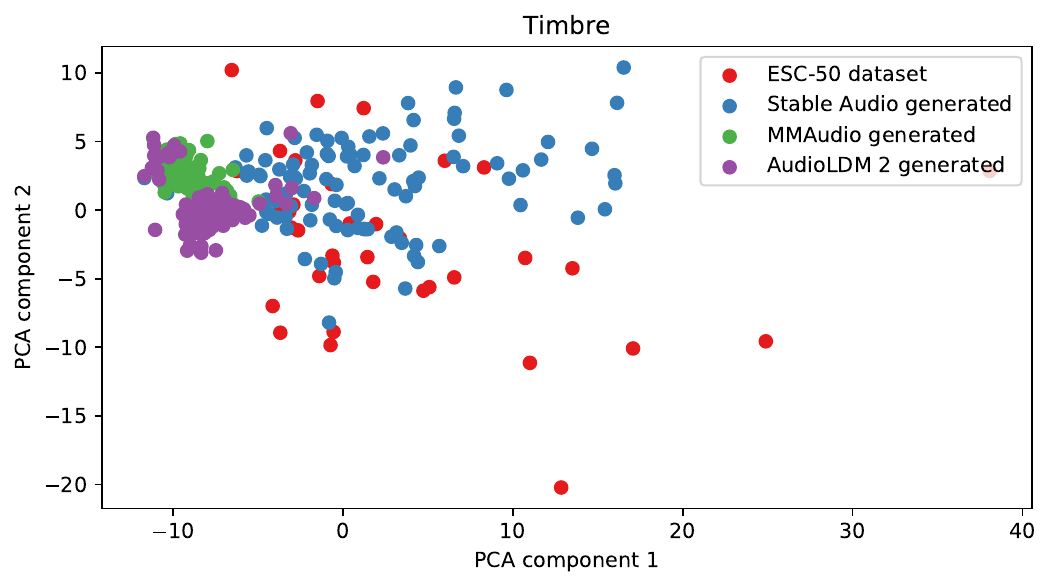}
\subsection{Prompt: Sound of sheep}\centering
\includegraphics[width=0.68\textwidth]{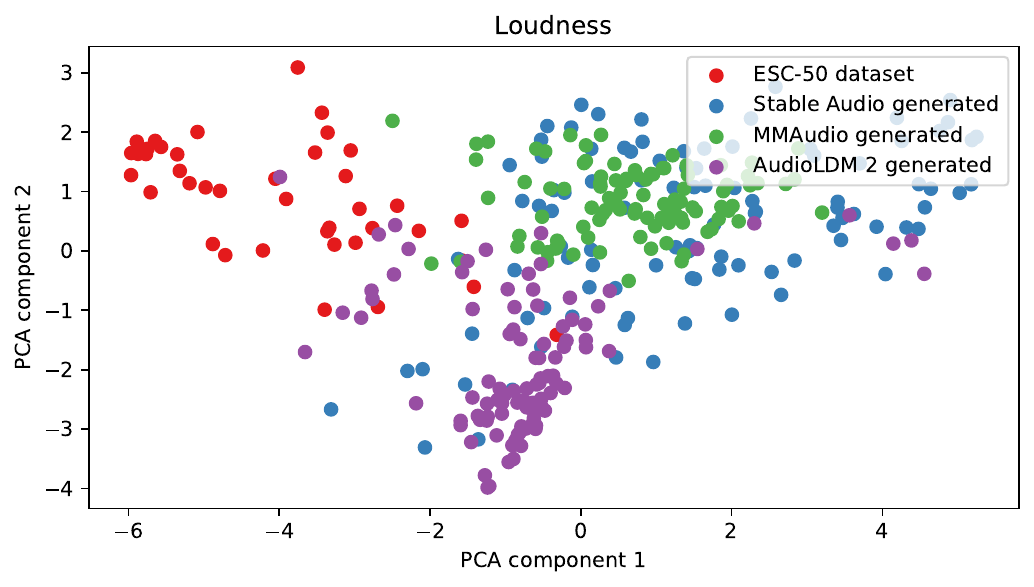}
\includegraphics[width=0.68\textwidth]{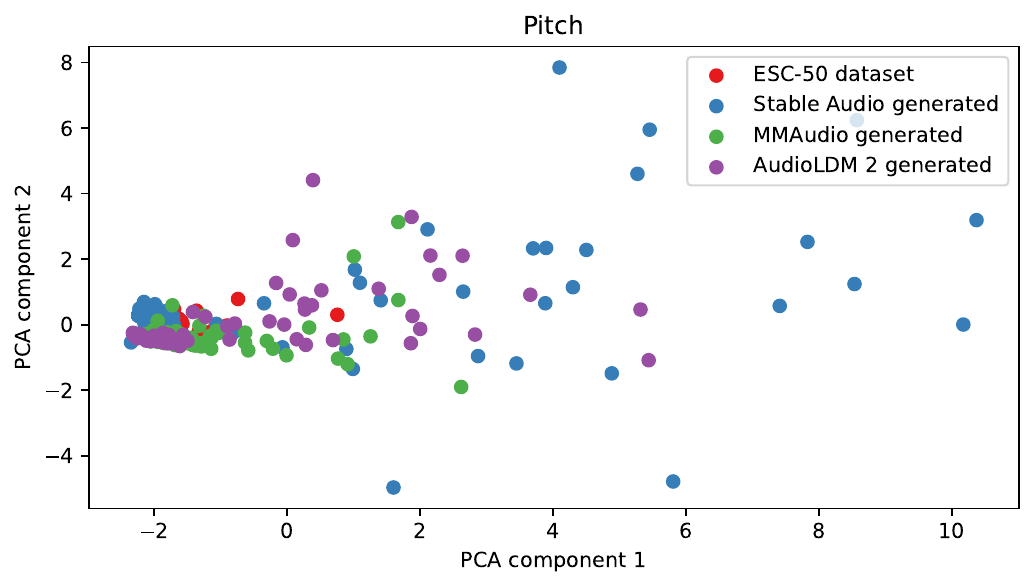}
\includegraphics[width=0.68\textwidth]{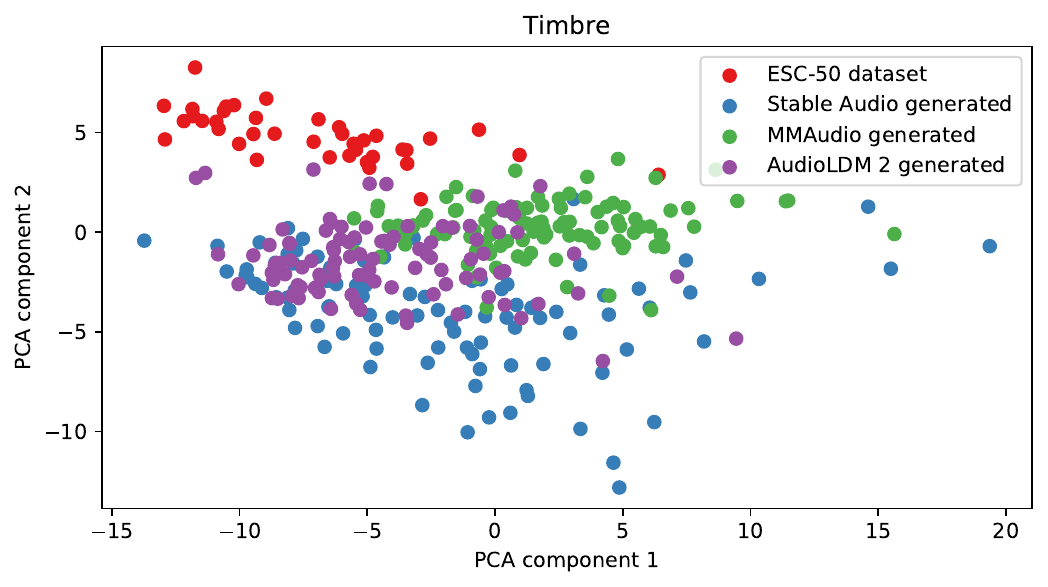}
\subsection{Prompt: Sound of siren}\centering
\includegraphics[width=0.68\textwidth]{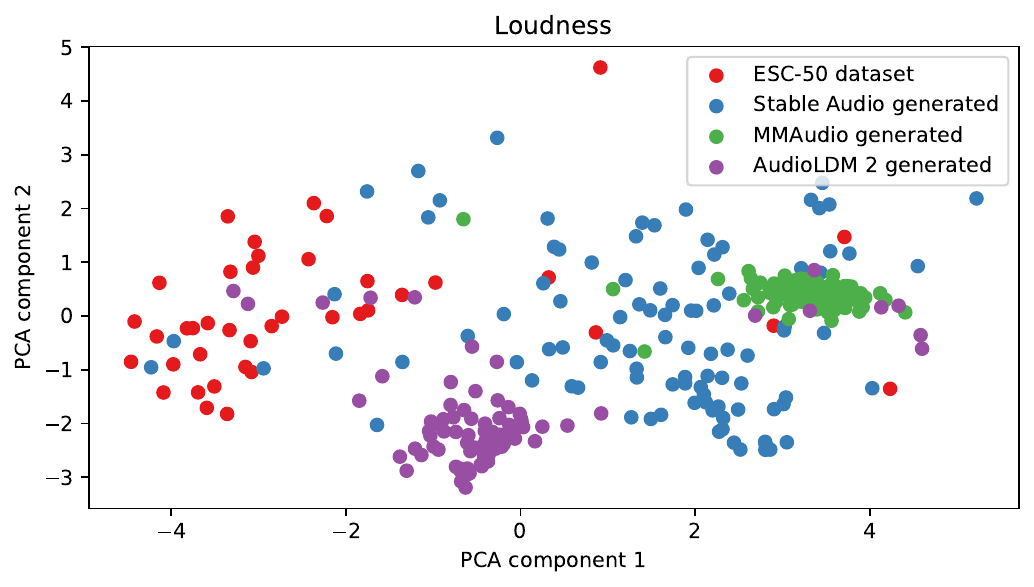}
\includegraphics[width=0.68\textwidth]{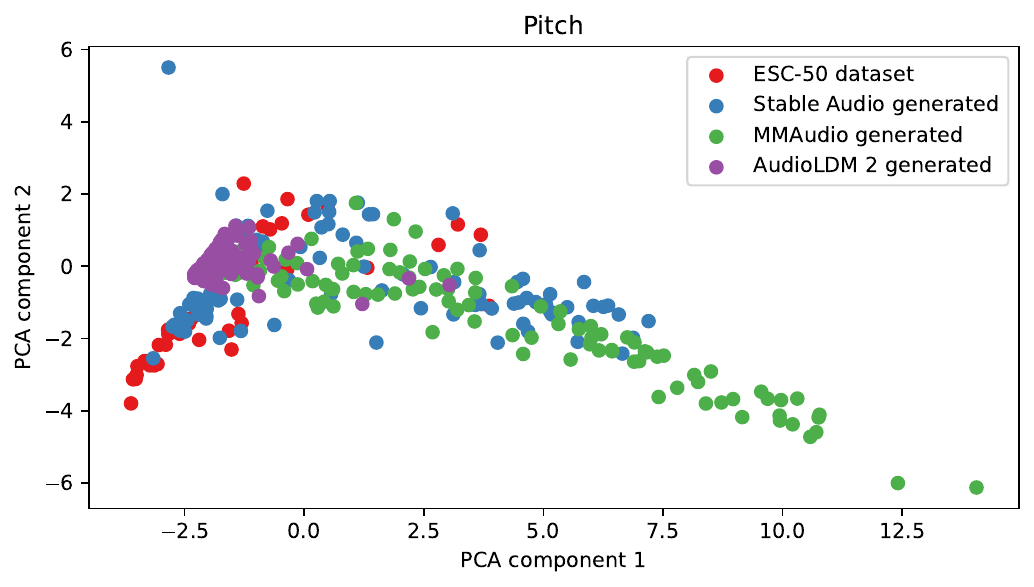}
\includegraphics[width=0.68\textwidth]{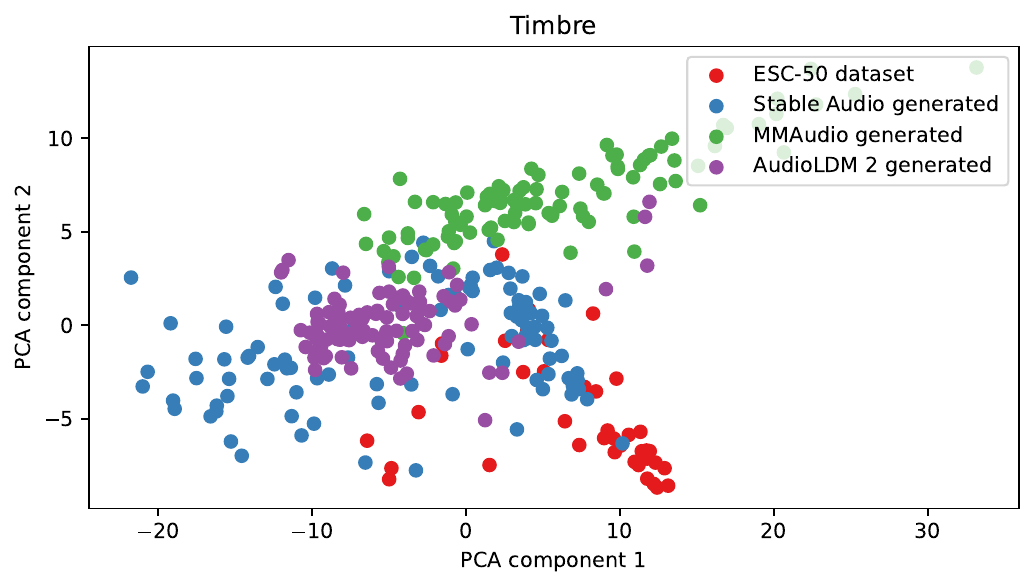}
\subsection{Prompt: Sound of sneezing}\centering
\includegraphics[width=0.68\textwidth]{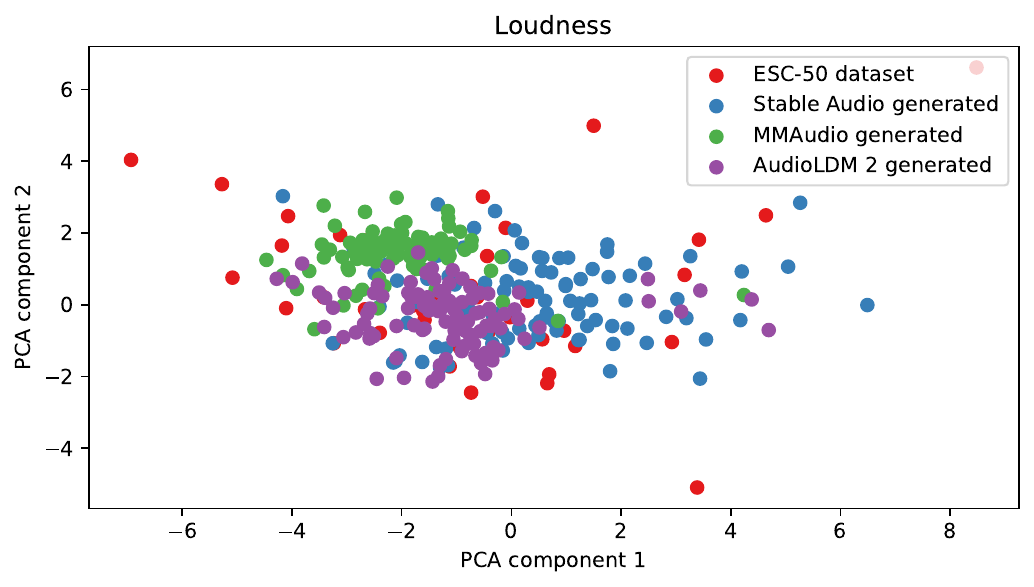}
\includegraphics[width=0.68\textwidth]{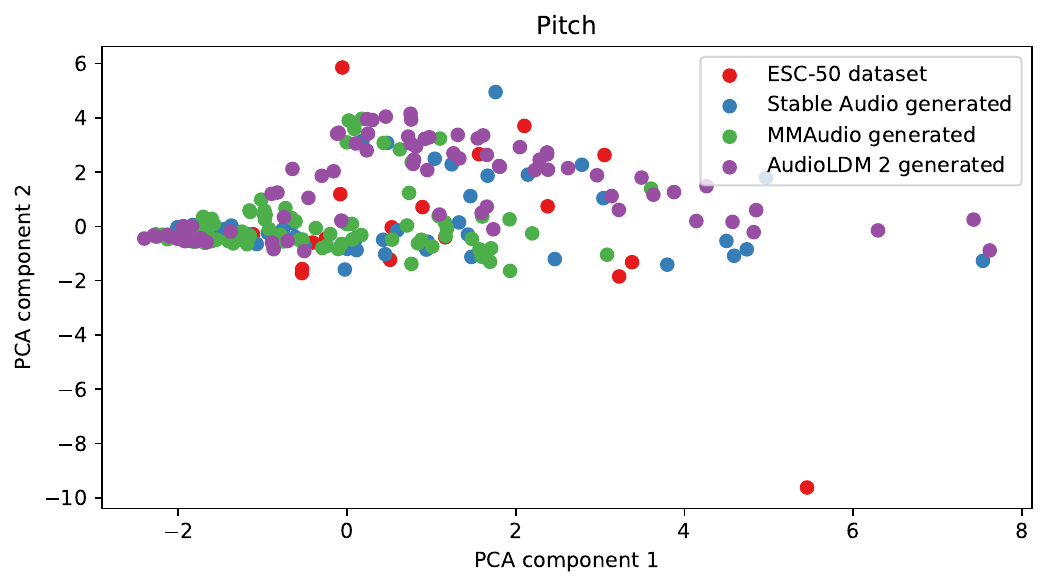}
\includegraphics[width=0.68\textwidth]{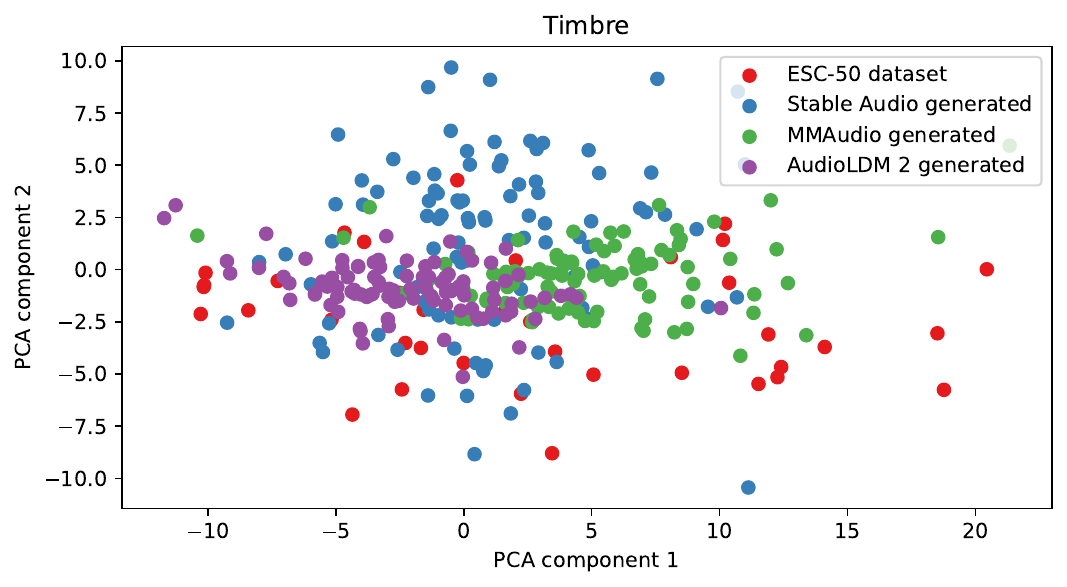}
\subsection{Prompt: Sound of snoring}\centering
\includegraphics[width=0.68\textwidth]{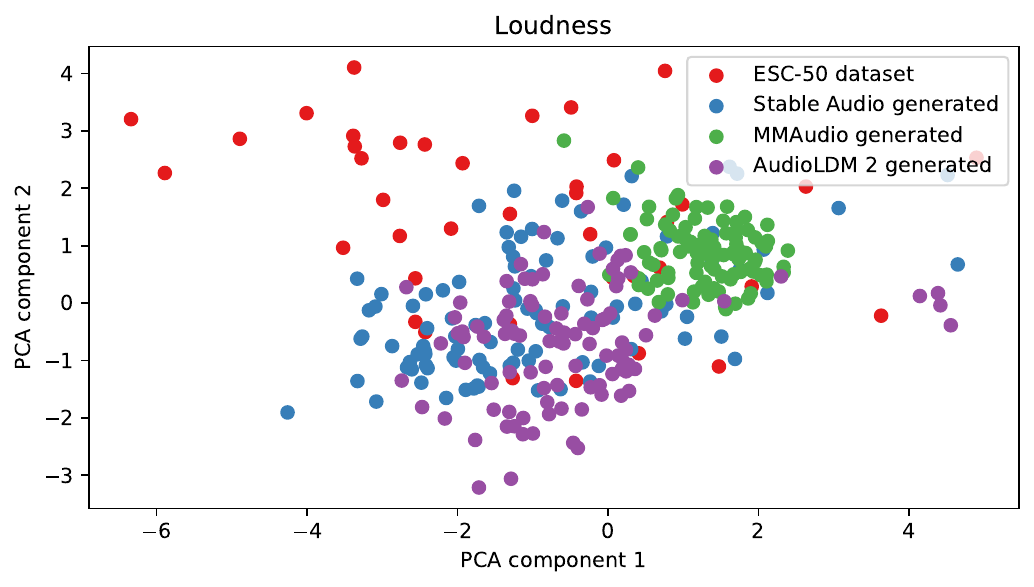}
\includegraphics[width=0.68\textwidth]{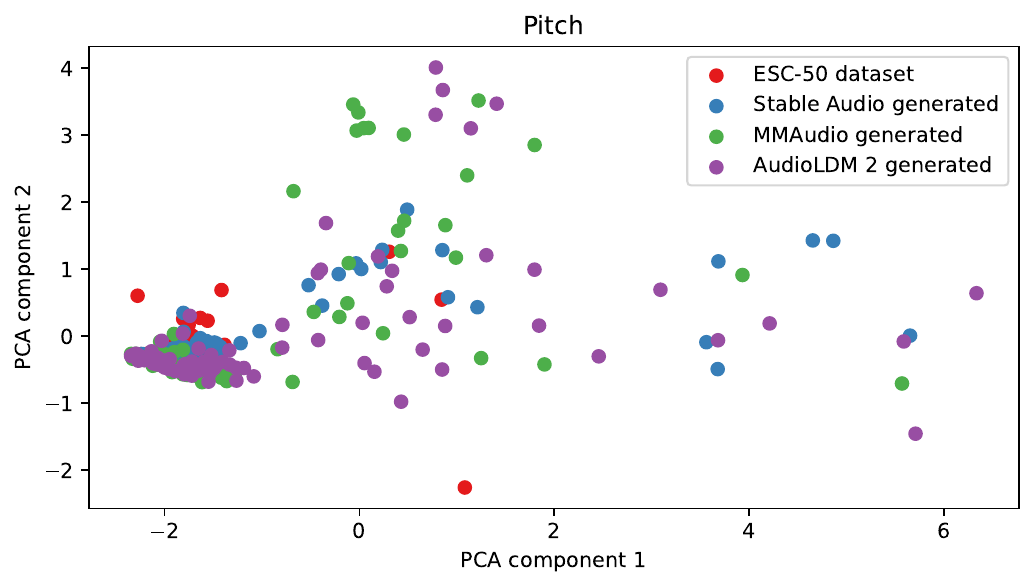}
\includegraphics[width=0.68\textwidth]{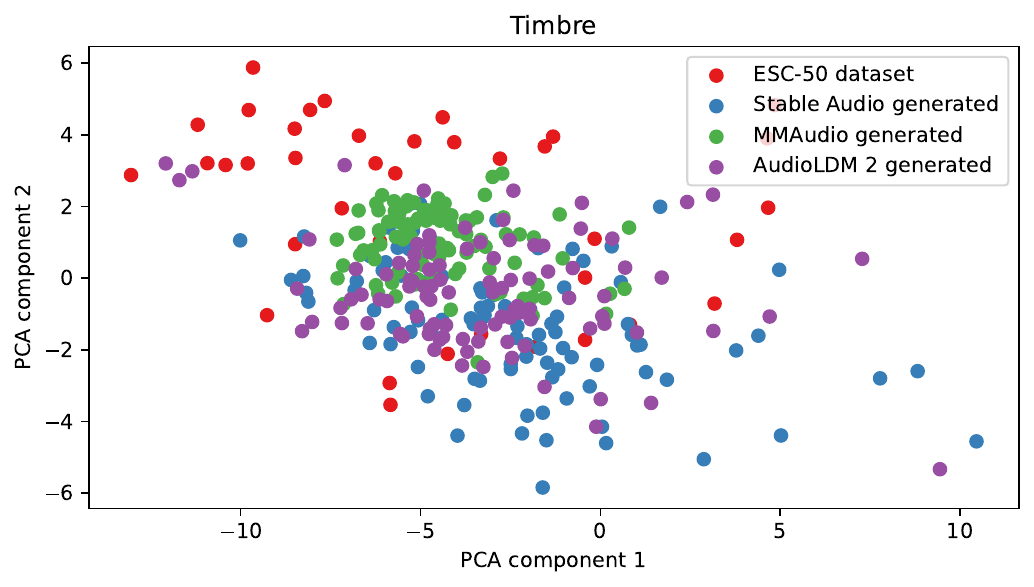}
\subsection{Prompt: Sound of thunderstorm}\centering
\includegraphics[width=0.68\textwidth]{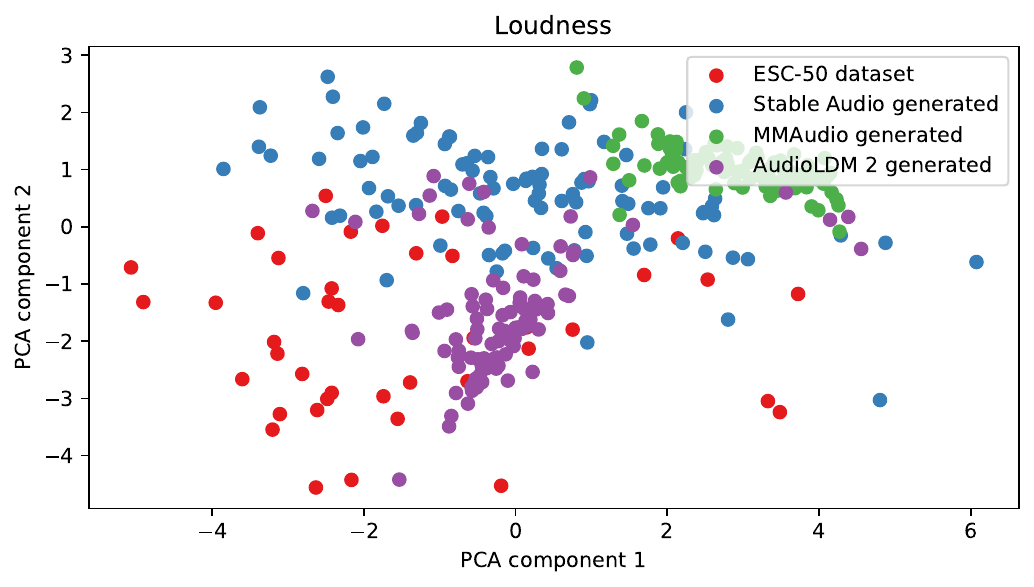}
\includegraphics[width=0.68\textwidth]{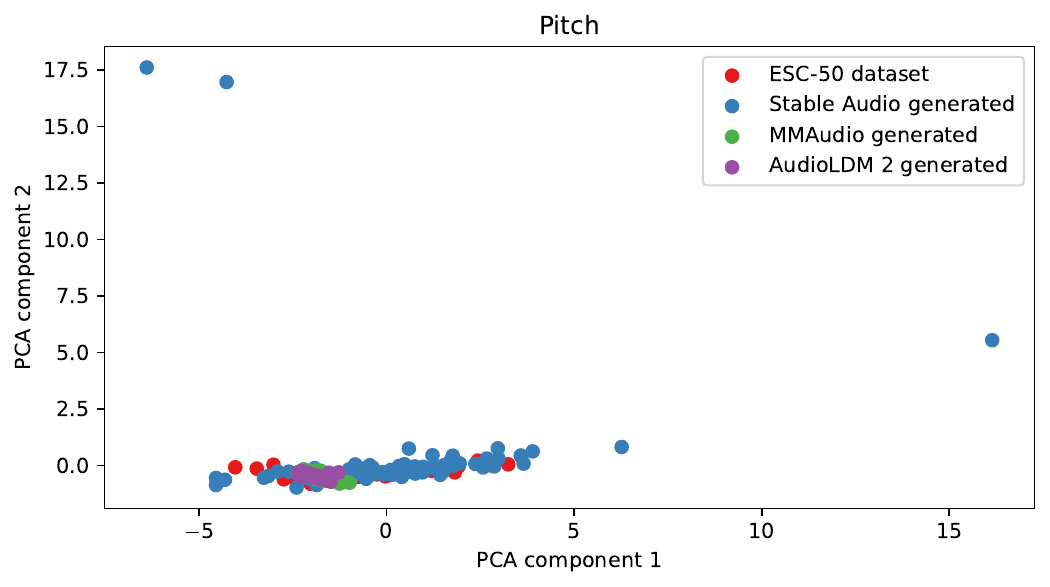}
\includegraphics[width=0.68\textwidth]{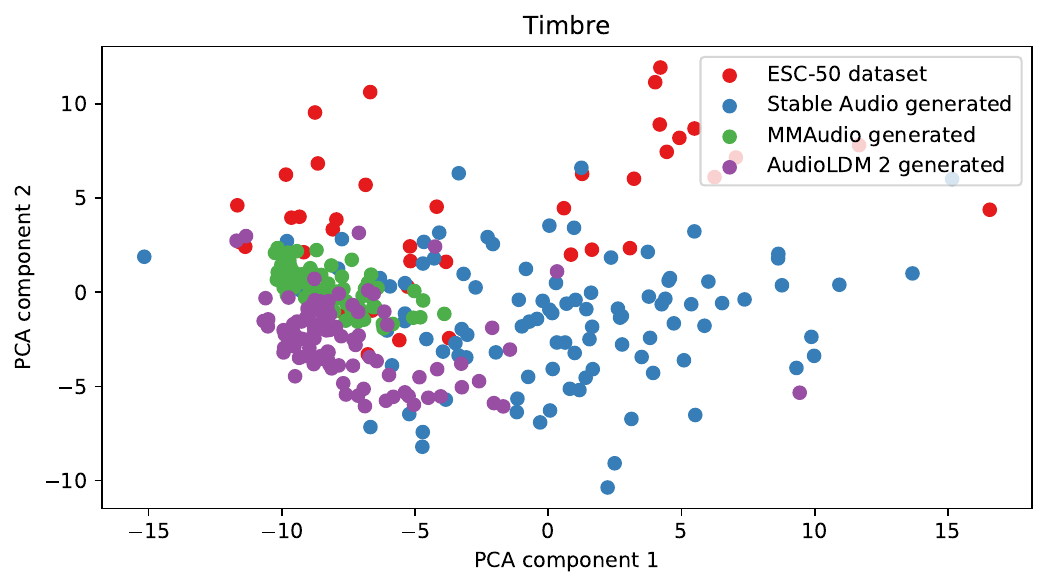}
\subsection{Prompt: Sound of toilet flush}\centering
\includegraphics[width=0.68\textwidth]{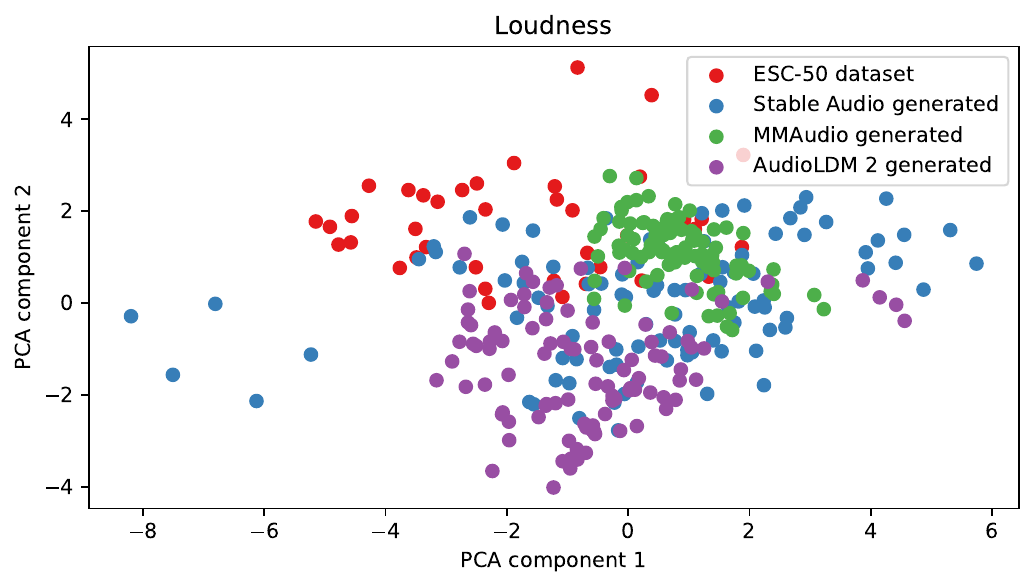}
\includegraphics[width=0.68\textwidth]{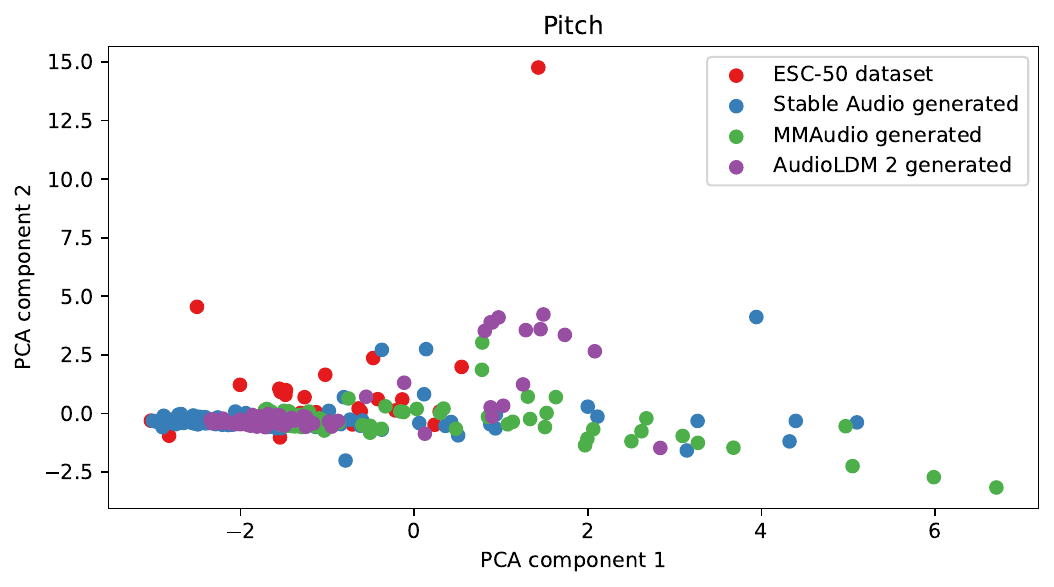}
\includegraphics[width=0.68\textwidth]{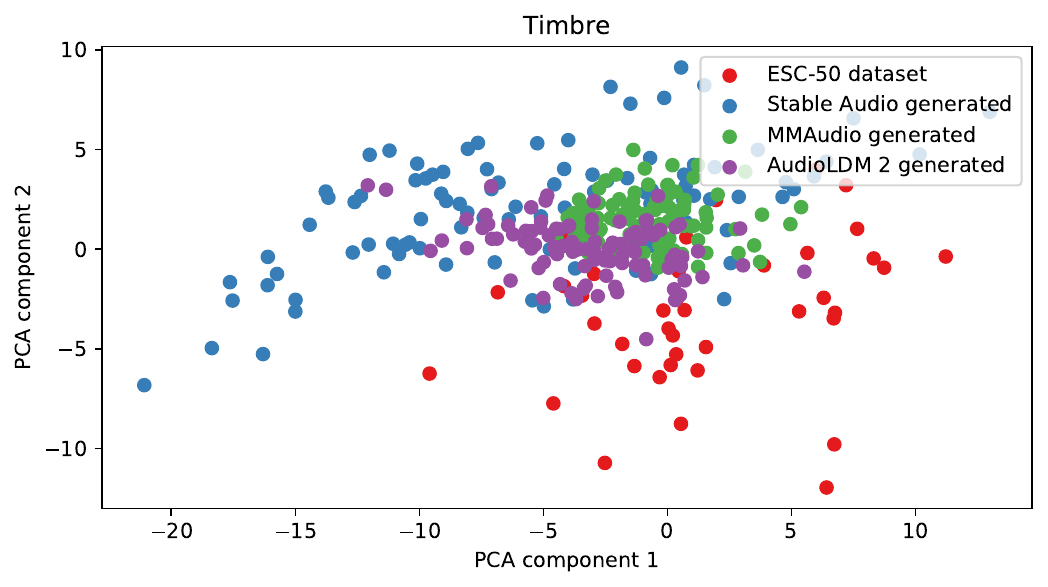}
\subsection{Prompt: Sound of train}\centering
\includegraphics[width=0.68\textwidth]{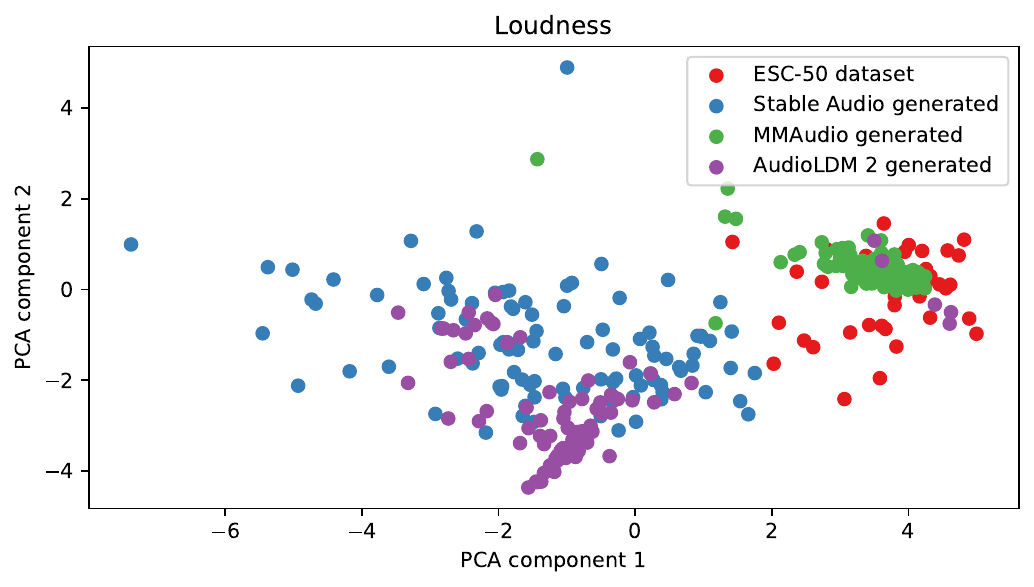}
\includegraphics[width=0.68\textwidth]{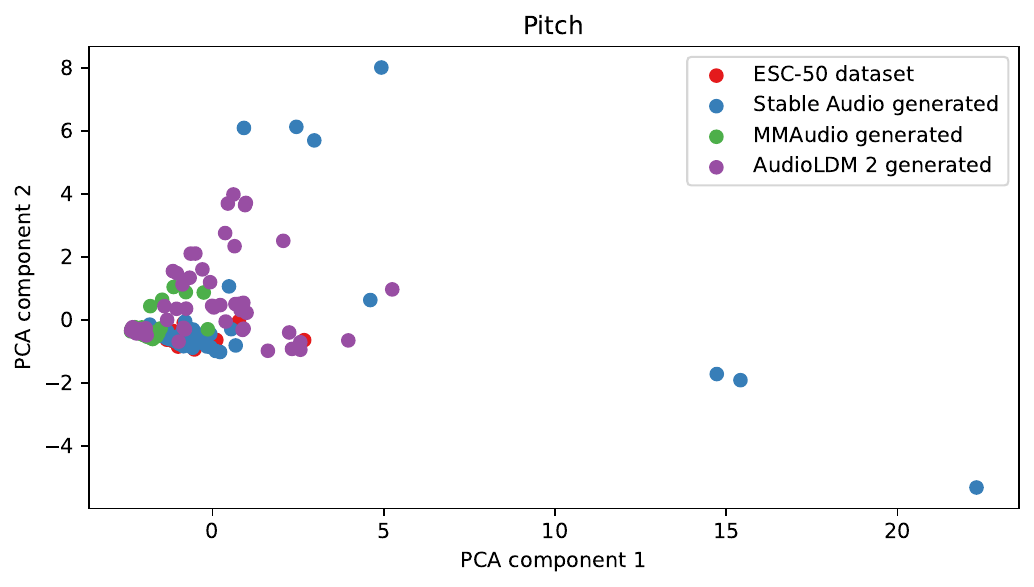}
\includegraphics[width=0.68\textwidth]{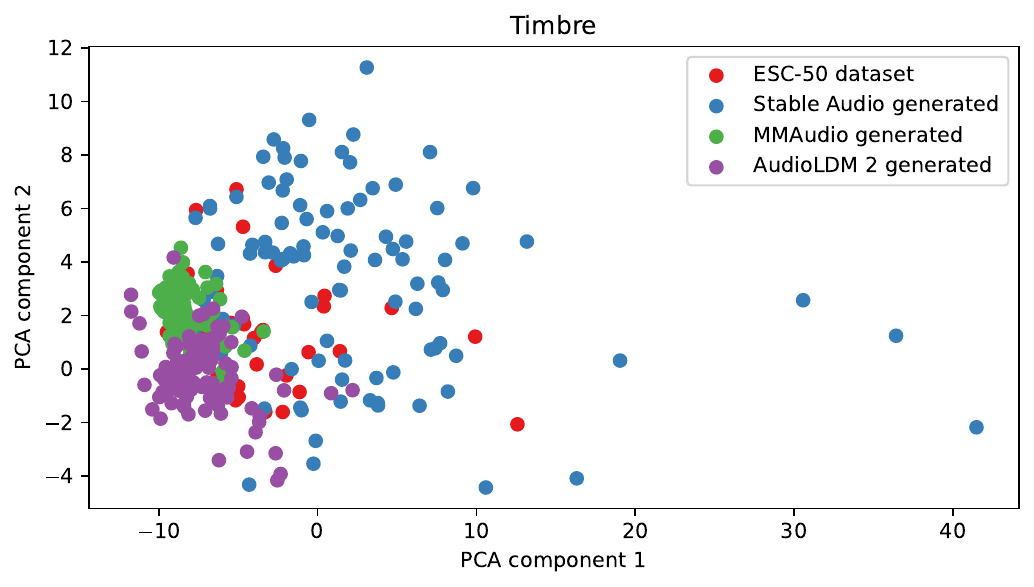}
\subsection{Prompt: Sound of vacuum cleaner}\centering
\includegraphics[width=0.68\textwidth]{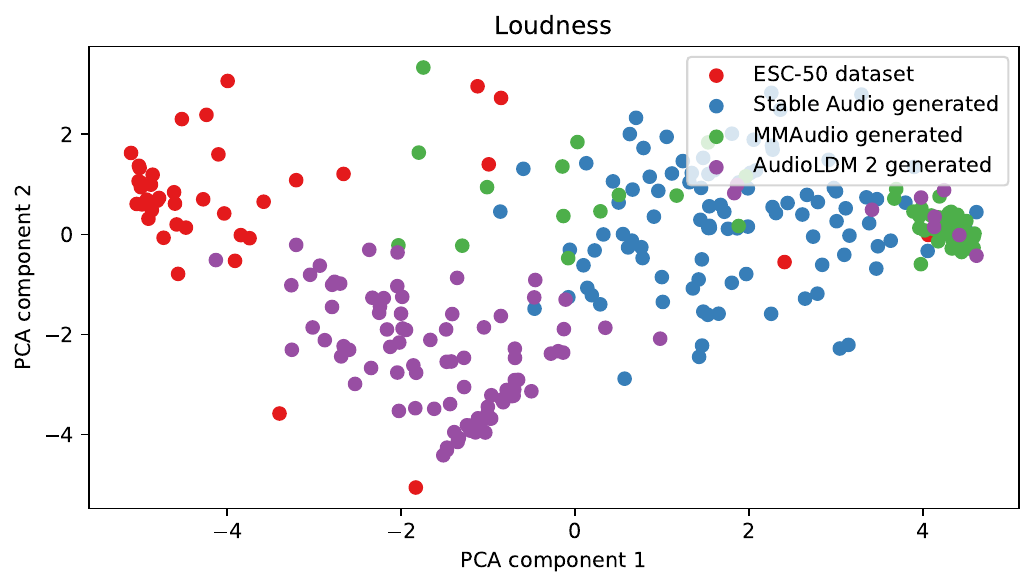}
\includegraphics[width=0.68\textwidth]{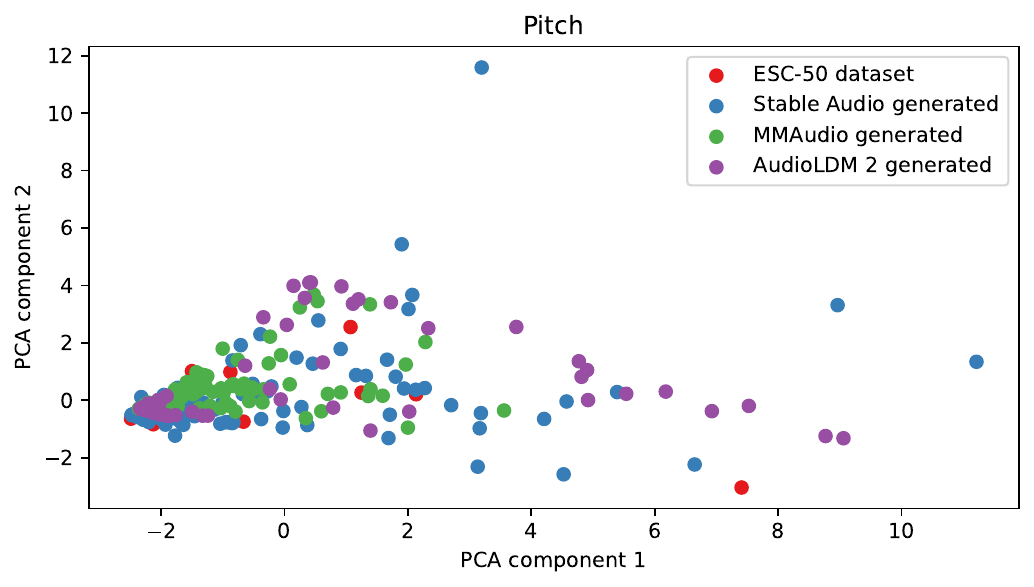}
\includegraphics[width=0.68\textwidth]{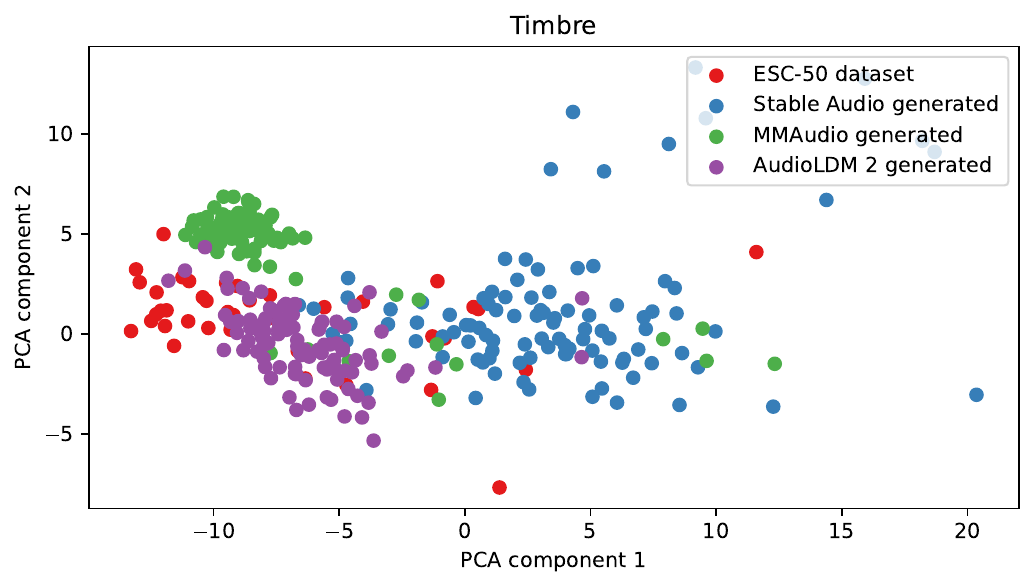}
\subsection{Prompt: Sound of washing machine}\centering
\includegraphics[width=0.68\textwidth]{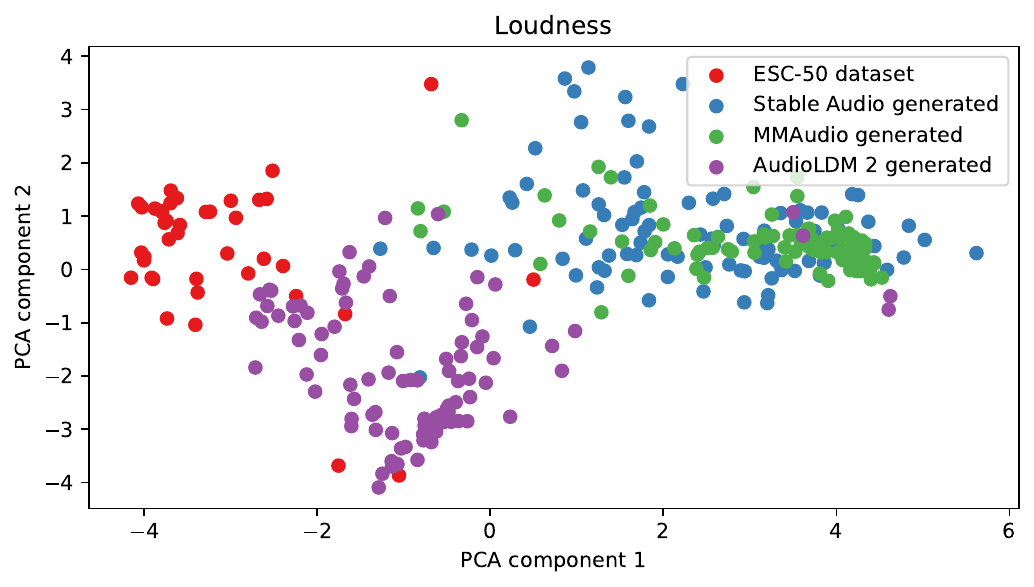}
\includegraphics[width=0.68\textwidth]{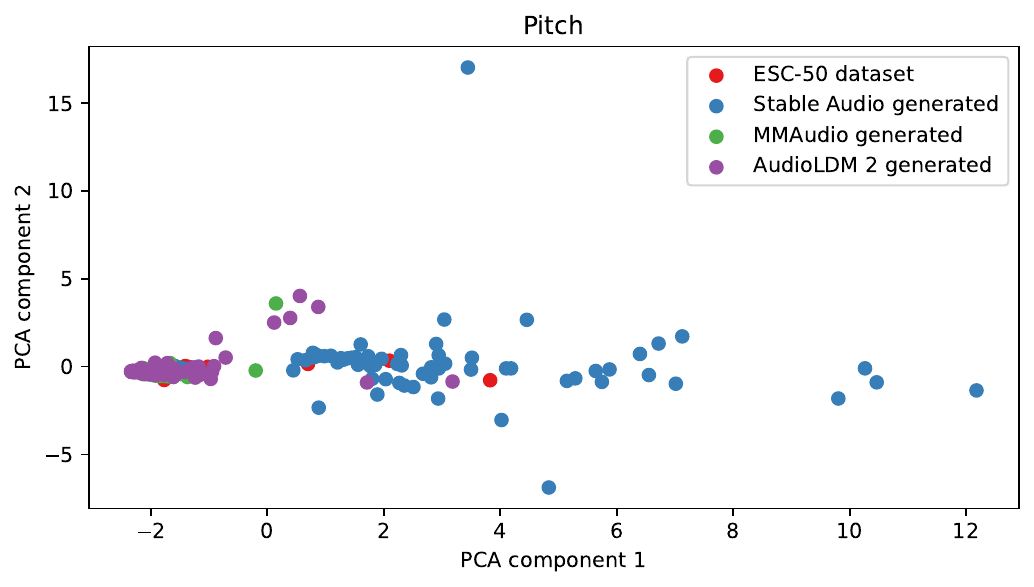}
\includegraphics[width=0.68\textwidth]{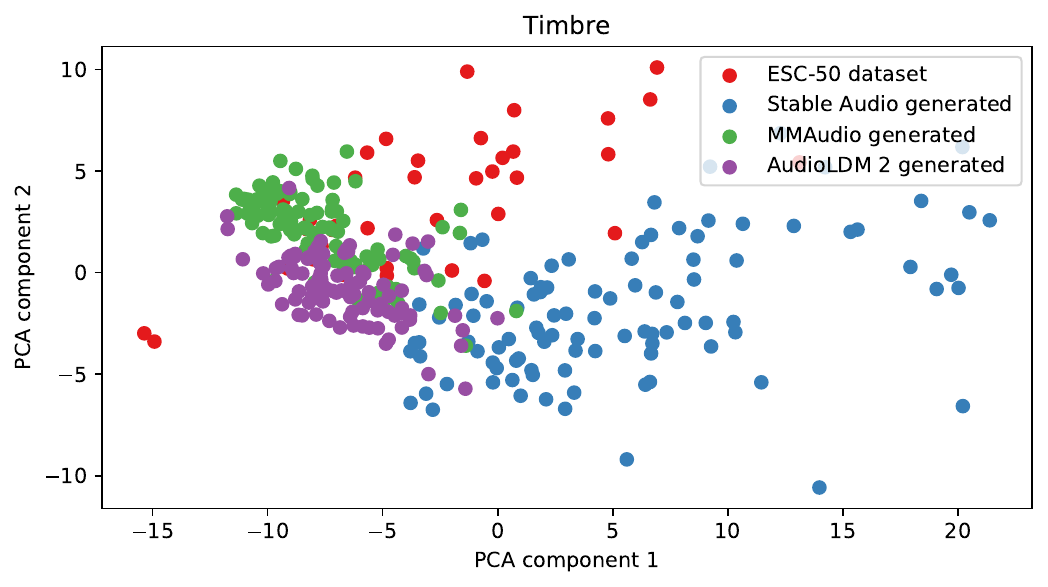}
\subsection{Prompt: Sound of water drops}\centering
\includegraphics[width=0.68\textwidth]{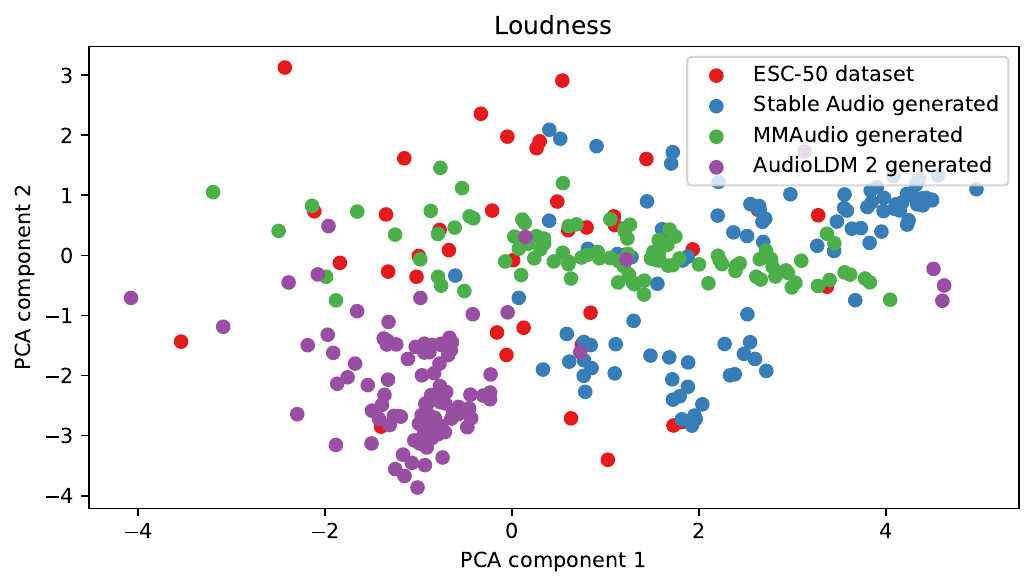}
\includegraphics[width=0.68\textwidth]{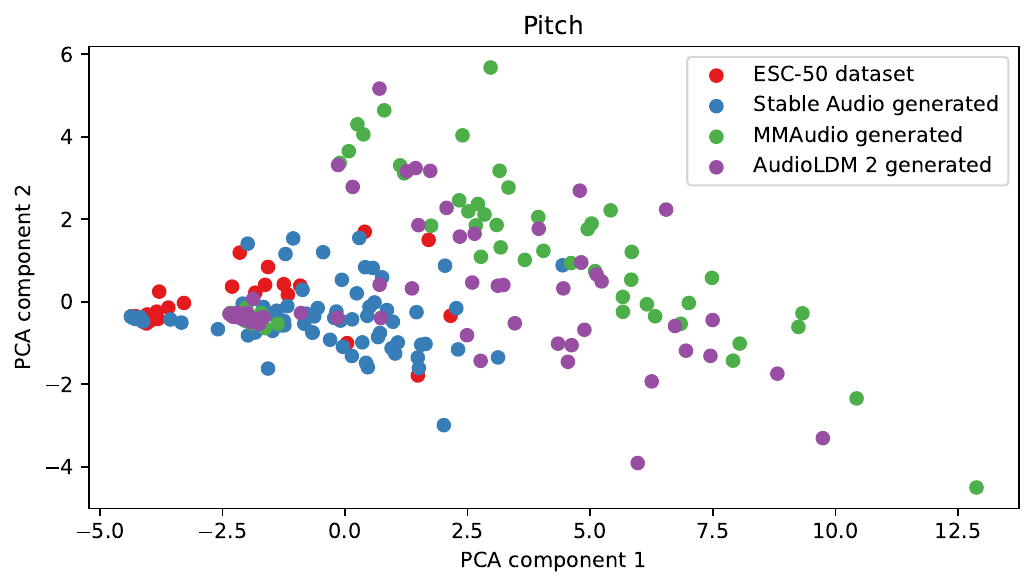}
\includegraphics[width=0.68\textwidth]{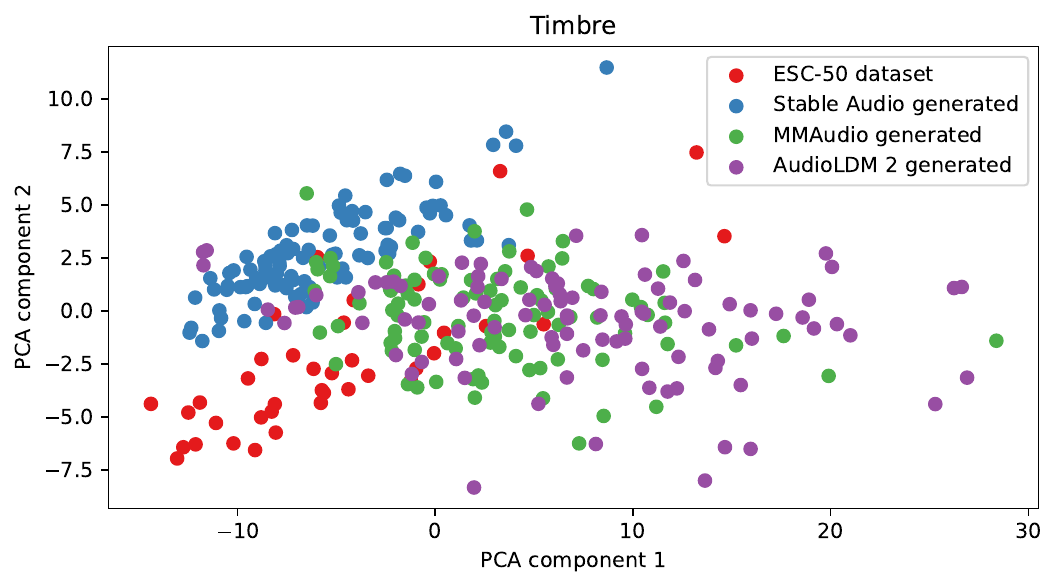}
\subsection{Prompt: Sound of wind}\centering
\includegraphics[width=0.68\textwidth]{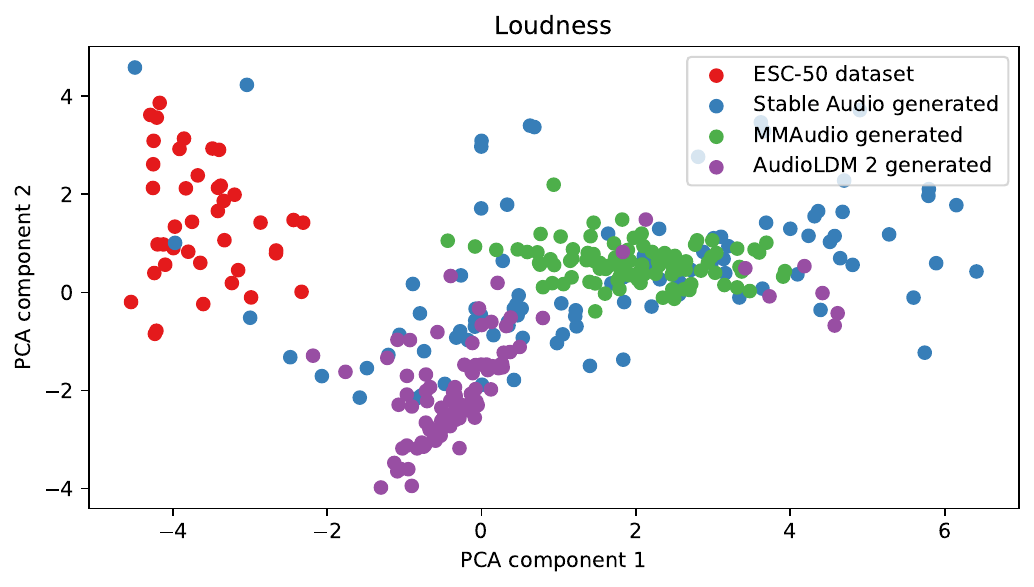}
\includegraphics[width=0.68\textwidth]{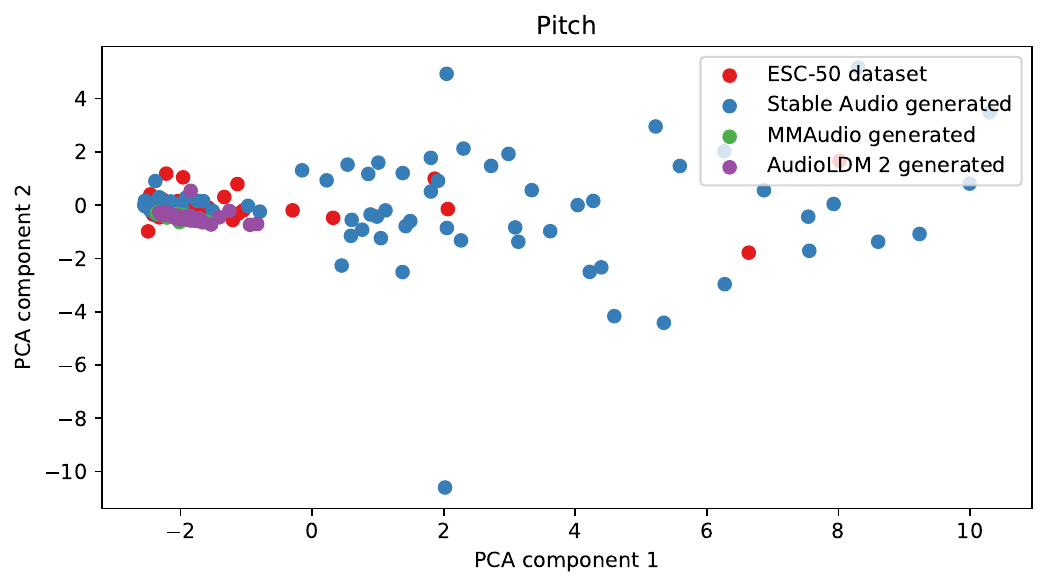}
\includegraphics[width=0.68\textwidth]{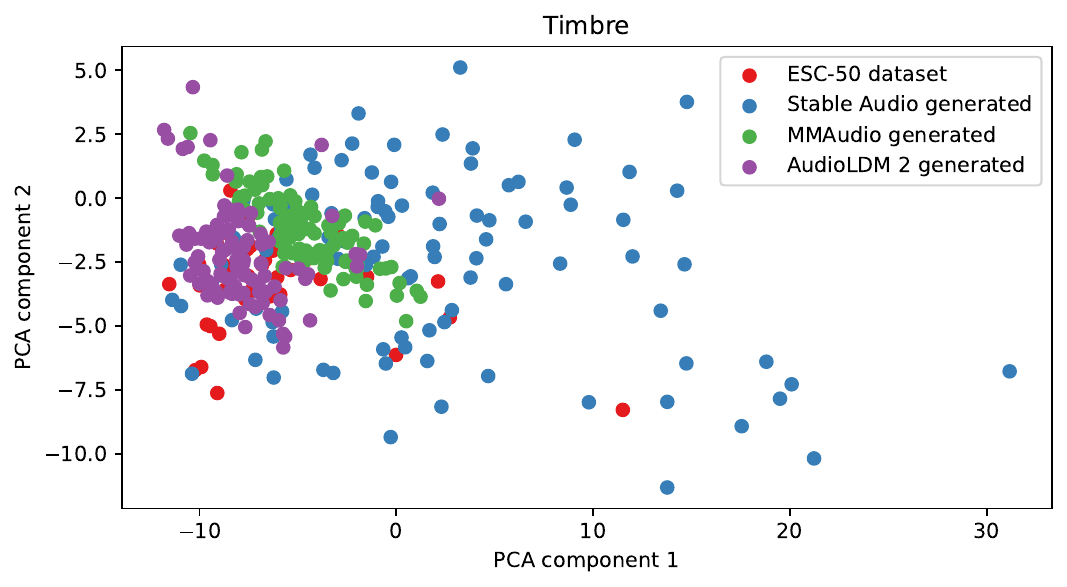}

\end{document}